%% file: ALEGRO18_rep_ADDENDUM.tex
\documentclass{cernrep} 
\usepackage{texnames}
\usepackage[T1]{fontenc}
\usepackage{units} 
\usepackage{texnames}
\usepackage[T1]{fontenc}
\usepackage[bookmarks, colorlinks=true, linktoc=page, pdftex, linkcolor=red, citecolor=red, urlcolor=red]{hyperref}
\input mymacros.tex

\begin{document}
\title{Towards an Advanced Linear International Collider}
 
\author{\textbf{ALEGRO collaboration}}
 

\begin{abstract}
This document provides additional information to support the ALEGRO proposal  for R\&D relevant to an Advanced Linear International Collider, ALIC, based on high gradient acceleration concepts. 
\end{abstract}


\maketitle 

\vspace{0.5cm}

\noindent \textbf{ Editing Board}
 
\noindent  Brigitte Cros, Patric Muggli, Carl Schroeder, Simon Hooker, Philippe Piot, Joel England, Spencer Gessner, Jorge Vieira, Edda Gschwendtner, Jean-Luc Vay,  Michael Peskin

\vspace{0.5cm}

\noindent  \textbf{ALEGRO collaboration members as of September 2018} :
Erik Adli$^1$, Weiming An$^2$, Nikolay Andreev$^3$, Oznur Apsimon$^4$, Ralph Assmann$^5$, 
Jean-luc Babigeon$^6$, Robert Bingham$^7$, Tom Blackburn$^8$, Christopher Brady$^9$, Michael Bussmann$^{10}$, 
Bruce Carlsten$^{11}$, James Chappell$^{12}$, Jian Bin Ben Chen$^{13}$, Sebastien Corde$^{14}$, Laura Corner$^{15}$, Benjamin Cowan$^{16}$, Brigitte Cros$^{17}$, 
Joel England$^{18}$, Eric Esarey$^{19}$, 
Ricardo Fonseca$^{20}$ , Brian Foster$^{5,21}$, 
Spencer Gessner$^{13}$, Leonida A Gizzi$^{22}$, Daniel Gordon$^{23}$ , Edda Gschwendtner$^{13}$, 
Anthony Hartin$^5$, Bernhard Hidding$^{24}$, Mark Hogan$^{18}$, Simon Hooker$^{21}$, T. Hughes $^{25}$, 
Alexei Kanareykin$^{26}$, Stefan Karsch$^{27}$, Valentin Khoze$^{28}$, Pawan Kumar$^{29}$, 
Wim Leemans$^{19}$,  Francois Lemery$^5$, Ang Li$^{30}$, R. Li$^{18}$, Vladyslav Libov$^5$, Emily Sistrunk Link$^{31}$, Michael Litos$^{32}$,  Gregor Loisch$^5$, Nelson Lopes$^{20,33}$, Olle Lundh$^{34}$, Alexey Lyapin$^{35}$, 
Edu Marin$^{13}$, Mattias Marklund$^8$, Timon Mehrling$^{19}$, Patric Muggli$^{13,27}$, Pietro Musumeci$^2$,
Zulfikar Najmudin$^{33}$,  Uwe Niedermayer$^{36}$, 
Jens Osterhoff$^5$, 
Marc Palmer$^{41}$, Rajeev Pattathil$^7$, Michael Peskin$^{18}$, Philippe Piot$^{38}$, John Power$^{39}$, Alexander Pukhov$^{40}$, 
Heather Ratcliffe$^{41}$, Marc Riembau $^{42}$,
Veronica Sanz$^{43}$, Gianluca Sarri$^{44}$, Yuri Saveliev$^{7}$, Levi Schachter$^{45}$, Lucas Schaper$^5$, Norbert Schoenenberger$^{30}$ , Carl Schroeder$^{19}$, Sarah Schroeder$^5$, Daniel Schulte$^{13}$, Andrei Seryi$^{46}$, Sergey Shchelkunov $^{56}$, Craig Siders$^{31}$, Evgenya Simakov $^{11}$,
Christophe Simon-Boisson$^{47}$, 
Michael Spannowsky$^{28}$, Christina Swinson$^{37}$, Andrzej Szczepkowicz$^{48}$, 
Roxana Tarkeshian$^{5}$, Johannes  Thomas$^{40}$,  Junping Tian$^{49}$, J.V. Tilborg$^{19}$, Paolo Tomassini$^{22}$, Vasili Tsakanov$^{50}$, 
Jean-Luc Vay$^{19}$, Jorge Vieira$^{20}$, Henri Vincenti$^{51}$, 
Roman Walczak$^{21}$, Dan Wang$^{52}$, Stephen Webb $^{53}$, Glen White $^{18}$
Guoxing Xia$^{4}$, 
Hitoshi Yamamoto$^{54}$, Tevong You$^{55}$, 
Igor Zagorodnov$^{5}$

\vspace{0.5cm}

\noindent 
$^1$ Univ Norway, Oslo, Norway\\
$^2$ UCLA, Los Angeles, California, USA\\
$^3$ IHED, Moscow, Russia\\
$^4$ Univ. Manchester, UK\\
$^5$ DESY, Hamburg, Germany\\
$^6$ LAL, Orsay, France\\
$^7$ STFC,  UK\\
$^8$ Chalmers, Sweden\\
$^9$ Warwick, UK\\
$^{10}$ HZDR, Germany\\
$^{11}$ LANL, Los Alamos, New Mexico, USA \\
$^{12}$ University College London, UK\\
$^{13}$ CERN, Geneva, Switzerland\\
$^{14}$ Ecole Polytechnique, Palaiseau, France\\
$^{15}$ Univ. Liverpool, UK\\
$^{16}$ Tech-X Corporation, Boulder, Colorado, USA\\
$^{17}$ CNRS LPGP, Orsay, France\\
$^{18}$ SLAC, Stanford, USA\\
$^{19}$ LBNL, Berkeley, USA\\
$^{20}$ IST, Lisbon, Portugal\\
$^{21}$ JAI and Dept of Physics, Univ. Oxford, Oxford, UK\\
$^{22}$ INO, Pisa, Italy\\
$^{23}$ NRL, USA\\
$^{24}$ Univ Strathclyde, Glasgow, UK\\
$^{25}$ Stanford Univ., Stanford, California, USA \\
$^{26}$ Euclid Tech labs, USA\\
$^{27}$ Max Planck Institute for Physics, Munich, Germany\\
$^{28}$ Univ. Durham, UK\\
$^{29}$ Raj Kumar Goel Institute of Technology, Ghaziabad,    India\\
$^{30}$ FAU, Germany\\
$^{31}$ LLNL, Livermore, California, USA \\
$^{32}$ Univ. of Colorado, Boulder, Colorado USA \\
$^{33}$ Imperial College London, UK\\
$^{34}$ University of Lund, Sweden\\
$^{35}$ RHUL, UK\\
$^{36}$ Tech Univ. Darmstadt,Darmstadt, Germany\\
$^{37}$ Brookhaven National Lab, USA\\
$^{38}$ North Illinois Univ. and Fermi National Accelerator Laboratory, USA\\
$^{39}$ ANL, USA\\
$^{40}$ Univ. Duesseldorf, Germany\\
$^{41}$ Warwick Univ, UK\\
$^{42}$ Universit\' e de Gen\`eve,  Switzerland \\
$^{43}$ Sussex Univ., UK\\
$^{44}$ Queen Univ. Belfast, UK\\
$^{45}$ Technion Israel Institute of Technology, Haifa, Israel\\
$^{46}$ JLAB, Newport news, VA, USA\\
$^{47}$ THALES LAS, France\\
$^{48}$ University of Wroclaw, Poland  \\
$^{49}$ Univ. Tokyo, Japan \\
$^{50}$ CANDLE SRI, Armenia     \\
$^{51}$ CEA Saclay , France \\    
$^{52}$ Tsinghua University, China  \\
$^{53}$ Radiasoft, Boulder, CO, USA \\
$^{54}$ Tohoku University , Japan  \\
$^{55}$ DAMTP, Univ. Cambridge, UK \\
$^{56}$ Yale University, New Heaven, CT, USA \\

\cleardoublepage

\chapter*{Introduction}

This document provides detailed information on the status of Advanced and Novel Accelerators techniques and describes the steps that need to be envisaged for their implementation in future accelerators, in particular for high energy physics applications.

It complements the overview prepared for the update of the European Strategy for particle physics,  and provides a detailed description of the field. The scientific priorities of the community are described for each technique of acceleration able to achieve accelerating gradient in the GeV~range or above.

ALEGRO working group leaders have coordinated the preparation of their working group contribution and contributed to editing the documents.

The preparation of this document was coordinated by the Advanced LinEar collider study GROup, ALEGRO. The content was defined through discussions at the ALEGRO workshop in Oxford UK, March 2018, and an advanced draft was discussed during a one day meeting prior to the AAC workshop in Breckenridge, CO, USA, August 2018.

This document was submitted as an addendum to the ALEGRO submission\footnote{arXiv:1901.08436} to the European Strategy for Particle Physics. %

\vspace{1cm}

\cleardoublepage

\newpage
\tableofcontents
\newpage

\chapter{ Physics considerations that motivate the parameters of the Collider energy and luminosity (30~TeV, $10^{36}$) (250~GeV), (1~TeV)}

\vspace{2cm}

\input{WG1physicsforALEGRO}
\cleardoublepage
 
 \chapter{ Laser Wakefield Acceleration}

\vspace{2cm}

\input{ALEGRO18_repWG4_LWFA}

 \cleardoublepage
 \chapter{Plasma Wakefield Acceleration}


\vspace*{2cm}



 \input{WG5PWFA}
 \cleardoublepage
 \chapter{Structure Wakefield Acceleration}








\vspace{2cm}

 \input{WG6SWFA}

 \cleardoublepage
 
\chapter{Dielectric Laser Acceleration}


 \vspace{2cm}

\input{WG7_DLA}

 \cleardoublepage
 \chapter{Positron Acceleration in Plasma: Challenges and Prospects}







\vspace{2cm}

 \input{WG8Positrons}
 \cleardoublepage
 \chapter{Theory, modelling and simulation}

\vspace{2cm}






 AAC concepts involve complex multiscale physics phenomena and will thus continue to require large scale high-performance simulation efforts. The infrastructure underpinning the simulation efforts will require sustained commitment to: 

\begin{itemize}
\item  recruitment and training of personnel in computational teams;
\item development of 'resource-saving' novel multi-scale models and numerical algorithms;
\item optimizations of those algorithms on constantly evolving computer architectures;
\item development of fast models guided by theory and Artificial Intelligence (AI) software fed by detailed simulation and experimental results.
\end{itemize}
 
The required level of design effort for a collider will drive development of enhanced simulation capabilities. With sustained support of simulation efforts, it is envisioned that computer hardware and software developments will reduce the time frame for detailed simulations of one collider stage from weeks to hours or even minutes on future exascale-capable supercomputers. By providing rapid turnaround time, high resolution with full physics packages enabled, simulations will transition from powerful tools used to understand and analyze our experiments into 'real-time' highly predictive tools used for the design and optimization of novel experiments with high fidelity. In addition, AI - in particular Machine Learning (ML) - will be useful to develop very fast models that can be used to guide large-scale parameter scans. 
 
The workhorse algorithm for AAC concepts modeling is the Particle-In-Cell (PIC) methodology, where beams and plasmas follow a Lagrangian representation with electrically charged macroparticles while electromagnetic fields follow a Eulerian representation on (usually Cartesian) grids. Exchanges between macroparticles and field quantities involve interpolation at specified orders. The standard 'full PIC' implementation is often too computationally demanding because of the large disparities of space and time scales between the driver beams (either laser or particle beam) and the plasma or structure. Speed-up is provided by either performing the simulation in a Lorentz boosted frame (which lowers the range of space and time scales) or using the quasistatic approximation to decouple fast and slow time scales. Additional approximations such as hybrid PIC-fluid, quasi-3D or laser envelope models enable additional savings at the cost of a reduction of domain of applicability. 
 
High-resolution, full three-dimensional standard PIC and quasistatic PIC (when self-injection is not involved) simulations are ultimately needed to capture potential hosing, misalignments, tilts and other non-ideal effects. Ensemble runs of simulations on large parameter space are required to estimate tolerances to those effects as well as study various designs. Yet, high-fidelity modeling of single stages in the 10-GeV range may necessitate several days or weeks on existing supercomputers, while modeling a 1 TeV ALIC collider will require the modeling of tens of 10 GeV stages. Hence, it is essential to pursue the development of better algorithms that improve the accuracy of existing plasma physics models (e.g. high-order Maxwell solvers, adaptive time-stepped particle pushers, adaptive mesh refinement, control of numerical instabilities) as well as to port the codes to the next generation of massively parallel supercomputers with multi-level parallelism. It will also be essential to develop efficient vectorization, threading and load balancing strategies for CPU-manycore (such as Intel KNL), GPU or other novel architectures that might arise. Interoperability of codes on the various computer architectures will be essential to reduce the cost and complexity of code development and maintenance.

PIC-based models are also crucial to (a) explore exotic wakefields with non-trivial topologies and geometries that may become pivotal to address the ongoing challenges of plasma accelerators (e.g. high quality positron acceleration, spin polarization) and (b) explore innovative approaches that exploit intrinsic and unique properties of plasma wakefields to pave novel pathways towards relativistic beams with unprecedented properties. 

In addition to the detailed PIC-based models, it is also important to develop fast tools that require far less computational resources and  that can be used to guide the parameter scans. The models used in these tools will be guided by theory and fits to the PIC-based simulations and experimental results. For the latter, the AAC community should start to take advantage of AI-ML and develop models based on accumulated data from simulations and experiments with advanced AI-ML algorithms. 

For PWFA/LWFA, the priorities for the modeling of the plasma physics for a single stage will be to (a) demonstrate that reduced models can be used to make accurate predictions for a single GeV-TeV stage (in the linear, mildly nonlinear and nonlinear regimes), (b) demonstrate that the PIC algorithm can systematically reproduce current experimental results (coordinated efforts with experiments) and (c) conduct tolerance studies and role of misalignments. It is also essential to integrate physics models beyond single-stage plasma physics in the current numerical tools, incorporating: plasma hydrodynamics for long term plasma/gas evolution and accurate plasma/gas profiles, ionisation \& recombination, coupling with conventional beamlines, production of secondary particles, QED physics at the interaction point, spin polarization. For DLA/SWFA, the priorities for the development of physics models for 3D photonic structures will be to (a) determine the material response to 10's as long electron bunches, (b) develop computational models based on material science experiments, (c)  include material erosion into available models and (d) establish automated and non-invasive mechanisms to determine which photonics structures are useful for acceleration after being produced.

The development of simulation tools needed for the design of an advanced multi-TeV collider requires robust and sustained team efforts. Teams will consist of individuals with various skills and responsibilities: code builders/developers, maintainers, and users. Members should have combined expertise in physics, applied mathematics and computational science. A multiple-team/code approach has many benefits, allowing for cross-checking of results and pursuit of varied approaches. Yet, while a multiplicity of teams and codes has positive effects, excessive duplications of the same or similar capabilities comes at a cost. Hence, it is important to foster the sharing of modules and codes interoperability via the development of libraries of algorithms and physics modules, as well as the development of standards for simulations input/output and for data structures. To enhance exchanges, interoperability, co-development and verification, the development of open source codes and libraries should also be encouraged.

%
%
%
%
%
%
%

%
\end{document}

%% file: mymacros.tex
\def\beq{\begin{equation}}
\def\eeq#1{\label{#1}\end{equation}}
\def\eeqn{\end{equation}}

\newenvironment{Eqnarray}%
   {\arraycolsep 0.14em\begin{eqnarray}}{\end{eqnarray}}
\def\beqa{\begin{Eqnarray}}
\def\eeqa#1{\label{#1}\end{Eqnarray}}
\def\eeqan{\end{Eqnarray}}
\def\CR{\nonumber \\ }

\let\bar=\overbar

\def\etal{{\it et al.}}

\def\lsim{\mathrel{\raise.3ex\hbox{$<$\kern-.75em\lower1ex\hbox{$\sim$}}}}
\def\gsim{\mathrel{\raise.3ex\hbox{$>$\kern-.75em\lower1ex\hbox{$\sim$}}}}

\def\half{\frac{1}{2}}

\def\del{\partial}
\def\Dslash{\not{\hbox{\kern-4pt $D$}}}
\def\dslash{\not{\hbox{\kern-2pt $\del$}}}

\def\Dlr{\mathrel{\raise1.5ex\hbox{$\leftrightarrow$\kern-1em\lower1.5ex\hbox{$D$}}}}

\def\Pl{{\mbox{\scriptsize Pl}}}

\def\ee{e^+e^-}
\def\sstw{\sin^2\theta_w}

\def\msb{{\bar{\scriptsize M \kern -1pt S}}}

\def\drb{{\bar{\scriptsize D \kern -1pt R}}}

\makeatletter
\def\section{\@startsection{section}{0}{\z@}{5.5ex plus .5ex minus
 1.5ex}{2.3ex plus .2ex}{\large\bf}}
\def\subsection{\@startsection{subsection}{1}{\z@}{3.5ex plus .5ex minus
 1.5ex}{1.3ex plus .2ex}{\normalsize\bf}}
\def\subsubsection{\@startsection{subsubsection}{2}{\z@}{-3.5ex plus
-1ex minus  -.2ex}{2.3ex plus .2ex}{\normalsize\sl}}

\renewcommand{\@makecaption}[2]{%
   \vskip 10pt
   \setbox\@tempboxa\hbox{\small #1: #2}
   \ifdim \wd\@tempboxa >\hsize     
       \small #1: #2\par          
     \else                        
       \hbox to\hsize{\hfil\box\@tempboxa\hfil}
   \fi}

 \def\citenum#1{{\def\@cite##1##2{##1}\cite{#1}}}
 
\newcount\@tempcntc
\def\@citex[#1]#2{\if@filesw\immediate\write\@auxout{\string\citation{#2}}\fi
  \@tempcnta\z@\@tempcntb\m@ne\def\@citea{}\@cite{\@for\@citeb:=#2\do
    {\@ifundefined
       {b@\@citeb}{\@citeo\@tempcntb\m@ne\@citea\def\@citea{,}{\bf ?}\@warning
       {Citation `\@citeb' on page \thepage \space undefined}}%
    {\setbox\z@\hbox{\global\@tempcntc0\csname b@\@citeb\endcsname\relax}%
     \ifnum\@tempcntc=\z@ \@citeo\@tempcntb\m@ne
       \@citea\def\@citea{,}\hbox{\csname b@\@citeb\endcsname}%
     \else
      \advance\@tempcntb\@ne
      \ifnum\@tempcntb=\@tempcntc
      \else\advance\@tempcntb\m@ne\@citeo
      \@tempcnta\@tempcntc\@tempcntb\@tempcntc\fi\fi}}\@citeo}{#1}}
\def\@citeo{\ifnum\@tempcnta>\@tempcntb\else\@citea\def\@citea{,}%
  \ifnum\@tempcnta=\@tempcntb\the\@tempcnta\else
  {\advance\@tempcnta\@ne\ifnum\@tempcnta=\@tempcntb \else\def\@citea{--}\fi
    \advance\@tempcnta\m@ne\the\@tempcnta\@citea\the\@tempcntb}\fi\fi}
\makeatother

%% file: WG1physicsforALEGRO.tex
%
%
%
%
%
%



\section{Theoretical background}

The situation of elementary particle physics today is  a difficult 
one.  We have a Standard Model of Particle Physics (SM) that seems to
describe 
all observations at  high-energy colliders, including measurements of
complex reactions at the Large Hadron Collider (LHC)  involving
quarks, 
leptons, gluons, and $W$ and $Z$
bosons.   However, this model is manifestly incomplete.  It does not
account for important aspects of fundamental physics observed in
cosmology, including the dark matter and dark energy and the excess of
matter of antimatter.  And, important features of the model, including
the spectrum of quark and lepton masses and the presence of a phase
transition that breaks its gauge symmetry,  are put in by hand rather than
being explained from physics principles.

The experimental results from the LHC have contributed to this
understanding, both positively and negatively.   Among the
positive contributions, the LHC has demonstrated that the SM applies
to high accuracy in quark- and gluon-initiated reactions up to
energies of  several TeV.   Further, the LHC experiments have
discovered the last particle predicted by the SM, the  Higgs boson.
This discovery, and the measurement of the Higgs boson mass, opens
the door to the study of this last crucial element of the SM  and defines the
parameters for high-precision measurements of the properties of the Higgs
particle.

On the negative side, the LHC experiments have carried out  extensive
searches for new particles associated with models of fundamental
physics  that generalize the SM.  Thus far, no new
particles have been found.  Though there is still space to explore
before reaching the ultimate limits of the LHC capabilities, the
window is closing.  

Before the start of the LHC experiments, many theorists found
supersymmetry an attractive idea to fill the gaps in the SM.  They hoped for the
discovery of supersymmetric particles with masses below 1 TeV, the
natural scale for a supersymmetric explanation of the Higgs field
potential,  and even projected  a
``desert'' extending above this mass scale to the grand unification
scale at $10^{16}$~GeV.  The
current limits on the masses of supersymmetric particles and the
observed value of the Higgs boson mass exclude this range of
parameters for supersymmetry.   Today, most viable supersymmetric
models put  the supersymmetric partners of quarks and leptons  at
multi-TeV energies~\cite{Poh:2015wta,Baer:2017pba,Chakraborty:2018izc}
or even, in 
``split SUSY'' models, energies as large as 
100~TeV~\cite{Arvanitaki:2012ps,ArkaniHamed:2012gw}.   
The hope to discover supersymmetric or other weakly coupled particles
at the LHC now focuses on states with only electroweak interactions.
These particles have much lower cross sections and, often, difficult signatures
in LHC experiments.

At the same time, the LHC exclusions have shifted attention to a
different class of models that extend the SM.   In these models, the 
Higgs boson is a composite state bound by new strong interactions whose
mass scale is 10~TeV
or above~\cite{Bellazzini:2014yua,Csaki:2015hcd}.
If a model of this type is the correct one, the new strong dynamics would be most
likely be
different from that in QCD and its elucidation would require many new
clues from experiment.

The results from the LHC have made the mysteries at the center of
particle physics sharper and more mysterious.   These results have
also moved higher our estimates of  the energy scales  at which these
mysteries can be resolved. There are immediate next steps to take to
study these questions.   But also, we need to confront the idea
that a full understanding will require concerted experiments at new
accelerators well above the energy of the LHC, and to plan a program
to realize them. 

In this report, we will explore the opportunities for $\ee$ and
$\gamma\gamma$ colliders from the 100~GeV energy scale to the 100~TeV
energy scale.   The physics considerations we have just reviewed
imply that these accelerators have important tasks at the
100~GeV energy scale in making precision measurements of the heaviest
particles of the SM.   They also imply that we need a route to the
construction of accelerators of much higher energies that have the power
to probe the electroweak interactions at those energy scales.   We
will explain how colliders based on advanced accelerator technology
can address both sets of challenges.

\subsection{$\ee$ and $\gamma\gamma$ physics from the $Z$ to TeV energies}

In the remainder of this chapter, we will discuss the physics
objectives of future $\ee$ and $\gamma\gamma$ colliders.  This section
will cover the energy region up to a few TeV in the center of mass.
The following sections will discuss physics in the energy region of
tens of TeV. 
Because some of the modes of
acceleration discussed in this report (in particular, plasma wakefield
accelerators) are not particle-antiparticle symmetric and the problem
of accelerating positrons in such accelerators is not solved, it is
interesting to consider $\gamma\gamma$ colliders, which require only
$e^-e^-$ acceleration, in addition to $\ee$ colliders.  We will point
out physics goals for which $\ee$, as opposed to $\gamma\gamma$
colliders, are essential.  We note that both $\ee$ and $\gamma\gamma$
linear colliders naturally allow large longitudinal beam
polarization.  This is an important physics tool for both types of machine.

There are many studies available that detail the physics opportunities
of $\ee$ colliders at energies  up to 3~TeV.   The $\ee$ physics
opportunities at the $Z$ resonance, the $WW$ threshold, and the peak
of the $\ee\to Zh$ cross section at 240-250~GeV have been described in
the CEPC Pre-CDR \cite{CEPCpre} and the FCC-ee reports
\cite{dEnterria:2016sca,Gomez-Ceballos:2013zzn}.
Physics opportunities for $\ee$ and $\gamma\gamma$ colliders in the
energy range 250~GeV--1~TeV have been described in the ILC
TDR~\cite{Behnke:2013xla,Baer:2013cma}. 
 The $\ee$ prospects up to 500~GeV have been updated more
recently in the reports  \cite{Fujii:2015jha,Fujii:2017vwa}.   The
physics prospects for an $\ee$  collider with up to 3~TeV in the
center of mass has
been documented in the CLIC Conceptual Design
Report~\cite{Linssen:2012hp}, with the Higgs boson prospects updated
more recently in 
\cite{Abramowicz:2016zbo}.  

The most important physics goals discussed in this literature are the following:

\begin{itemize}

\item {\bf Higher-precision measurement of the mass and couplings of
    the $Z$ boson:}   At the moment, our precision knowledge of the
  properties of the $Z$ boson is generally accurate to the third
  decimal place.   Most of the limitations are statistical, so
  improvement by another factor of 10 would require an increase of 100
  in the total number of $Z$ bosons studied.   Since the LEP 1 program
  accumulated 20 M  $Z$ bosons, the goal for a next stage precision
  electroweak  program would be 1 billion ($10^9$) $Z$ bosons or more
  (``GigaZ'').    The CEPC and FCC-ee reports envision collecting $10^{12}$
  $Z$ bosons (``TeraZ'').  

There are still gaps that limit the current power of precision
electroweak measurements.  The  two most important are the values of the
$b$ and   leptonic couplings of the $Z$.   There is a longstanding 3
$\sigma$ discrepancy between the values of the leptonic polarization asymmetry $A_\ell$ or the associated value of $\sstw$ measured from
leptonic observables and from the $b$ quark forward backward asymmetry 
at the $Z$,  $A_{BF}(b)$~\cite{ALEPH:2005ab}.
This could result from an anomaly in the $b_R$ coupling
to the $Z$. The issue needs to be resolved experimentally.  The quantity $R_b =
\Gamma(Z\to b\bar b)/\Gamma(Z \to \mbox{hadrons})$ is now known to
0.5\% accuracy. 
This is a theoretically important quantity that tests the coupling of
the $b_L$   quark (the $SU(2)$ partner of the $t_L$) to a strongly
interacting Higgs sector.  The measurement of $R_b$ also limits
deviations from the SM value  in the $b_R$ couplings.  Ideally, three quantities---$A_{FB}(b)$, 
$R_b$, and $A_b$, the polarization asymmetry in $Z$ decay to
$b$---should be 
remeasured and analyzed
jointly~\cite{Takeuchi:1994zh}.  All of these issues can be addressed straightfowardly 
by
increasing the total sample of $Z$ bosons.

Improving the overall precision of $Z$ pole measurements can improve
the power of these measurements to constrain (or discover) physics
beyond the SM.  A factor 10 improvement in the current situation would
require computation of the theoretical SM predictions at the 2-loop
level, but fortunately this has just been
completed~\cite{Dubovyk:2018rlg}.
The extension of these calculations to the 3-loop level, as would be
required for a TeraZ program, seems feasible but is beyond the current
state of the art.  The report \cite{Blondel:2018mad} is optimistic about achieving this in the next decade with a dedicated effort.

The improvement of precision electroweak measurements
requires $\ee$ beams to produce the $Z$ boson as a
resonance.   Certain observables with high sensistivity to the
electroweak parameters, in particuar,  the polarization
asymmetries in $Z$ decay $A_b$, $A_c$, and $A_\ell$, can be measured
directly only with polarized beams.

\item {\bf Precision measurement of the $W$ boson mass at the $WW$
    threshold:}   The $W$ boson mass is another observable with an
  important role in 
  the precision electroweak fit to data.  Currently, this mass is
  known to $0.2\%$ precision.   Measurements from the LHC have the
  potential to improve this precision by a factor of 3, and perhaps
  study of $W$ boson  production at an $\ee$ or $\gamma\gamma$
  collider can contribute a further factor of 2~\cite{Baak:2013fwa}.
  A possible strategy for a significantly better measurement would be
  to devote 1~ab$^{-1}$ or more of integrated luminosity to a
  measurement of the $WW$ threshold region~\cite{dEnterria:2016sca}.
Since the accuracy of this measurement derives from precision
knowledge of the beam energies, this method requires an $\ee$ collider.

\item {\bf  Precision measurement of couplings associated with Higgs
  boson decays:}   The Higgs boson couplings to $W$, $Z$,
$b$, $\tau$, $c$, $\mu$, $\gamma$ can be measured in Higgs decay. 
  It is now understood that
measurements of $\ee\to Zh$ at 250~GeV with 2~ab$^{-1}$ and beam polarization (or
5~ab$^{-1}$ without beam polarization) can measure the first three of
these couplings in a model-independent way to 1\% accuracy or
 better.
The search for deviations from the SM predictions in these couplings
gives a new window into physics beyond the SM, orthogonal to
that of particle searches at the LHC. This is a must-do for particle
physics.  It is important to find an affordable technology and begin
this program as soon as possible.   Any anomalies discovered in this
program could be confirmed by running at 500~GeV, which would 
add measurements of $W$ fusion production of the Higgs boson
and cut the individual coupling  errors in
half~\cite{Barklow:2017suo}.   

The measurement of the absolute scale of Higgs boson couplings
requires that the Higgs bosons be produced in a way that their
production is tagged by the recoil $Z$ and so the observation of a
Higgs event does not depend on the final state of Higgs decay.   This
requires an $\ee$ initial state.   Some particular couplings of the
Higgs boson---in particular, the couplings to $b$ and to
$\gamma$---can be measured at the Higgs resonance in $\gamma\gamma$
collisions, but only as a measurement of the set of $\sigma \times BR$ for
$\gamma\gamma$ production and decay~\cite{Asner:2001ia}.  At higher
energies, it may be possible to measure more of the Higgs boson
couplings using the $WW$ fusion subprocess of the $\gamma\gamma$ cross
section.  This reaction needs further study.

\item {\bf Search for exotic decays of the Higgs boson:}   The Higgs
  boson may couple to new particles with no SM gauge interactions,
  including the particle that makes up dark matter.   Measurements in
  $\ee$ at 
  250~GeV allow comprehensive searches for possible exotic modes of
  Higgs decay~\cite{Liu:2016zki}.

\item {\bf Precision measurement of the remaining couplings of the
    Higgs boson:}   Two important Higgs boson couplings require higher
  energy experiments.   These are the Higgs boson coupling to the top
  quark and the Higgs self-coupling.  In both cases, measurements of
  limited precision can be made at 500~GeV, while higher-precision
  measurements require even higher energies.  At 1~TeV in the center of
  mass, with 5~ab$^{-1}$ of data, it is possible to measure the $htt$
  coupling to 2\% accuracy and the $hhh$ coupling to
  10\%~\cite{Asner:2013psa}.   
 The
  latter analysis uses the complementary processes $\ee\to Zhh$ and
  $\ee\to \nu\bar\nu hh$.  It is argued in \cite{Barklow:2017awn},
  that, although it is possible to measure double Higgs production in
  $pp$ collisions, it is very difficult to attribute a deviation from
  the SM prediction to a shift in the $hhh$ coupling.   In fact, it is
  likely that, for both of these couplings, a precise,
  model-independent determination is possible only with an $\ee$
  collider.   

\item {\bf Precision measurement of the top quark mass:}    The top
  quark mass is a fundamental parameter of the SM, and so it is
  important to measure it as accurately as possible.   For most
  applications, what is needed is a short-distance mass parameter, for
  example, the $\msb$ mass.   At this moment, it is not understood
  how to convert the top quark mass quoted by the LHC experiments to a
  short-distance mass, or even to the top quark pole mass.  The
  conversion from the pole mass to the $\msb$ mass brings in a
  perturbative theoretical uncertainty of
  200~MeV~\cite{Marquard:2015qpa},
 plus less well
  characterized nonperturbative uncertainties.  On the other hand, the
  position of the $t\bar t$ threshold in $\ee$ annihilation is
  controlled by a mass parameter that is very close to the $\msb$
  mass.  Measurement of this threshold  would give the $\msb$ top
  quark mass to an accuracy of 40~MeV~\cite{Fujii:2015jha}.  

\item {\bf Precision measurement of the top quark electroweak couplings:}
 Models in which the Higgs boson is composite typically predict
 sizable deviations from the SM expectations for the  $W$ and $Z$
 couplings of the top quark.   At an $\ee$ collider, the $Z$ vertices
 appear in the production reaction, where the individual contributions
 of $s$-channel $\gamma$ and $Z$ exchange can be disentangled using
 beam polarization.  This allows measurements of the $Z$ vertices with 
precisions of better that 1\%~\cite{Fujii:2015jha}.  These effects are
proportional to $s/m_t^2$   and so are increasingly visible at higher
energies. 

\item {\bf Search for pair-production of invisible particles:}
Though much attention is being given today to searches for
pair-production of invisible particles (including dark matter
particles) at the LHC, this is a difficult endeavor.   If the dark
particles are connected to SM particles by a hard operator or a
heavy-mass exchange, it is possible to search for very high $p_T$
monojet
events.  However, if the production is by electroweak Drell-Yan
production, the mass reach is limited by QCD and parton distribution
uncertainties estimation of the irreducible background process,
Drell-Yan 
production of $Z\to
\nu\bar\nu$. For example, for
Higgsinos, the discovery reach of LHC is expected to be less than
200~GeV~\cite{Low:2014cba}.
In contrast, the pure Higgsino is thermally produced as dark matter with the correct
relic density for a mass of about 1~TeV~\cite{Kowalska:2018toh}. 

 In $\ee$ annihilation, the
corresponding process of photon plus missing momentum is much more
precisely understood, allowing searches for invisible particles almost
to the kinematic limit~\cite{Chae:2012bq}.   Thus, an $\ee$ collider
operating above 2~TeV might be the unique way to search for this
special  dark matter candidate and other high-mass invisible
particles. 

\item {\bf Search for new electroweak gauge bosons and lepton
    compositeness}
The processes $\ee\to f\bar f$ can be used to search for new
electroweak gauge bosons and for 4-fermion contact interactions
including those signalling fermion compositeness.   For 3~TeV in the
center of mass and 1~ab$^{-1}$ of data, an $\ee$ collider would be
sensitive to $Z'$ masses of 15~TeV, more than double the eventual LHC
reach.  Such a collider would be sensitive to compositeness scales in
the range of 60-80~TeV~\cite{Linssen:2012hp}. 

\end{itemize}

Thus, there are strong physics arguments, already well
documented in the literature,   that motivate new $\ee$ colliders in the
energy region from 91~GeV to several TeV.

\subsection{$\ee$ and $\gamma\gamma$ physics at 10~TeV and above}

If new particles are not discovered at the LHC, this will not mean
that the problems of the SM are resolved.  It is possible that dark
matter problem might be resolved by a light particle that has escaped
detection because it is very weak coupled to ordinary matter.
However, the problems of electroweak symmetry breaking, the spectrum
of fermion masses, and CP violation can only be
explained by new interactions at shorter distances.   For problems
involving flavor, though we might obtain clues to the answers by
discovering anomalies at low energy, these clues can only point to new
particles of very high mass that should eventually be discovered and
characterized.

For this reason, it is imperative to begin thinking already about
accelerators that can extend our knowledge of physics to the
10~TeV--100~TeV energy region.  In this section, we will discuss in
particular the physics opportunities of a 30~TeV $\ee$ or
$\gamma\gamma$ collider based on advanced
 accelerator technology (ALIC).   We will first discuss the general
 parameters of particle searches 
that set the luminosity goals for ALIC, then discuss some
 novel aspects of experimentation at 30~TeV, and finally present some  
scenarios that go beyond simple particle searches to confront deeper
issues. 

\subsubsection{Pair-production in $\ee$ and $\gamma\gamma$ at very high 
energy}

Experience at previous $\ee$ colliders suggests that it is
straightforward to discover or exclude pair-production of new
particles with masses almost up to the kinematic limit of
$\sqrt{s}/2$.    This will continue to be true in $\ee$ collisions at
higher energies, and in $\gamma\gamma$ collisions at a dedicated
$\gamma\gamma$ collider, as long as a sufficient number of events are
produced.   Unfortunately, the pair-production cross sections decrease
steeply with energy.    The natural unit is the point cross section
\beq
   1~{\bf R} = { 4\pi \alpha^2\over 3 s}  = { 100~{\mbox{fb}}\over
 (E_{CM}({\mbox TeV}) )^2 }  \ ,
\eeq{Rdefin}
using the value $\alpha = 1/128$ applicable at high energy. In
practical terms, using a Snowmass year of $10^7$~sec, this is 
\beq
   1~{\bf R} = ( 10^5~\mbox{events} /yr )\cdot ( {{\cal L}\over
     10^{35}~\mbox{/cm$^2$ sec} } )~/~  (E_{CM}({\mbox TeV}) )^2
\eeqn
For the
illustrative case of a vectorlike heavy lepton (with $I = \half, Y =
\half)$, 
\beqa
\sigma (e^-_Le^+_R \to L\bar L ) &=&   {\bf R} \cdot (4.0) \cdot
({p\over E}) \cdot (1 + {m^2\over 2E^2}) \CR
\sigma(e^-_R e^+_L\to L\bar L) &= &    {\bf R} \cdot (0.9) \cdot
({p\over E}) \cdot (1 + {m^2\over 2E^2}) 
\eeqan
and, for (unpolarized) $\gamma\gamma$ collisions,
\beq
\sigma (\gamma\gamma \to L\bar L ) =  {\bf R} \cdot (6) \cdot
({p\over E}) \cdot  \bigl({E^2 + p^2\over 4 E^2} \log {E+p\over m} - \half
(1 + {m^2\over E^2})\bigr) \ .
\eeqn
Thus, a luminosity of  $10^{36}$/cm$^2$ sec is the minimum even for
an exploratory physics program.   

 It is important to note that, at
hadron colliders, although the relevant pair production cross sections
are larger
(with $\alpha_s$ instead of $\alpha$), they are still sharply
decreasing with the new particle mass, while  SM backgrounds increase
relative to physics signals, and pileup increases linearly with
increased luminosity.   To explore the full spectrum of new particles in the 10~TeV 
region, it is of great advantage to use a lepton collider.  

For the purpose of this study, we will assume that our model collider
ALIC has an energy of  30~TeV and a luminosity of $10^{36}$/cm$^2$
sec, giving an event sample of 30 ab$^{-1}$ over a few years of running.

\subsubsection{New features of the SM  at 30~TeV in the center of mass}

At energies in the multi-10-TeV region of $\ee$ and $\gamma\gamma$
collisions, it remains true that pair-production yields events with
pairs of narrow jets that are clearly distinguishable from events
with new particle production.   However, there are three new features of
SM physics at these energies that play an important role.

The first is the presence of electoweak emissions in the final state
of lepton and quark production.    It is true at any energy that
pair-production of a particle with charge $Q$ is accompanied by
collinear photon emission with probability
\beq
           P_\gamma(z) dz  =   Q^2\ {\alpha\over \pi} \ \log {E_{CM}\over m_e}
           \cdot  {1 + (1-z)^2\over z}\ dz \ . 
\eeqn
For 30~TeV, this is
\beq
           P_\gamma(z) dz  \approx  0.09 \ Q^2\  {1\over z} dz  \ .
\eeqn
At 30~TeV, the center of mass energy is also well above the mass of
the $W$, and so collinear $W$ emission becomes an important effect.
The probability of a collider $W$ emission from a left-handed lepton
or quark is 
\beq
           P_W(z) dz  =   {\alpha_w\over 2\pi} \ \log {E_{CM}\over m_W}
           \cdot  {1 + (1-z)^2\over z}\ dz \ . 
\eeqn
For 30~TeV, this is
\beq
           P_W(z) dz  \approx  0.03  {1\over z} dz  \ .
\eeqn
Then approximately 40\% of events with  directly produced pairs of  left-handed leptons or
quarks  will contain, on one side or the other, a change of flavor (for example, $\mu \to
\nu_\mu$), accompanied by a $W$ boson that, typically, will decay
hadronically to two jets.   The analysis of events from the ALIC
collider will need to take this into account. 

The second effect comes from the correspondingly high probability of
$W$ emission in the initial state.  In $\ee$ collision and even more
strongly in $\gamma\gamma$ collisions, the initial particles can exchange a
$W$ boson in the $t$-channel.   Radiation from this $W$ propagator 
will lead to multiperipheral production of 
multiple $W$ and $Z$ bosons, with a total cross section 
that grows as a small power of the center of mass
energy. The $W$ and $Z$ bosons are produced with transverse
momenta $m_W$ with respect to the beam axis.  
Multiple  Higgs bosons  can also be emitted in this cascade.
The most important effect of this multiple boson emission is to
provide additional  SM backgrounds to new physics reactions that one might
wish to study.   To reconstruct these multiply produced vector bosons,
it will be very important for the detector at the ALIC collider 
to reconstruct hadronically decaying $W$ and $Z$ bosons with high efficiency and also
to distinguish these from one another.

The third effect is a simple consequence of the boosting of 
emitted quarks and leptons.  Typical decay lengths become
\beq
           \begin{tabular}{ccc}
           $b$ & $c$ & $\tau$ \\ \hline
      40~cm    &   20~cm &   74~cm 
\end{tabular}
\eeqn
assuming that $\tau$ leptons are directly produced from the hard
reaction and $B$ and $D$ mesons are produced with a typical momentum
fraction $z$ of 0.2 in $b$ and $c$ jets.   Thus, an important task of
the detector at the ALIC collider will be to identify highly displaced
secondary vertices and to measure their properties. 

We are preparing a model detector that takes these features 
of the collisions into account.   This detector will have a tracking region for 
charged particle sign selection and the identification of displaced vertices and a 
tracking calorimeter that will have the bulk of the burden for energy measurement.
Identification of heavy flavor SM particles does not require a vertex detector within 
cm of the interaction point.  This is fortunately, since we
expect that this region will be taken up by final focus magnets and, in the case of a
$\gamma\gamma$ collider based on  $e^-e^-$ beams, 
the equipment for the forward Compton collisions.

\subsubsection{Novel physics scenarios at 30~TeV}

Beyond the possibility of discovering new particles that hint at new
physics,  a 30~TeV $\ee$ or $\gamma\gamma$ collider
offers the possibility that the qualitative nature of the processes
that dominate the total cross section may change from the SM
expectation.  In this section, we will present several
possible scenarios with this property.

\begin{itemize}

\item {\bf Thermalization:}    The electroweak sector of the SM is a
  weakly-coupled field theory, and so one might expect that the
  qualitative features of  electroweak scattering would remain the
  same up to very high energy.   However, this is not obvious, and
  actually the point has been debated since the 
1980's~\cite{Ringwald:1989ee,Espinosa:1989qn,Goldberg:1990qk,Mattis:1991bj,Rubakov:1996vz,Tye:2015tva}. The
electroweak theory contains a nonperturbative solution, the {\it
  sphaeleron},  which has a mass of about 9~TeV.   It has been argued
that processes
that naively would be suppressed by the small value of the Higgs boson
self-coupling actually become dominant at energies above the
sphaeleron mass.   This idea has recently been revived in
\cite{Khoze:2017tjt}, with the suggestion that the decay of a
sufficiently heavy electroweak object of mass $M$ is dominated by
final states with large numbers of Higgs, $W$, and $Z$ bosons with
momentum of order $m_W$ in the frame of the decaying heavy particle.
The appearance in $\ee$ or $\gamma\gamma$ events of any substantial
subsample of events with 300 or more heavy bosons would be remarkable.
Even if this possibility is not theoretically favored at the moment,
it could happen.

\item {\bf Composite Higgs models:}   A more mainstream idea with
  dramatic implications at these high energies is the idea that the
  Higgs boson is composite.   Just after the formulation of the SM,
  Weinberg and Susskind proposed that the Higgsboson was a composite
  of fermions bound by new strong interactions similar to the QCD
  interactions but with a strong interaction scale of
 1~TeV~\cite{Weinberg:1975gm,Susskind:1978ms}.   This specific model was found
 to be inconsistent with the results of precision electroweak
 measurements and is now certainly excluded by the fact that it predicts a
 heavy Higgs boson.   However, the idea that the Higgs boson is not an
 elementary scalar remains an attractive one.

Kaplan and Georgi suggested another approach to the construction of
models in this class.    A new strong interaction theory at 10~TeV
could give rise to a multiplet of composite scalar Goldstone bosons
that might be
identified, or might contain, with a complex $SU(2)$ doublet with the
quantum numbers of the Higgs doublet field of the SM.  These Goldstone
bosons can then receive mass by a small perturbation of the model.
This idea is realized in ``Little Higgs''
models~\cite{ArkaniHamed:2002qy,Schmaltz:2005ky},
in which the interactions with the top
quark and a heavy vectorlike top quark partner generate a
negative (mass)$^2$ term for this Goldstone multiplet.    The top
quark partner is expected to have a mass of 1--3 TeV and might be
discovered at the LHC.   However, full structure of the model,
including, in particular, its strong interaction dynamics, can only be
studied by experiments at higher energy.   As with QCD, the new
interactions would be understood by producing the analogue of the
$\rho$ meson and the other elements of its hadron spectrum.   This
would be most readily achieved by an accelerator capable of sweeping
through the spectrum of resonances of the model that is produced from
electroweak currents. 

\item {\bf Extra space dimensions:}   Another realization of the
  Georgi-Kaplan idea is to consider its  AdS/CFT
  dual~\cite{Maldacena:1997re}.   In this approach, the strong
  interaction theory is represented by a gauge theory in a finite
  slice of  5-dimensional
  anti-de Sitter space, according to the construction of Randall and
  Sundrum~\cite{Randall:1999ee}.  In this representation, the strong
  interaction resonances correspond to 
  the higher-dimensional Kaluza-Klein resonances, and the Goldstone
  bosons correspond to zero-mass modes of the 5th components of gauge
  fields.   Realistic models of the Higgs sector and electroweak
  symmetry breaking have been constructed along these lines, for
  example, in  \cite{Agashe:2004rs}.   In this approach, the structure of the new
  strong interactions is encoded in the spectrum of $s$-channel
  resonances in  electroweak reactions.

There are many other proposals for  new physics models 
based on extra space dimensions~\cite{Hewett:2002hv}.   In all cases,
the proof that the theory is correct would come from recognizing the
pattern in the masses of  Kaluza-Klein resonances that is generated by
the geometry of the extra dimensions.    We can only guess how many resonances must be
discovered to  make this pattern clear.  However, the lower limits on
such resonances are now set by LHC exclusions 
at  2-4~TeV~\cite{Aaboud:2017buh,Sirunyan:2018exx}.  Thus, observing a
series of resonances from new physics will require quark or lepton
center of mass energies above 20~TeV.

\item  {\bf Dynamics of flavor:}   The SM with one fundamental Higgs
  boson is unique among theories of electroweak symmetry breaking in
  that all flavor and CP violation can be removed by changes of
  variables except for the the CKM mixing matrix with its single
  Kobayashi-Maskawa CP-violating phase.   The model naturally has
  lepton universality and zero lepton-flavor-changing amplitudes.  All
  of these features are in good agreement with experiment. 
  More general models, especially those that seek to explain the quark
  and lepton masses, must somehow suppress flavor-changing amplitudes
  that correct the SM predictions.

Recently, the LHCb collaboration has obtained results that suggest
deviations from the SM in $B \to K^* \ell^+\ell^- $ decays~\cite{Oyanguren:2018huo}.  Most
strikingly, these results suggest that  lepton universality is violated in this
decay.  To explain these effects, it is necessary to introduce new
particles with flavor-violating interactions that cannot appear in the
SM while maintaining the absence of flavor-changing and lepton number
violation in other strongly constrained processes.
 This can be done in one of two
  ways.   The first is to include new scalar particles with TeV masses
  and couplings of the order of Yukawa couplings (for example,
  \cite{Bauer:2015knc}).  The second is to include new vector
  particles with flavor-dependent couplings due to gauge symmetry
  breaking (for example, \cite{Gripaios:2014tna}).  In this approach,
  the lightest vector boson may be at 1~TeV, but other members of the
  vector multiplet will be at 10~TeV or higher.   The introduction of
  a new scalar introduces a new hierarchy problem whose solution
  would also require new particles at higher energy.

If the LHCb results are confirmed and the problem becomes sharp, the
proof of a solution to the problem would require the discovery of a
new 
particle whose couplings, measured directly, show flavor violation of
the required form.  The exploration of this physics with a 
very high energy hadron collider has been discussed in \cite{Allanach:2017bta,Allanach:2018odd}. 
The same measurements, and more, 
can be made at the ALIC collider. It will eventually necessary to measure the complete decay
profile of the new leptoquarks, and this  can typically be achieved only at
$\ee$ colliders.

\end{itemize}

\subsection{Conclusions} 
 
Thus, there are many pressing physics motivations for the construction
of $\ee$ and $\gamma\gamma$ colliders of higher energy, up to center
of mass energies
of 30~TeV and beyond.


%% file: ALEGRO18_repWG4_LWFA.tex
\section{Introduction}




Laser wakefield acceleration (LWFA) holds much promise for compact acceleration owing to the ultra-high gradients supported in the plasma. A linear collider based on LWFA concepts would be composed of a series of laser-driven accelerating modules or stages \cite{leemans2009}.  Components such as electron injector, accelerating structures, electron lenses, or laser coupling devices all could potentially be plasma based, making this scheme extremely compact.

There are several regimes of LWFA that may be accessed, based on the laser-plasma parameters.  Two regimes that have attracted attention for collider applications are the quasi-linear regime and the bubble (or blow-out) regime. The quasilinear regime is characterized by regular plasma wave buckets and nearly-symmetric regions of acceleration-deceleration and focusing-defocusing, for both electron and positron beams.  Multiple laser driver pulse trains are possible in this regime, as well as loading multiple accelerating buckets with a particle bunch train.  These features make the quasilinear regime of interest for collider applications. Most experiments to date have focused on the nonlinear bubble regime, and the quasilinear regime thus needs to be thoroughly explored in the near future to determine its suitability as the elementary brick to build an LWFA-based ALIC.

Electrons from the plasma can be trapped and accelerated to relativistic energies in the laser-driven plasma structures, typically hundreds of MeV, over a few mm.  Several methods of electron trapping have been investigated
to generate electron beams with characteristics suitable for injection and further acceleration in subsequent plasma structures.

Compact positron sources need to be developed in the environment of laser-plasma accelerators. Compact methods to generate and cool positron beams require exploration.  Positron beams could be created from the interaction of LWFA generated electrons with a converter target or from the radiation generated in nonlinear acceleration regimes of LWFA (e.g., betatron emission in the plasma wave).

Many of the elements of a laser-driven plasma collider can be studied separately at presently available facilities, but dedicated facilities will be needed to demonstrate the reliability and beam quality necessary for coupling and operating multiple LWFA stages toward the collider application.

\section{ Status of LWFA R\&D}

LWFA R\&D status is described in the ANAR2017 Report~\cite{cros2017} and is briefly summarized here.

For more than a decade, most R\&D effort has been dedicated to exploring the limits of acceleration of electrons from the plasma in nonlinear regimes. The achieved accelerated electron energy has increased by almost 2 orders of magnitude, from approximately 100 MeV in 2004 to nearly 10 GeV today. This has been made possible by the progress in laser development, increasing laser peak power, and repetition rate, and by the development of laser guiding techniques for extending the length over which acceleration can be driven.
 
Plasma control and the development of laser guiding techniques have made tremendous progress, resulting in plasma tailoring and channel formation over longer distances, with increased frequency of repetition. Numerous diagnostics are now available to monitor plasma creation, laser guiding, and plasma wave evolution on micron and femtosecond scales. 
 
In parallel, as the characterization of the electron properties progressed, it became necessary to address beam quality improvement. This gave rise to numerous studies of the laser-plasma interaction regimes, indicating that the high quality of the laser pulse is a strong requirement to achieve a high quality electron beam. These studies also resulted in  new concepts for controlling the injection of electrons into the plasma wave as alternative techniques to self-injection, enabling additional control of the properties of the injected electron beam.

The external injection of electrons into an accelerating plasma structure in the quasilinear regime is one of the most challenging tasks to achieve experimentally and also the most promising scheme for scaling up electron energy. This has been achieved with moderate coupling efficiency in recent LWFA-staging experiments \cite{Steinke16}. 

\section{ Facilities using advanced acceleration}

 
 
As relativistic electrons in the 100 MeV range are readily produced with a relatively compact, 10 TW-scale laser system, this type of experiment has been carried out by many  research groups worldwide. These independent research groups have contributed to the wealth and variety of results on the subject, published in refereed journals, and advanced the field over a short time scale. The achievement of electron beams with reproducible properties is now an important goal and large scale facilities, such as those hosted by accelerator laboratories or in the environment of conventional accelerators, are being developed.

\subsection{U.S. LWFA R\&D Facilities}

There is an active research program in laser-driven plasma acceleration in the United States.  LWFAs accelerate electron beams using a plasma wave driven by the ponderomotive force of short-pulse, high-intensity laser, and therefore LWFA R\&D requires high-peak power laser facilities.  Advances in laser technology and the availability of short-pulse (tens to hundreds of fs) laser systems delivering high peak powers ($\sim$100~TW, up to PW) have enabled progress in this field.  Presently, LWFAs are able to generate multi-GeV electron beams (with tens to hundreds of pC) using cm-scale plasmas \cite{Leemans14,Wang13}.  Table \ref{tab:LWFA-facilities} lists the major laser facilities (with peak laser power $>$100~TW) performing LWFA R\&D presently operating in the US.  

In the US, plasma acceleration of electron beams is primarily funded by the US Department of Energy, Office of Science, Office of High Energy Physics.  In 2016, the US DOE published the {\it Advanced Accelerator Development Research Strategy Report} \cite{US-Roadmap} describing the R\&D roadmaps towards the goal for constructing a multi-TeV e$^+$e$^{-}$ collider.  In this report, the R\&D goals for the next ten years were presented.  Critical to the collider application is demonstration of multi-GeV LWFA staging with independent, drive beams.  Following demonstration of staging at the $\sim$100~MeV level \cite{Steinke16}, experimental preparation for a multi-GeV LWFA staging experiment is on-going at the BELLA (Berkeley Lab Laser Accelerator) Center of LBNL.  Also identified in this report was the need for laser technology development.  In particular, the need for development of high peak and high average power laser systems, operating at high efficiency.  
 
First applications of LWFAs considered over the next decade are betatron sources for x-ray generation, free-electron lasers for ultra-fast science applications, and gamma-ray sources (generated using Thomson scattering) to produce MeV photons for nuclear detection applications. These light sources would initially operate at the few Hz repetition rate, and following initial demonstrations and the development of short-pulse kW-class average power laser systems, these sources would operated at kHz repetition rates. 
 \begin{table}[h]
\caption{US laser facilities ($>$100~TW) performing LWFA R\&D.}
\label{tab:LWFA-facilities}
\centering
\begin{tabular}{lllllll}\hline\hline
Facility & Institute  & Location & Gain & Energy & Peak power  & Rep. rate 
\\
&&& media & (J) &  (PW) &  (Hz) 
\\ \hline
BELLA \cite{bella} & LBNL & Berkeley, CA & Ti:sapphire & 42 & 1.4 & 1  \\
Texas PW \cite{texasPW} & U. Texas & Austin, TX & Nd:glass & 182 & 1.1 & single-shot \\
Diocles \cite{diocles} & U. Nebraska & Lincoln, NE & Ti:sapphire & 30 & 1 & 0.1 \\
Hercules \cite{hercules} & U. Michigan & Ann Arbor, MI & Ti:sapphire & 9 & 0.3 & 0.1\\
Jupiter \cite{jupiter} & LLNL &  Livermore, CA &  Nd:glass &  150 &  0.2 &  single-shot \\
 \hline\hline
\end{tabular}
\end{table}

\subsection{LWFA R\&D Facilities in Europe}

Table \ref{tab:eulwfafac} provides an overview of the facilities in Europe where LWFA is studied using laser systems with peak powers of order 100~TW or above. All of these facilities currently have laser systems based on Ti:sapphire as amplifying material and operate with pulse durations of the order of 30~fs, except CILEX which should reach pulse durations as short as 15~fs. Some of these, e.g.,  LLC and laser systems at the LOA and RAL facilities, have been in operation since the mid-1990s with continuous laser upgrades and have explored, and sometimes pioneered, various regimes of electron acceleration.  Facilities marked with * in Table \ref{tab:eulwfafac} have not demonstrated operation performing electron acceleration with the stated parameters  at the time of writing. The most recent facilities have PW peak power capabilities, sometimes with multiple beams. These will be useful to carry out tests of multi-stage acceleration provided enough beam time can be obtained at user facilities to implement and perform such complex experiments.

All facilities, except the Extreme Light Infrastructure (ELI), were built with national funding. They are opened to users for various fractions of their operation time, depending on the facility policy and source of funding for operation. European projects, such as LASERLAB and ARIES, include funding for some selected facilities to provide transnational access.
Discussions within the European Network for Novel Accelerators (EuroNNAc2) \cite{euronnacrep} 
have led to propose EUPRAXIA \cite{eupraxia},
a Horizon2020 design study, supported by 16 partners and 22 associated partners. 

\begin{table}[h]
\caption{Laser facilities ($\gsim$100~TW) performing LWFA R\&D in Europe.}
\centering
\begin{tabular}{llllll}\hline\hline
Facility & Institute  & Location  & Energy & Peak power  & Rep. rate 
\\
&& & (J) &  (PW) &  (Hz) 
\\ \hline
ELBE \cite{elbe} & HZDR&  Dresden, Ge   & 30 & 1 & 1 \\
GEMINI \cite{gemini} & STFC, RAL & Didcot, UK  & 15 &0.5  & 0.05  \\
LLC\cite{llc} & Lund Univ &  Lund, Se &    3  & 0.1 &  1 \\
Salle Jaune \cite{LOA} & LOA  & Palaiseau, Fr &  2 & 0.07 & 1 \\
UHI100 \cite{uhi100} & CEA Saclay & Saclay, Fr &  2 & 0.08 & 1\\
CALA*\cite{cala} & MPQ &  Munchen, Ge &    90  & 3 &  1 \\
CILEX*\cite{cilex} & CNRS-CEA& St Aubin, Fr  & 10-150  & 1-10 & 0.01\\
ELIbeamlines*\cite{eli} & ELI & Prague, TR  & 30  & 1 & 10 \\
ILIL*\cite{ilil} & CNR-INO &  Pisa, It &    3  & 0.1 &  1 \\
SCAPA* \cite{scapa} & U Strathclyde & Glasgow, UK  & 8 & 0.3  & 5 \\
 \hline\hline
\end{tabular}
\label{tab:eulwfafac}
\end{table}

\subsection{LWFA R\&D Facilities in Asia}

In addition to several multi-petawatt facilities (see Table~\ref{tab:asialwfafac}), several 100~TW class facilities are in operation in Asia: 200~TW, 1-5~Hz laser at SIOM in China, 110~TW, multi-pulse laser at NCU in Taiwan, 200~TW (Thales) laser at Peking University in China, 200~TW (Amplitude Technology) laser at Shanghai Jiao Tong University in China, and 100~TW (Amplitude Technology) laser at the Tata Institure in Mumbai, India.
\begin{table}[h]
\caption{Laser facilities ($\gsim 100~$TW) performing LWFA R\&D in Asia}
\centering
\begin{tabular}{llllll}\hline\hline
Facility & Institute  & Location & Energy & Peak power  & Rep. rate 
\\
&& & (J) &  (PW) &  (Hz) 
\\ \hline
CLAPA  & PKU &  Beijing, PRC   & 5 & 0.2 & 5 \\
CoReLS \cite{corels} & IBS &  Gwangju, Kr   & 20-100 & 1-4 & 0.1 \\
J-Karen-P* \cite{kpsi} & KPSI & Kizugawa, Jn  & 30 &1  & 0.1  \\
LLP \cite{lpp} & Jiao Tong Univ &  Shanghai, PRC   & 5 & 0.2 & 10 \\
SILEX* & LFRC &  Myanyang, PRC &    150  & 5 &  1 \\
SULF* \cite{sulf} & SIOM & Shanghai, PRC &  300 & 10& 1 \\
UPHILL\cite{tata} & TIFR &  Mumbai, In &   2.5 & 0.1 &  \\
XG-III & LFRC &  Myanyang, PRC &    20 & 0.7 &  \\
\hline\hline
\end{tabular}
\label{tab:asialwfafac}
\end{table}


\section{Our long-term goal: the Advanced Linear International Collider, ALIC}


 

Laser-driven plasma acceleration uses a short-pulse, high-intensity laser to ponderomotively drive a large electron plasma wave (or wakefield) in an underdense plasma. The electron plasma wave has relativistic phase velocity (approximately the group velocity of the laser) and can support large electric fields in the direction of propagation of the laser.   When the laser pulse is approximately resonant (duration of the order of the plasma period), and the laser intensity is relativistic (with normalized vector potential $eA/mc^2\sim 1$), the magnitude of the accelerating field is of the order of $E_0[{\rm V/m}] \simeq 96 (n_0[{\rm cm}^{-3}])^{1/2}$, and the wavelength of the accelerating field is of the order of the plasma wavelength $\lambda_p [{\rm mm}] \simeq 3.3 \times 10^{10}(n_0[{\rm cm}^{-3}])^{-1/2}$, where $n_0$ is the ambient electron number density.  For example, $E_0 \simeq~30$~GeV/m (approximately three orders of magnitude beyond conventional RF technology) and $\lambda_p \simeq~100~\mu$m for a plasma density of $n_0 = 10^{17}~{\rm cm}^{-3}$.  Our long-term goal is to take advantage of these LWFA characteristics, namely ultra-high gradients and intrinsically ultra-short bunches, for the collider application. 


In general, the energy gain in a single LWFA stage may be limited by laser diffraction effects, dephasing of the electrons with respect to the accelerating field phase velocity (approximately the laser driver group velocity), and laser energy depletion into the plasma wave. Laser diffraction effects can be mitigated by use of a plasma channel (transverse plasma density tailoring), guiding the laser over many Rayleigh ranges. Dephasing can be mitigated by plasma density tapering (longitudinal plasma density tailoring), which can maintain the position of the electron beam at a given phase of the plasma wave. Ultimately, the single-stage energy gain is determined by laser energy depletion.  After a single LWFA stage, the laser energy is depleted and a new laser pulse must be coupled into the plasma for further acceleration.  Development of LWFA staging technology is critical for the collider application.

\begin{table}[h]
\caption{LWFA single stage parameters operating at a plasma density of $n_0 = 10^{17}$~cm$^{-3}$.}
\label{tab:LWFA}
\centering
\begin{tabular}{lc}\hline\hline
Plasma density (wall), $n_0$[cm$^{-3}$] & $10^{17}$ \\
Plasma wavelength, $\lambda_p$[mm] & 0.1  \\
Plasma channel radius, $r_c$[$\mu$m] & 25\\
Laser wavelength, $\lambda$[$\mu$m] & 1 \\
Normalized laser strength, $a_0$ & 1\\
Peak laser power, $P_L$[TW] & 34 \\
Laser pulse duration (FWHM), $\tau_{L}$[fs] & 133 \\
Laser energy, $U_{L}$[J] & 4.5\\
Normalized accelerating field, $E_z/E_0$ & 0.14\\
Peak accelerating field, $E_L$[GV/m] & 4.2 \\
Plasma channel length, $L_c$[m] & 2.4\\
Laser depletion, $\eta_{pd}$ & 23\% \\ 
Bunch phase (relative to peak field) & $\pi/3$\\
Loaded gradient, $E_z$[GV/m]& 2.1 \\
Beam beam current, $I$[kA] & 2.5\\ 
Charge/bunch, $e N_b=Q$[nC] & 0.15\\
Length (triangular shape), $L_b$[$\mu$m] & 36 \\
Efficiency (wake-to-beam), $\eta_b$ & 75\% \\
e$^{-}$/e$^{+}$ energy gain per stage [GeV] & 5\\
Beam energy gain per stage [J]& 0.75\\ 
 \hline\hline
\end{tabular}
\end{table}

A detailed design of an LWFA collider has not been carried out at this time.  Many collider components (e.g., beam damping and cooling, beam delivery systems, final focus, etc.) are in the conceptual stage, or have not been considered in detail.  The goal of the outlined R\&D will be to explore the LWFA physics and collider components to enable a detailed design.  Preliminary attempts \cite{Schroeder16} to outline collider parameters and the requirements on the laser system powering a future laser-plasma collider have been performed, based on order-of-magnitude scaling laws that govern some of the important physics considerations for an LWFA, as well as assumptions on the efficiencies (energy transfer from laser to plasma and from plasma to electron beam) that could be obtained in principle. 

An example of single LWFA stage parameters operating at a plasma density of $10^{17}~{\rm cm}^{-3}$ is shown in Table~\ref{tab:LWFA}.  Operating at a plasma density of $n_0 \sim 10^{17}~{\rm cm}^{-3}$ reduces the power costs $\propto n_0^{1/2}$ while providing for a high accelerating gradient $\propto n_0^{1/2}$ and acceptable beamstrahlung background $\propto n_0^{-1/2}$.  These designs are based on the use of hollow-plasma channels (to control emittance growth via scattering and ion motion) \cite{Schroeder16}.   Novel methods for beam-breakup instability mitigation will be required in this configuration.  Table~\ref{tab:LWFA-collider} shows possible configurations for 0.25, 1, 3, and 30~TeV center-of-mass collider designs based on the LWFA stage parameters in Table~\ref{tab:LWFA}. 
These examples assumed a wall-to-laser efficiency of 40\% for the 250~GeV and 1~TeV examples, as well as a laser energy recovery efficiency of  90\%.  The 3~TeV and 30~TeV examples assumed  a wall-to-laser efficiency of 50\% and a laser-energy recovery efficiency of 95\%.
\begin{table}[h]
\caption{Example parameter sets for 0.25, 1, 3, 30~TeV center-of-mass
LWFA-based colliders.}
\label{tab:LWFA-collider}
\centering
\begin{tabular}{lcccc}
\hline\hline
Energy, center-of-mass, $U_{\rm cm}$[TeV]  & 0.25 & 1 & 3  & 30\\
Beam energy, $\gamma mc^2=U_b$[TeV]  & 0.125 & 0.5 & 1.5 & 15\\
Luminosity, $\mathcal{L}$[$10^{34}$~s$^{-1}$cm$^{-2}$ ]  & 1 & 1 & 10 & 100\\
Beam power, $P_b$[MW]  & 1.4 & 5.5 & 29 & 81 \\
Laser repetition rate, $f_L$[kHz]  & 73 & 73 & 131 &  36 \\
Horiz.\ beam size at IP, $\sigma_x^*$[nm]  & 50 & 50 & 18 & 0.5 \\
Vert.\ beam size at IP, $\sigma_y^*$[nm]  & 1 & 1 & 0.5 & 0.5 \\
Beamstrahlung parameter, $\Upsilon$  & 0.5 & 2 & 16 & 2890\\
Beamstrahlung photons, $n_\gamma$  & 0.6 & 0.5 & 0.8 & 2.8\\
Beamstrahlung energy spread, $\delta_\gamma$  & 0.06 & 0.08 & 0.2 & 0.8\\
Disruption paramter, $D_x$ & 0.07 & 0.02 & 0.05 & 3.0 \\
Number of stages (1 linac), $N_{\rm stage}$  & 25 & 100 & 300 & 3000\\
Distance between stages [m]  & 0.5 & 0.5 & 0.5 & 0.5 \\
Linac length (1 beam), $L_{\rm total}$[km]  & 0.07 & 0.3 & 0.9 & 9.0\\
Average laser power, $P_{\rm avg}$[MW]  & 0.3 & 0.3 & 0.6 & 0.17 \\
Efficiency (wall-to-beam)[\%] & 9 & 9 & 13 & 13\\
Wall power (linacs), $P_{\rm wall}$[MW]  & 30 & 120 & 450 & 1250 \\
\hline\hline
\end{tabular}
\end{table}

The collider requirements on average laser power, rep-rate, and laser efficiency are beyond the current state-of-the-art for short-pulse, high-peak-power laser systems (e.g., currently at the $\sim 100$~W average power level). However, high-efficiency diode-pump lasers and fiber lasers are rapidly evolving technologies that offer some promise in closing this large technology gap within the upcoming years.

Realizing the parameters shown in Table~\ref{tab:LWFA-collider} will require the development of sources of ultra-low emittance, ultra-short (sub-100 fs) electron and positron beams, with shaped current distributions, and novel beam manipulation and cooling methods will need to be developed.  Development of methods to stage LWFAs with high beam emittance preservation is critical.   Beyond state-of-the-art methods to deliver, focus, and align beams, with sub-nm size, at the IP will also need to be developed. 


\section{ALIC Machine components}



\subsection{e-/e+ sources, cooling}

The injector parameters are determined by the choice of main accelerator stage and by the performance of the beam transport from the injector to the main accelerator. 
This section gives an overview of established  methods to achieve electron sources based on mechanisms injecting a fraction of plasma electrons into the laser generated plasma wave, or ``injection methods'', and a summary of their current performance. Laser-plasma-based electron sources are expected to reach the 10~pC per MeV charge density for an energy of the order of a few hundred MeV and bunch duration of a few fs, which would make them suitable for injection in a plasma accelerator provided their emittance can be controlled.
The main limitation in the progress of any plasma-based injection method is currently the low average power of existing laser and linac facilities, which severely impairs the luminosity of a future plasma-based collider. However, existing facilities can be used to test injection strategies until appropriate drivers are available.

\subsubsection{Electron sources} 


Several strategies exist for moving background plasma electrons onto trapped orbits, allowing energy exchange between the plasma wakefield and the electrons.    

\textbf{Self-injection:} In the nonlinear blow-out regime, which occurs typically for strong laser pulses with a normalized amplitude $a_0 \gtrsim 2.5$,
the plasma wave may ``break'' and a portion of plasma electrons 
can be injected into the bubble (co-moving ion cavity).  Due to its dependence on nonlinear laser evolution, which might be subjected to shot-to-shot variations and the nonlinear interaction between the bunch's space-charge and the wakefield, it is difficult to control the injected beam parameters in self-injection. In particular, the continuous nature of the process leads to different final energies of electrons injected at different positions along the accelerator and hence a broad spectrum. 


\textbf{Ionisation assisted injection}: This scheme has attracted attention recently as a means for injecting into a nonlinear wave without the need for wavebreaking (operating in the self-trapping regime), and much more stable electron beams can be generated. To that end, typically, the Hydrogen or Helium target gas is doped with a small amount of intermediate-Z atoms such as Oxygen or Nitrogen. While the low-Z gas is ionized in the foot of the drive laser pulse, the high-Z component is ionized near its peak, such that the N or O electrons are born inside the accelerating region of the wakefield and are therefore trapped, leading to stable, high-charge beam. However, for a homogeneously doped target, the injection is not localized, and the output beam energy spectrum is typically broad \cite{clayton2010,pak2010}
Localized doping, plasma density tailoring \cite{lee2018}
or beam-loading control via dopant concentration may yield a peaked electron spectrum \cite{couperus2017}.
There are also possibilities to localize the ionisation injection using additional laser pulses \cite{Bourgeois2013,yu2014}.

\textbf{Colliding laser pulses}: In order to overcome the uncontrolled nature of self-injection, this approach avoids driving the wakefield into self-trapping by reducing the laser amplitude and/or the electron density. Injection is now performed by exploiting the enhanced ponderomotive force in a beat wave of two colliding laser pulses to locally heat some background electrons enough to overcome the trapping velocity \cite{Esarey97}. 
This scheme allows injection and trapping as well as further acceleration of the electrons in the wakefield, and  provides independent tuning for the electron energy via the injection position and the charge/spectral bandwidth via the colliding pulse intensity \cite{faure2006}. 
However, The generated electron charge is typically small.

\textbf{Shock-front injection}: As in colliding pulse injection, this scheme separates the injection and acceleration processes in the plasma wave. By obstructing the flow from a supersonic gas jet, a downward density step is introduced in the target. While the plasma wave crosses this step, its wavelength abruptly becomes longer, enabling trapping of electrons.  The location of the density step provides control over the injection position and hence the electron energy, and while the laser power can be used to control the charge and optimize the energy bandwidth \cite{buck2013,swanson2017}.  
Moreover, the abrupt change in plasma wavelength causes an intrinsic dephasing step at the injection position, which leads to lower final energy. This drawback can however be overcome by following the shock front with a density up-ramp to counter electron beam dephasing~\cite{schmid2010}.

\subsubsection{Positron sources}

Experimental study of positron acceleration in a plasma is challenging due to the difficulty of providing an injector of suitable quality, synchronised with the positron-accelerating region of a wakefield. 
Proposals exist to provide positron beams suitable for test studies of subsequent wakefield acceleration at the FACET-II facility. 

The generation of a positron beam of high spectral and spatial quality for injection in further acceleration stages is experimentally challenging and two main routes are being pursued. 
The first possiblity is to generate a high-charge population of low-energy positrons collected from the interaction of an electron with a solid target, stored and cooled. This scheme has the advantage of ensuring a high number of positrons with very low emittance. However, it requires expensive and bulky cooling systems and, most importantly, results in long beam durations. 

The second possiblity is to consider laser-based generation of positrons and recent experimental results on laser-based generation of high-quality ultra-relativistic positrons are promising. For instance, fs-scale and narrow divergence positron beams in a plasma-based configuration have been recently reported~\cite{sarri2013,sarri2015}
In this mechanism, the positrons are generated as a result of a quantum cascade initiated by a laser-driven electron beam propagating through a high-Z solid target. For sufficiently high electron energy and thin converter targets, the generated positrons present properties that resemble those of the parent electron beam, hence the fs-scale duration, mrad-scale divergence, and micron-scale source size. The maximum positron energy attainable in this scheme is naturally dictated by the peak energy of the parent electron beam. Positrons with energy up to 0.5 GeV were produced in recent experiments. These beams have durations comparable to the positron-accelerating region of wakefield and are naturally synchronised with a high-power laser. However, other characteristics still need to be carefully optimised, such as their non-negligible normalised emittance and the relatively low charge. 

In order to optimise these sources, it is desirable to increase the charge in the positron beam and minimise its emittance. Both goals can be achieved if the charge and maximum energy of the primary electron beam is increased, together with dedicated studies of beam manipulation and transport. In this respect, the technological and scientific developments in laser technology and electron acceleration will naturally benefit also the generation of such positron beams, a necessary ingredient towards the high-energy applications of laser-wakefield electrons and positrons.

\subsubsection{Beam Cooling}

Beam cooling is a crucial step for obtaining the required luminosity for a linear collider. This is conventionally achieved through radiation damping based on synchrotron radiation from beam particles moving along curved trajectories in a storage ring (i.e., damping ring) with a circumference of a few hundred meters. Particles emitting synchrotron radiation under lateral acceleration lose momentum in all directions. Longitudinal momentum is restored by radio-frequency cavities located in the ring, while transverse momentum is damped until an equilibrium value is reached (typically of the order of milliseconds) through the balance of radiation damping and quantum excitation. Faster damping can be achieved by increasing the energy loss per turn by adding high field periodic magnetic structures (wigglers or undulators). For ultra-short bunches generated in laser-plasma interaction, novel cooling methods might be necessary to maintain the short bunch length, including single-pass cooling concepts.

\subsection{Accelerating structures}

The core component of the plasma accelerator comprises a region of plasma in which the species, ionization state, density, uniformity, and length are well defined.  This plasma may be created by the driving beam itself, or by auxilliary beams and/or electrical discharges. 

Some requirements of the plasma source are particular to the driver. For laser-driven plasma accelerators the intensity of the driving laser is usually sufficient to create the plasma by ionisation of a target gas. For LWFAs the length of the plasma source should usually match the shorter of the dephasing or pump-depletion lengths, although in advanced schemes, dephasing can be overcome partially by controlling the longitudinal profile of the plasma density. The length of the transition region between the body of the plasma and the surrounding vacuum is important for laser-driven plasma accelerators since it determines the extent to which the laser is defocused as it is coupled into the plasma.  In some regimes of operation (the nonlinear or bubble regime) the driving laser is self-guided through the plasma, whereas in others (the linear or quasi-linear regime) the laser pulse must be guided by an external waveguide. 


Important practical considerations include the operating lifetime and shot-to-shot reproducibility of the plasma source. In many cases, diagnostic access to the plasma is required and isolation it from other parts of the beam line and/or vacuum system.

\subsubsection{Status}
\paragraph{Gas jets}
Gas jets have been used in LWFAs, with plasma densities in the range $n_e = \unit[10^{18} - 10^{19}]{cm^{-3}}$ and lengths from  approximately $\unit[100]{\mu m}$ to \unit[10]{mm}. Gas jet targets are simple, reliable, with excellent diagnostic access, and could operate at high repetition rates; they are also necessary for some schemes for controlling electron injection, such as shock-front injection. However, variations of the plasma density within the gas jet are often relatively high (of order 10\%), and they produce a high gas load for the vacuum system. Moreover, the typical supersonic flow of gas typically produces random fluctuations in the gas density distribution, which in turn might lead to a significant degree of shot-to-shot fluctuations in the main electron beam properties.

\paragraph{Gas cells}
Gas cells comprise a small, gas-filled volume in which a pair of coaxial pinholes allows the driver pulse and electrons to enter and leave. The pinholes have a diameter of order $\unit[100]{\mu m}$ which considerably reduces the gas flow into the vacuum system. A significant advantage of gas cells is that the density within the cell is very uniform, whilst maintaining a reasonably sharp transition from the cell density to vacuum. Multi-chamber, variable length gas cells have been developed and successfully deployed in experiments; these can provide a short region of doped, high density gas in which ionisation injection can occur, followed by a long region of low density gas for acceleration to high energies.

A variant of the gas cell is the capillary gas cell which comprises a capillary with a diameter of approximately \unit[1]{mm}, into which the source gas is introduced by one or more gas feeds. Capillary gas cells produce very uniform regions of plasma, with low flow rates into the surrounding vacuum systems. Structured capillary gas cells are being developed for controlling electron injection and electron dephasing,




\paragraph{Waveguides}
Laser-plasma accelerators operating in the linear or quasi-linear regimes require a waveguide to channel the driving laser pulse over the length of the plasma stage. Two geometries have been used to date: hollow capillary waveguides, and plasma channels.

In a hollow capillary waveguide the laser pulse is guided by grazing-incidence reflections from the capillary wall. This approach has been shown to guide joule-level pulses with peak intensities above $\unit[10^{16}]{W cm^{-2}}$ over lengths of \unit[100]{mm}, and to generate electron beams in the \unit[100]{MeV} range with improved stability over the non-guided case.

A plasma channel comprises a cylinder of plasma in which the electron density increases with radial distance from the axis to form a gradient refractive index waveguide. Plasma channels have been produced by: slow electrical discharges in evacuated plastic capillaries; fast capillary discharges; open-geometry discharges; hydrodynamic expansion of laser-heated plasma columns; and gas-filled capillary discharges. Of these, only the last two methods have been used to accelerate electrons, generating electron beams with energies up to \unit[120]{MeV} and \unit[4.2]{GeV}, respectively.

\paragraph{Hollow and near-hollow plasma channels}
A hollow plasma channel comprises an evacuated (or unionized) cylindrical core surrounded by an annulus of plasma. A laser beam propagating through the core excites plasma waves in the annulus which in turn generate highly uniform longitudinal electric fields in the core region. In hollow and near-hollow the focusing fields are weak.  Emittance growth by Coulomb scattering is greatly reduced. Hollow plasma channels are particularly attractive for positron acceleration since defocusing by the background plasma ions is eliminated.

Hollow plasma channels (albeit with neutral gas in the core) up to \unit[1.3]{m} long, with $n_\mathrm{annulus} \approx \unit[8 \times 10^{16}]{cm^{-3}}$, have been generated by ionizing Li vapour with a high-order Bessel beam. 

\subsubsection{Next milestones}

The following milestones may be identified for the development of accelerating structures:

\begin{itemize}
\item Development of hollow or near-hollow plasma channels.

\item Development of accelerating structures capable of high repetition rate operation ($> \unit[100]{Hz}$) for extended periods.

\item Development of energy recovery systems for extracting (and potentially recycling) energy remaining in the wakefield after particle acceleration; these systems will be particularly important as the mean energy of the accelerator is increased, both to improve the machine efficiency and to avoid damage of the accelerating structure.

\item Development of long ($>\unit[100]{mm}$), low-density ($n_\mathrm{e} < \unit[10^{18}]{cm^{-3}}$) plasma stages providing $>\unit[10]{GeV}$ energy gain.

\item Development of accelerating structures with tailored longitudinally density profiles to overcome dephasing. 
\end{itemize}




\subsection{Coupling/transport components between stages}


LWFA Staging is required to achieve high energy for collider applications. The LWFA staging must preserve bunch charge, in order to achieve efficiency, and preserve normalized transverse emittance on $\sim$10 nm level, to reach a small spot size at the IP and the required luminosity. It should be compativble with operation at high repetition rate and high average power. The inter-stage distance should be minimized to keep the effective gradient larger than 1 GV/m, implying length smaller than 1 km/TeV to limit construction costs.

\subsubsection{Interstage particle beam transport}

In order to transport the beam between stages, the beam should have a small energy spread at the exit of the plasma and an adiabatic release is required, achieved by plasma density tailoring.
Beam optics are necessary to transport and shape the beam to the next plasma module. 
To conserve emittance and keep coupling distances in the range of 1 to 10~$L_{acc}$, where $L_{acc}$ is the length of an accelerating module, 1 kT/m $\times 1$~m scale focussing optics are needed. Plasma-based lenses offer the possibility to achieve these requirements, as they produce strong symetrical magnetic fields. Further investigation of plasma-based lenses, and the effects of beam-driven wakefields, is needed. 

\subsubsection{Compact laser in coupling}


The laser in-coupling distance is linked to  mirror technology and the required laser focal spot for a given value of laser intensity.  Given the required focal spot size, conventional mirrors must have a focal length of the order of 10 m in order to prevent damage by a PW-class laser pulse. 
The use of plasma mirrors could reduce the coupling distance to the 10 cm range. In these, a high laser intensity (of the order of $10^{16} $~W/cm$^2$) generates an optically flat, critical-density plasma surface of high reflectivity, which can be positioned operating only a few centimeters from the accelerator stage.  Since the material of the plasma mirror can be thin, it is possible to place the mirror in the beam axis of the accelerator.

Several options are under study for plasma mirrors,  based on liquid jets,
or tape drives.
For each option, the effects on the electron beam going through the mirror need to be examined, as for example, the  emittance growth due to beam scattering in mirror and collective plasma effects, and the impact on using mirrors with reflectivity lowered to about 80 \% on efficiency.

An alternative method for coupling in fresh drive laser pulses into each stage is the use of a ``plasma on-ramp,'' i.e., a section of bent plasma channel which guides an off-axis laser pulse into the next plasma stage. 
Detailed studies including experiments need to be performed to evaluate the performance of this scheme.

\subsection{Drivers }

The LWFA scenarios outlined in this report require laser performance which is well beyond the current state of the art. Developing suitable laser sources for LWFA applications requires considerable R \& D to improve current performance in the following areas:

\textbf{Power}: the average power performance of laser drivers for accelerators need to be improved from currently the 10--100 W level to the 100 TW  -- 1 PW peak power level and to reach mean powers of 10s to100s of kW per laser.  For driving the injection module, typical operating parameters would be of the order of 1 - 40 J, < 50fs duration pulses at 10s of kHz repetition rate. For the subsequent accelerating modules, longer pulses (100 - 150 fs) with similar  pulse energy and repetition rate would be required.

\textbf{Efficiency}: Current high power (>100 TW) laser sources have wall-plug efficiency at the sub-percent level.  To be viable drivers for planned accelerators, the efficiency must be above the 30\% level, i.e. an improvement of two orders of magnitude over current, flashlamp-based systems. Diode pumping and the use of materials with a small quantum defect will be needed, as well as other methods to maximize electrical to optical efficiency.

There are a number of approaches for developing laser technology that can fulfill these requirements. Detailed discussion of these is beyond the scope of this report, but comprehensive surveys of the requirements and potential laser technology roadmaps can be found in the following documents \cite{ICFA-ICUIL,k-BELLA}.  It is anticipated that kW average power (1 kHz, few J) short pulse laser systems will become available in the near to medium term \cite{k-BELLA}, and such systems will be an important step toward achieving the required laser parameters.
 
Additionally, it is clear that the performance of optics (e.g., damage threshold, longevity, efficiency, heat handling), frequency conversion materials, pulse compressors etc. will need to be improved to be able to allow the reliable operation of these laser sources for a particle physics collider scenario, i.e., 24/7 for months or years at a time, without sustaining optical or radiation damage. Of particular concern for both efficiency and longevity are the optical gratings used in compressor systems and the average and peak power handling capability of routing mirrors. Significant R\&D is needed to address this urgent need over a 5 to 10 year period.




\subsection{Instrumentation}


In future plasma-based colliders, electron beams may be focused to sub-micrometer transverse beam sizes to maximize the luminosity of the collider and ultrashort beam durations will aid to minimize the amount of Beamstrahlung that would impact the performance of accelerator and detectors. Measuring and controlling the properties of these beams is a major challenge and novel advanced diagnostic techniques are required to setup and run such accelerators. 

The beam-integrated \textbf{charge and energy distribution} are measured with integrated current transformers, YAG and phosphor screens, and dipole magnets as magnetic spectrometers.

For measuring \textbf{beam emittance}, source fluctuations, mrad-level divergences, and $\%$ level energy spread make traditional scanning-based methods unreliable. However, single-shot concepts have been recently demonstrated, taking advantage of the energy spread. Under the assumption of an energy-independent source size and divergence over a sub-percent energy spread range (or alternatively, having knowledge of the divergence through other means), the source emittance can be recorded in a single shot.

The \textbf{longitudinal phase space} characterization in laser-plasma accelerators (LPAs) has been largely limited to measuring coherent transition radiation (CTR), either from the intrinsic plasma-vacuum boundary or from inserted foils. Schemes employing electro-optic crystals placed near the beam focus have been demonstrated, although they are limited in resolution to more than tens of fs. Due to space constraints and cost, deflecting RF structures have not been applied to LPAs. However, we anticipate a growing role for alternative deflecting high-gradient structures, such as deflection by a diagnostic laser-driven plasma wave, or through self-streaking by having the beam drive a strong wakefield in, for example, a corrugated structure. 

Combining temporal streaking/deflection in one plane with energy dispersion in the other allows for great progress towards transverse and longitudinal phase space characterization, including source correlations. We emphasize the opportunity for time-sliced properties, such as pointing, current, energy, and energy spread (perhaps emittance as well), to become accessible to new diagnostics. At the same time, basic energy selection and manipulation in the dispersive plane of a chicane will become a critical component of advanced beam lines.

Plasma based concepts need to be explored for characterizing few-femtosecond electron beams. For example, a plasma based technique has been proposed to measure the volumetric charge density of the beam by ionizing a neutral gas with the intense radial space charge fields of the beam, and characterizing the secondary species.
The large unipolar fields of the charged particle beam imparts a significant momentum to the plasma electrons. As they escape with high radial velocities, they leave the ions unshielded. This non-neutral plasma undergoes Coulomb explosion and the resulting dynamics offers new avenues for direct particle beam characterization. The properties of the particle bunch can be retrieved by characterizing the tunnel ionization induced species.  The technique may extend towards other particle beam characterization such as transverse beam size and transverse emittance with high resolution.


\subsection{Simulation}


 
The required level of design effort for a collider will drive development of enhanced simulation capabilities. With sustained support of simulation efforts, it is envisioned that computer hardware and software developments will reduce the time frame for detailed simulations of one collider stage from weeks to hours or even minutes on future exascale-capable supercomputers. By providing rapid turnaround time, high resolution with full physics packages enabled, simulations will transition from powerful tools used to understand and analyze our experiments into `real-time' highly predictive tools used for the design and optimization of novel experiments with high fidelity. In addition, AI --- in particular Machine Learning (ML) --- will be useful to develop very fast models that can be used to guide large-scale parameter scans. 
 
The workhorse algorithm for modeling novel accelerator concepts is the Particle-In-Cell (PIC) methodology, where beams and plasmas follow a Lagrangian representation with electrically charged macroparticles while electromagnetic fields follow a Eulerian representation on (usually Cartesian) grids. Exchanges between macroparticles and field quantities involve interpolation at specified orders. The standard ``full PIC'' implementation is often too computationally demanding because of the large disparities of space and time scales between the driver beams (either laser or particle beam) and the plasma or structure. Speed-up is provided by either performing the simulation in a Lorentz boosted frame (which lowers the range of space and time scales) or using the quasi-static approximation to decouple fast and slow time scales. Additional approximations such as hybrid PIC-fluid, quasi-3D or laser envelope models enable additional savings at the cost of a reduction of domain of applicability. 
 
High-resolution, full three-dimensional standard PIC and quasi-static PIC (when self-injection is not involved) simulations are ultimately needed to capture potential hosing, misalignments, tilts and other non-ideal effects. Ensemble runs of simulations on large parameter space are required to estimate tolerances to those effects as well as study various designs. Yet, high-fidelity modeling of single stages in the 10-GeV range may necessitate several days or weeks on existing supercomputers, while modeling a 1 TeV AAC collider will require the modeling of tens of 10 GeV stages. Hence, it is essential to pursue the development of better algorithms that improve the accuracy of existing plasma physics models (e.g. high-order Maxwell solvers, adaptive time-stepped particle pushers, adaptive mesh refinement, control of numerical instabilities) as well as to port the codes to the next generation of massively parallel supercomputers with multi-level parallelism. It will also be essential to develop efficient vectorization, threading and load balancing strategies for CPU-manycore, GPU or other novel architectures that might arise. Interoperability of codes on the various computer architectures will be essential to reduce the cost and complexity of code development and maintenance.

PIC-based models are also crucial to (a) explore exotic wakefields with non-trivial topologies and geometries that may become pivotal to address the ongoing challenges of plasma accelerators (e.g. high quality positron acceleration, spin polarization) and (b) explore innovative approaches that exploit intrinsic and unique properties of plasma wakefields to pave novel pathways towards relativistic beams with unprecedented properties. 

In addition to the detailed PIC-based models, it is also important to develop fast tools that require far less computational resources and  that can be used to guide the parameter scans. The models used in these tools will be guided by theory and fits to the PIC-based simulations and experimental results. For the latter, the ANA community should start to take advantage of AI-ML and develop models based on accumulated data from simulations and experiments with advanced AI-ML algorithms. 

For LWFA, the priorities for the modeling of the plasma physics for a single stage will be to (a) demonstrate that reduced models can be used to make accurate predictions for a single GeV-TeV stage (in the linear, mildly nonlinear and nonlinear regimes), (b) demonstrate that the PIC algorithm can systematically reproduce current experimental results (coordinated efforts with experiments) and (c) conduct tolerance studies and role of misalignments. It is also essential to integrate physics models beyond single-stage plasma physics in the current numerical tools, incorporating: plasma hydrodynamics for long term plasma/gas evolution and accurate plasma/gas profiles, ionisation \& recombination, coupling with conventional beamlines, production of secondary particles, QED physics at the interaction point, spin polarization.

The development of simulation tools needed for the design of an advanced multi-TeV collider requires robust and sustained team efforts. Teams will consist of individuals with various skills and responsibilities: code builders/developers, maintainers, and users. Members should have combined expertise in physics, applied mathematics and computational science. A multiple-team/code approach has many benefits, allowing for cross-checking of results and pursuit of varied approaches. Yet, while a multiplicity of teams and codes has positive effects, excessive duplications of the same or similar capabilities comes at a cost. Hence, it is important to foster the sharing of modules and codes interoperability via the development of libraries of algorithms and physics modules, as well as the development of standards for simulations input/output and for data structures. To enhance exchanges, interoperability, co-development and verification, the development of open source codes and libraries should also be encouraged.

\section{ What is needed? What do we support?}

 \textbf{Accelerating structures: } 
The most severe challenges to be overcome in developing accelerating structures suitable for ALIC are likely to be: (i) the development of hollow channels for positron acceleration; and (ii) the development of structures capable of handling high average powers.

Overcoming the first of these will require considerable research and development at universities and national facilities, and should be considered a priority.  Overcoming the second will require greater involvement of teams with expertise in advanced materials, handling high thermal loads, and interfacing vacuum systems with a high gas density environment. The formation of one or more dedicated test facilities in this area would therefore likely be extremely beneficial in this regard.

To date the development of accelerating structures has largely been undertaken as an ancillary task by groups investigating other aspects of plasma accelerators. Many groups continue to develop novel accelerator structures, but there are no facilities specifically dedicated to this purpose.

\textbf{Coupling between stages:}
Several developments need to be achieved for coupling successive plasma modules driven by laser, such as  plasma sources with tailored boundaries,  kT/m scale, emittance conserving focussing optics, and  highly stable alignment between laser and electron beam. Alternatives to co-linear laser-electron-beam coupling, need to be studied, as well as the  scalability of technologies to high repetition rate and high average power
 The community should work on a detailed conceptual collider-ready staging design.

 \textbf{Simulations:}
The infrastructure underpinning the simulation efforts will require sustained commitment to: 
(i) recruitment and training of personnel in computational teams;
(ii) development of `resource-saving' novel multi-scale models and numerical algorithms;
(iii) optimizations of those algorithms on constantly evolving computer architectures;
(iv) development of fast models guided by theory and Artificial Intelligence (AI) software fed by detailed simulation and experimental results.

\textbf{Laser technology:}
Significant progress in laser technology will be required for application of LWFAs to ALIC. LWFAs providing energy gain close to 10 GeV have been demonstrated, using chirped pulse amplification (CPA) and titanium-doped sapphire lasers pumped by flashlamp-pumped lasers. However, these lasers have low wall-plug efficiency and, at the pulse energies required, are limited to repetition rates of a few hertz. Fortunately, substantial collaborative efforts between research laboratories and industry are underway to develop efficient multi-joule, diode-pumped short-pulse lasers operating at kHz repetition rates and beyond, with several options being pursued.  Longer wavelength operation of some of these options presents additional design space over which to evaluate and optimize LWFA system performance.

\textbf{LWFA test facilites:}
Multiple facilities with different parameters facilitate the emergence of new ideas and the development of innovation from other fields, particularly from ultra-fast science, plasma or atomic physics. The synergy with accelerators physics and engineering should be encouraged, in particular through the development of instrumentation and suitable fedback systems.

LWFA accelerator test facilities dedicated to or including  beamlines dedicated to accelerator R\&D are absolutely necessary to achieve progress in electron beam quality and reproducibility.
Two flagship facilities are under design for the future development of LWFA, EuPRAXIA in Europe and k-BELLA in the US.

EuPRAXIA is a Horizon2020 design study. It will produce a conceptual design report for the first 5 GeV plasma-based accelerator with industrial beam quality and user areas.  EuPRAXIA is the required intermediate step between proof-of-principle experiments and ground-breaking, ultra-compact accelerators for science, industry, medicine, or, on the long term, the energy frontier (plasma linear collider). The study is designing accelerator technology, laser systems and feedbacks for improving the quality of plasma-accelerated beams. Two user areas are being developed for a novel Free Electron Laser and High Energy Physics detector science. EuPRAXIA, if constructed, would be a new large research infrastructure with an estimated footprint of about 250 m. The design study is laying the foundation for a possible decision on start of construction in 2020. 

k-BELLA (kW Berkeley Lab Laser Accelerator) is a proposed short-pulse, high-field laser facility operating at kHz repetition rate and kW average power levels. High-peak power, short-pulse laser systems, required to drive LWFAs, are available; however, present laser systems operate with average powers up to 10--100 W (at 1 to 10 Hz), and LWFA-based colliders require of the order of 100 kW.  The k-BELLA facility is envisioned as an intermediate step, providing more than an order of magnitude increase in average laser power compared to today's state-of-the-art laser systems.  Such a facility is critical for advancing understanding and development of high-flux LWFAs.  In addition to demonstrating the laser technology, including the high-average power laser components, this facility would address the required development of high-repetition rate targetry, sources, diagnostics, and heat management systems.  The k-BELLA facility, operating at a few J, 100 TW peak power, would be capable of generating 1 GeV electron beams at 1 kHz.  These beams could be used to generate secondary radiation sources for a wide variety of ultrafast science experiments and applications.


%% file: WG5PWFA.tex
%
%
%
%
%




\section{Introduction}

There are currently two PWFA concepts studied for application to HEP. %
The first one considers acceleration of e$^-$ and e$^+$ in wakefields driven by an e$^-$ bunch for application to an e$^+$/e$^-$ collider. %
The second one considers acceleration of e$^-$ in wakefields driven by an p$^+$ bunch for application to what is known as beam dump experiments (e$^-$ on solid targets) in the mid-term an e$^-$/p$^+$ collider in the long term.

The requirements for the main beam parameters for the two schemes are quite different, e.g., ultra-low emittance, e$^-$ and e$^+$ for the first one, mm-mrad emittance for e$^-$ only in the second case. %

The acceleration of the main e$^-$ beam is very similar in both cases, at least at facilities that will be available in the near future. %
The case of acceleration of e$^+$ is described in the WG8 document and is thus only shortly mentioned in this document. %

Since the drive bunch for an e$^-$/p$^+$ collider is expected to carry only a few hundreds of Joules and an energy per particle on the order of 25GeV, this scheme will be based on staging of many plasma acceleration stages interleaved with beam optics to separate the spent drive bunch from the main bunch, refocus the main beam out one plasma into the next and in-couple the new drive bunch. %
Development of such a concept can therefore occur around a single stage (the first one) meeting the specifications for the collider. %
This will be the main focus of experiments in the coming years at exiting and upcoming facilities. %
The next step is to address the staging of two plasmas. %
In order to reach the luminosity required to counter the decrease of the reaction cross section with the square of the energy, the luminosity of such a collider must reach the 10$^{36}$cm$^{-2}$s$^{-1}$. %
Besides large average beam power, this also requires extremely low beam normalized emittance, at the nm-mrad level, for both the e$^-$ and e$^+$ bunches, %
High level of beam polarization is also a must. %

Relativistic proton bunches carry large amounts of energy, many kilo Joules in the case of the CERN SPS or LHC beams and can therefore in principle drive wakefields over very long plasma lengths. %
However, these bunches are 6-12cm-long and therefore too long to drive large amplitude wakefields when matching the plasma skin depth to their length: $\sigma_z\cong\sqrt{2}c/\omega_{pe}$. %
Therefore one has to rely on a self-modulation process to transform the long bunch into a train of short bunches driving wakefields at or beyond the GV/m level. %
Since p$^+$ bunches are usually focused only to micron transverse sizes at the collision point, the requirement on the electron bunch emittance in order to match this transverse focused size is relatively modest, at the mm-mrad level. %
The luminosity of the collider is limited by the characteristics of the p$^+$ bunch, that serves as wakefield driver on one side and collision beam on the other side, and thus requirements on the e$^-$ bunch are also relaxed when compared to the other case. %

We put emphasis on the first application in the following text. %

A compact high-gradient, high-energy e$^-$ accelerator could find application to a $\gamma/\gamma$ collider. %
This scheme would require beam parameters similar to those of the e$^-$ beam of a e$^+$/e$^-$ collider and is thus not specifically described here below. %

\section{ALIC machine components}

The PWFA concept presents many similarities with the CLIC X-band accelerator scheme. %
Therefore, many of the optimizations performed for CLIC can be applied to the PWFA. %
Also, some of the CLIC components can potentially be used for the PWFA, in particular the drive beam complex could possibly be adapted for the PWFA. %
Innovative solutions exist for the main beam source. In particular the plasma-based one could alleviate the need for an electron beam damping ring. %
These are outlined below. %

\subsection{e$^-$/e$^+$ sources (witness)} 
For the PWFA, both the main and drive beam can be produced by sources similar to those that were developed for example for CLIC. %
However, new ideas have be put forward to produce ultra-low emittance electron bunches out of a plasma-based source (see Sect.~\ref{sec:plasmaesource}), therefore suppressing the need for an electron damping ring. %
Positrons can also be produced by very intense laser pulses impinging on target (see Sect.~\ref{sec:positronsource}). %

\subsubsection{Conventional main bunch sources}
The witness must contain $\sim$10$^{10}$ particles with an emittance at the nm-rad level. %
Such bunches have similar characteristics as those envisaged for example for CLIC~\cite{bib:clic}. %
One can take advantage of the designs that have been developed and consider that these could be used as a baseline for the PWFA-based accelerator. %
These include damping rings to reach low emittance values. %
Polarized electron bunches can be produced using semi-conductor photo-cathodes. %
The positron source is based on pair production, particles collimation and cooling in a damping ring. %
Polarized gamma-rays can also produce polarized positrons. %

\subsubsection{Plasma-based main bunch sources~\label{sec:plasmaesource}}

Several techniques have been identified for producing high-brightness beams by direct injection inside the strong longitudinal electric field of a beam-driven plasma wake. The strong longitudinal plasma wakefield causes the injected beam to be accelerated to ultra-relativistic speed in a short duration of time, suppressing emittance growth due to space charge. This allows the ultra-compact initial beam phase space to be maintained as the beam is further accelerated and transported. Most of these beam-driven plasma wakefield injection techniques share general characteristics with injection techniques applicable to laser-driven plasma wakes. The major themes are: 1) rapid plasma density transitions, 2) laser-induced ionization inside the wake, and 3) drive-beam-induced ionization.

{\bf Rapid Plasma Density Transitions.}
The common principle of the rapid plasma density transition-based injection techniques is that they rely upon a brief deceleration of the phase velocity of the rear of the nonlinear blowout wake. This is achieved by forming a negative plasma density gradient along the propagation direction, and it allows nonrelativistic electrons that are able to enter the accelerating phase of the wake (in the rear) sufficient time to be accelerated to relativistic speed without encountering the collapsing plasma electrons forming the rear of the wake. After the initial acceleration, they will propagate near the speed of light, in phase with the driver and the front of the wake. The variations on this theme primarily differ on the method of generating the plasma density gradient, which must occur over a length scale of $\sim c/\omega_p$. Examples of these techniques include density downramp injection~\cite{PB:densityDownramp} and plasma torch injection~\cite{GW:plasmaTorch}.

{\bf Laser-Induced Ionization Inside the Wake.}
These techniques rely upon laser-triggered ionization of a neutral high ionization threshold (HIT) gas. It is assumed that the primary plasma that sustains the accelerating wake is formed from a separate low ionization threshold (LIT) gas that has been fully pre-ionized. The neutral HIT gas may have a static, uniform density distribution throughout the volume of the plasma accelerator, or it may be localized through the use of a gas jet. A major advantage that these techniques provide is the ability to precisely control the initial phase space of the injected beam through the optics of the injection laser. Examples of these techniques include Trojan Horse injection~\cite{BH:trojanHorse}, and colliding pulse injection~\cite{FL:collidingPulse}.

{\bf Drive Beam-Induced Ionization Inside the Wake.}
This method bears some resemblance to the laser-induced ionization techniques, in that a highly localized region of neutral HIT gas is ionized directly inside the plasma wake to produce a high-brightness beam. In this case, the natural betatron oscillation of the drive beam envelope in the plasma is leveraged to create a longitudinally periodic pattern of high field where the beam pinches to a minimum spot size~\cite{NV:beamIonizationInjection}, the pattern of betatron oscillations can be readily predicted, and a localized HIT gas jet can be strategically placed to produce ionization and injection in a single location. A major advantage of this technique is the potential for incredibly robust operation, in that it does not require a precisely shaped plasma density ramp, nor does it require the sophisticated integration of an ionizing laser pulse.

\subsubsection{Plasma-based positron source~\label{sec:positronsource}}

The generation of high-quality laser-driven positron beams is mainly motivated by the necessity of overcoming the intrinsic constraints in the amplitude of the accelerating fields sustainable by radio-frequency accelerators. These constraints currently pose significant challenges for, on one side, the construction of compact and relatively cheap accelerators in the MeV and GeV range and, on the other, the next-generation of TeV-scale electron-positron colliders, the next frontier in experimental particle physics. While significant effort is now put by the research community in studying and optimising plasma-based electron accelerators, plasma-based positron acceleration is still at its infancy. This is predominantly due to the scarcity of positron test facilities worldwide in conjunction with the difficulty of accessing the region of a plasma wakefield that can efficiently accelerate positrons. To date, only FACET-I - together with its current upgrade, FACET-II - can provide short and high-energy bunches of positrons that can be used for post-acceleration studies. It is this scarcity of facilities that is significantly slowing down progress in this area. Experimental work in FACET-I has explored two alternative routes of acceleration, both within the remit of PWFA. The first option uses an electron beam as a driver and a positron beam as a witness~\cite{bib:lotov123} both propagating inside a hollow channel to avoid the positron-defocussing fields. The temporal synchronisation is ensured by propagating the electron drive through a thin converter to generate the positron beam. Alternatively, the self-loaded plasma wakefield acceleration has been proposed~\cite{bib:corde123}, in which a high-charge positron beam is propagated through a hollow channel. The front of the positron bunch is in charge of generating the wakefield, whose accelerating fields are experienced by the back of the bunch. Recently, the transverse fields in a hollow channel have been experimentally characterized using off-axis positron beams~\cite{bib:lind123}. To date though, no experimental results on laser-driven wakefield acceleration of positrons has been reported.  \\ \\ 
A possible solution to this impasse is given by the direct generation of high-energy positron beams following the interaction of an LWFA electron beam with a solid high-Z target~\cite{bib:sari123}. Pioneering experimental work in this area has recently resulted in the generation of high-energy and fs-scale duration positron beams of high spatial quality,  together with the first ever generation of a neutral matter-antimatter plasma in the laboratory~\cite{bib:sari456}.  
In a nutshell, the positrons are generated as a result of a quantum cascade initiated by a laser-driven electron beam propagating through the target. For sufficiently high electron energy and thin converter targets, the generated positrons present properties that resemble those of the parent electron beam, hence the fs-scale duration, mrad-scale divergence, and small source size. These beams present unique advantages for being injected in further wakefields, when compared to more conventional sources. For instance, they have durations comparable to the positron-accelerating region of a wakefield and are naturally synchronised with a high-power laser. 
The main drawback in this approach is the broad energy spectrum that these positrons present at source, essentially extending from zero all the way up a cut-off energy corresponding to the maximum energy of the primary electron beam driving the cascade. However, the relatively low emittance that these positron beams present at the GeV level indicate that passive energy selection can be efficiently achieved, with expected multi-pC positrons beams with 5\% bandwidth at the GeV level for a 5 to 10 GeV primary electron beam. This positron source would be ideal for the test studies of wakefield positron acceleration, an area of research that can currently be experimentally studied only in FACET. \\ \\

To further progress in this area, it is thus necessary to extend the maximum electron energy achievable by a laser-driven electron accelerator, since this will ultimately dictate the maximum positron energy that can be achieved, the emittance of the positron beam, and its charge. Moreover, it is desirable to minimise the duration and divergence of LWFA electron beams, in conjunction with maximising their overall charge. 
We must stress here that, due to the energy broadening induced by the cascade within the solid target, this particular application of LWFA electron beams does not have any requirement on their monochromaticity.  
Currently, positron beams from a laser-driven configuration with the following characteristics have been demonstrated: maximum energy of the order of 0.5 GeV, durations in the fs-range, energy-dependent divergence that can be as small as a few mrad in the high-energy part, overall charges of the order of 100s of pC, and 100\% energy bandwidth. The relatively low emittance of these beams suggest that a few pC of positrons can currently be achieved at the GeV level within a bandwidth of a few percent~\cite{bib:alejo}. \\ \\
For the final goal of a TeV electron-positron collider, it is desirable to achieve sub-micron focussing of TeV-scale positrons with 0.1\% bandwidth. The desired peak luminosity of the collider should exceed $10^{35}$ cm$^{-2}$ s$^{-1}$, an extremely challenging goal also for conventional accelerator technology. In order to fulfill these stringent requirements, it is then necessary to increase the maximum energy, charge, and repetition rate of laser-driven electron beams. Achieving all these intermediate goals is necessary also for the electron arm of a collider, thus implying that developments of the two arms can proceed hand-in-hand. In order for plasma-based accelerators to progress, a laser-driven positron source can prove invaluable as a test beam for positron-wakefield studies.

\subsection{Phase-space control}

The optimization and control of transverse and  longitudinal phase space is a challenge in  any plasma wakefield accelerator. Transverse phase space is defining the emittance and is crucially  important for the brightness of electron beams, and longitudinal phase space is defining the energy spread and chirp and, jointly with current they define the 6D brightness of electron beams.

\subsection{Accelerating structures}\label{accelelstructures}
An energy gain on the order of 5 to 25\,GeV per stage and an accelerating gradient of 5 to 10\,GeV/m require a meter long plasma. %
Three plasma source options are available: the alkali metal vapor source~\cite{bib:muggliLi}, the capillary discharge 
and the plasma discharge. %

One limitation of plasma sources may be the use at high repetition rate. %
In particular energy deposition by the drive bunch in the low density plasma/gas not extracted by the accelerated bunch can lead to gas expansion and displacement. %
Therefore, the bunch train time format of warm colliders (long trains of ns-spaced bunches a few times a second) is not suitable to drive  plasma-based collider modules. %
Preliminary designs~\cite{bib:adli} thus assumed the same number of bunches per second as in collider designs, but equidistantly separated in time. %
This has implications on the detector and other systems of the collider. %

\subsubsection{Metal vapor sources}
An alkali metal vapor source provides a very uniform density ($\delta n/n<1\%$) and very large radius vapor ($\sim$\,cm, no critical alignment) that can be ionized either by a laser pulse\cite{bib:green} or directly by the drive bunch field~\cite{bib:oconnell}. %
Temperature regulation provides the density value and its uniformity. %
It also provides stability over time and fine control and adjustment of the density value, particularly important for a multi-stage system. %
Due to their thermal inertia, these sources are very stable and reproducible in time. %
This can be particularly advantageous for systems with a large number of stages whose density must be tuned precisely. %
Such a source naturally has a meter-scale length and can extend over ten meters~\cite{bib:oz} or more. %
It is a continuously heated system with temperature between 200 and 1000$^\circ$C depending on the desired vapor and thus plasma density (10$^{14}$-10$^{17}$cm$^{-3}$). %
Sources were operated in the 10$^{14}$-10$^{15}$\,cm$^{-3}$ density range with rubidium and in the 10$^{14}$-10$^{17}$\,cm$^{-3}$ range with lithium. %
They naturally provide a density ramp at each end with a characteristic length of a few centimeters~\cite{bib:plyushchev}. %
However, the profile of the density ramp is difficult to control, for example for beam transverse matching to the plasma. %

Alkali metal vapors have low ionization potential (rubidium: 4.17713\,eV, lithium: 5.39172\,eV) and can be field-ionized with low laser intensities (I$_{appear}\cong$1.7$\times$10$^{12}$\,Wcm$^{-2}$ for RbI to RbII) or by the $\sim$6\,GV space-charge field of $\sim$100\,fs, relativistic bunch with 2$\times$10$^{10}$ electrons focused to $\sim$10\,$\mu$m rms transverse size~\cite{bib:oconnell}. %
Ionization by the drive bunch greatly simplifies the plasma source. %

Alkali metals such as rubidium have large ion mass, which is advantageous in mitigating ion motion~\cite{bib:ionmotion}. %
While the source uses the metal, it can be recycled. %
Since the source can be relatively large transversely, precise alignment is not necessary. %
Even when a laser pulse is used for ionization, the pulse intensity remains low ($<$10$^{14}$\,Wcm$^{-2}$) and the pulse does not hit the source wall. %
Since there is no discharge and no laser pulse interception, the life time of the source can in principle be very long. %

The PWFA features large transverse wakefield amplitudes. %
These pause a challenge when needing to match the beam exiting the plasma strong plasma focus to the interstage with weak focus magnetic optics. %
However, the plasma usually has density ramps at each end that can in principle be used to assist the matching by gentle transition from strong to weak focusing. %
This may require tapering of the density ramp to optimally achieve the matching. %
However, previous results have already shown the effectiveness of the naturally occurring ramps. %
The use of active plasma lenses as interstage optics may ease the matching by increasing the interstage focusing and thereby also reducing the interstage length. %
However, their application to collider-like beams is still an open question. %


\subsubsection{DC Discharges} 
Pulsed direct-current discharge plasmas in low pressure ($\sim$\,mbar) gases with plasma diameters around %
1 cm allow for plasma sources of simple setup and low power consumption, which can be scaled to lengths of several meters. %
The discharge current is forced between two ring electrodes separated by a dielectric tube (glass) containing the gas. The symmetry of the electric field in the tube is important to keep the plasma uniformity, for long tubes ($> 30$ cm),
 a metal cage (can be as simple as a few metal wires placed in circle along the discharge) can be used to keep the field uniformity avoiding hot spots in the discharge. 
The metal cage is attached to the anode and extended along the tube up to a distance $L_{gap}$ of the cathode. This distance depends on the gas and pressure and should be longer than the electron mean free path during the beginning of the ionisation process. Additionally, for long plasma tubes, the radius of the cage can be tapered to geometrically force an electric field from where any electron released in the tube (e.g. by photoionisation) can acquire enough energy to start a charge multiplication process. 

By keeping the duration of the pulses short (typically $< 10 \, \mu$s) 
 and keeping the tube diameter constant along the discharge we avoid the development of plasma striations that would result in short scale density fluctuations. 
If the PWFA driving pulse intensity is low enough to avoid ionisation, we can use any ionisation level in the preformed plasma and adjust the pressure to attain the target plasma density. However, to improve the uniformity and reproducibility, we should choose an ionisation level with low sensitivity to small variations in current. For argon, at pressures $< 1 \,$ mbar,
a plasma temperature close to 1 eV
 ensures an ionisation level in the range $90-95$\% corresponding to a maximum of the $Ar^{+}$ population and a negligible amount of $Ar^{++}$ (assuming thermodynamic equilibrium). 
Typically argon is used as a discharge gas, as it provides low breakdown potential, easy handling and rather large ion mass, which helps to mitigate ion motion effects with long duration driving pulses~\cite{bib:ionmotion}.
Preliminary demonstration experiments of these plasma sources suggest the necessary ionisation level can be reached, for plasma tubes in the range 0.1 to 3 m,
with current densities close to 1~kA/cm$^2$
  for 10 $\mu$s
   pulses.

Plasma densities in the 10$^{12}$-10$^{16}$\,cm$^{-3}$ range have been achieved in 0.1~m plasma cells. %
In these short cells shot-to-shot density stability on the order of the measurement resolution of 0.5\,\% has been measured at a plasma density of 10$^{15}$\,cm$^{-3}$. %
Energy deposition on glass and electrode surfaces is low in such sources due to the large diameter\,/\,low gas density and no degradation in discharge performance was observed after several 10$^6$ discharges. %
At the ends of the plasma column density ramps are naturally present. %
On the cathode side the length of the ramp was measured to be on the order of a few cm. %
The ramp length on the anode side was found to be significantly shorter. %

The production of several meter long plasmas requires progressively higher voltage pulses. Besides the cage described above, long plasmas can take advantage of a more advanced igniter-heater circuit with capability to initially impose a short rise time and very high voltage pulse (typically with $> 25\, $kV/m) 
 with a modest current but enough to bring the tube to a clear arc with the consequent reduction in the voltage between electrodes (typically 5-20 A).
 When this arc reaches a considerable current it quickly drains the energy accumulated in the igniter circuit but the heater circuit, with at least an order of magnitude lower voltage, can increase the arc current to the level compatible with the target ionisation level.
A plasma tube with an igniter-heater circuit is under development for the AWAKE experiment (Imperial College, London and IST, Lisbon). Preliminary results with 1~m to 3~m
long plasmas suggest this type of plasma sources can be used to produce uniform plasmas of at least 10 m long.

A critical aspect of the use of this type of plasma sources in accelerators and colliders is the possibility of combining multiple plasma sources to obtain the required beam energy or accelerating two colliding beams. This brings two main problems: how to combine the different plasma sources in the accelerator and how to fine tune the density of each plasma. Although this was not tested yet these plasma sources seem to be able to be grouped in several ways with compact interfaces. By joining electrodes of the same sign and using magnetic isolation (e.g. common mode chokes) we can combine many individual plasma sources with a reduced number of thin vacuum windows in the beam path. In long tubes, the fine tuning of the plasma density (including the production of small density gradients and compensating ion flow) can be achieved with imposing a precise temperature profile along the plasma tube. The desired fine density control requires a precise transversal density diagnostic which can be based on two color interferometry with near infrared lasers.

\subsubsection{Laser-ionized gas} 
Laser-ionized gas plasma sources offer dynamic and precise control over the plasma density profile, which is crucial for beam matching into and out of the plasma source in order to preserve beam emittance. A compressed (10-50 fs) laser pulse of relatively low energy (10-100 mJ) is focused by diffractive optics to form a Bessel function radial intensity profile with high peak magnitude that is sustained over a centimeters-to-meters of propagation distance~\cite{bib:green}, a room-temperature neutral gas is ionized to form a column of plasma. This design allows for maximum flexibility in the density profile of the plasma source, which can be controlled by two means: 1) the optical focal pattern of the laser itself can determine the relative fraction of the gas that is ionized at a given location, and 2) the shaping of the neutral gas density profile, for example through the use of a shaped, elongated gas jet~\cite{HS:elongatedGasJet}, can determine the maximum possible plasma density at a given location.

As stated, the ability to control the plasma density profile is critical to beam matching and emittance preservation. For high energy beams (>10 GeV) with realistic vacuum beta functions (10-100 cm), an adiabatic plasma density ramp required to match into a plasma source of $10^{16} -  10^{17}\,{\rm cm^{-3}}$ needs to be at least 10-30 cm in length. The strength of adiabatic matching is that it is relatively insensitive to the macroscopic shape of the ramp, so long as the average longitudinal density gradient remains below a particular threshold~\cite{KF:adiabaticMatching}. It is also possible to match with much shorter plasma density ramps that are non-adiabatic~\cite{XX:nonadiabaticMatching}, but the matching quality (thus degree of emittance preservation) in such cases is highly sensitive to the exact plasma density profile. It is plausible that an optically determined plasma density profile could meet such a demand, though it would be significantly more challenging for other means of density ramp shaping to reliably do so.

A major advantage of the laser-ionized-gas plasma source is its accessibility: it can be deployed inside a large, optically accessible vacuum chamber for maximum diagnostic access. This allows for the possibility of density profiling through direct optical and spectral observation of the plasma~\cite{DS:starkBroadening}, density diagnostics based on probe laser phase retrieval~\cite{ZL:phaseContrastImaging}, and direct measurement of the plasma temperature through the use of a floating Langmuir probe~\cite{SC:tripleProbe}

Another advantage of the laser-ionized-gas plasma source is that it has no inherent restrictions on the species of gas that can be used. This may be important in the face of undesired additional ionization of many-electron species (e.g. Argon) by the strong field of the electron beam. Instead, a gas that can be completely ionized by the laser, such as Hydrogen, might be used to prevent further ionization by the electron beam. A previously perceived disadvantage of Hydrogen is the potential for the onset of deleterious ion motion within the wake, but recent studies have shown that emittance growth due to ion motion may be significantly smaller than was previously expected~\cite{WA:ionMotion}

The laser-ionized-gas source can handle high repetition rate through high-velocity transverse gas flow. The ionized plasma column need be only 0.5-1 mm in diameter, and thus if the gas can traverse this distance between shots, a freshly ionized and unperturbed region of gas will be ionized on every shot. If operating at 1 kHz, the gas velocity need only be 1 m/s; if operating at 1 MHz, the gas velocity must be 1 km/s. The limiting factor for the rep-rate thus becomes the laser system. If each laser pulse is utilized only once, cutting-edge kHz terawatt Ti:sapphire systems can suffice at 1 kHz rep-rate. Beyond this, optical schemes to recycle each laser pulse multiple times will be required, which could simultaneously increase the energy efficiency of the plasma creation process.

\subsection{Coupling/transport components between stages}
The beam optics requirements between plasma cells include injection
and extraction of drive beams, matching the main beam beta functions
into the next cell, canceling dispersion as well as constraining bunch
lengthening and chromaticity. Non-linear effects must be minimized
to avoid further sources of emittance growth. For a PWFA-LC with many
stages the acceptable emittance growth would be on the percent-level
or below. To maintain a high effective acceleration gradient, this
must be accomplished in the shortest distance possible. The optimal
layout of an interstage optics will depend strongly on the beam and
plasma parameters on a collider, and no working designs that fulfills
all the constraints listed have so far been documented. Some preliminary
ideas and concepts for interstage optics are discussed in \cite{EA:PWFA_LC_ISTAGE}.
We summarize some of the key findings below. Experiments have performed
first interstage demonstrations, however with low energy beams (order
100 MeV) and low charge coupling (order few $\text{\%}$) \cite{EA:BELLA_STAGING}.

Assuming that the plasma stages makes use of the blow-out regime,
the beam exiting the plasma will necessary be strongly focused, with
$\beta^{*}=\sqrt{2\gamma}/k_{\text{pe}}.$  The injection of fresh
drive bunches requires a minimum amount of longitudinal space in the
lattice (order of meters), $L^{*}$ until the first quadrupole. In
this distance the beam propagates with a divergence of $\sim\text{1/}\beta^{*}$,
and for a single lens, the magnitude of chromatic effects scales as
$\sim L^{*}/\beta^{*}$. For the beam and plasma parameters in \cite{EA:PWFA_LC},
the projected emittance growth due to chromatic errors is far above
the acceptable limit if a single quadrupole is used to refocus (match)
the beam into the subsequent plasma cell. Plasma ramps \cite{EA:PLASMA_RAMP}
may be used to generate an effective $\beta^{*}$ several factors
higher at the output of the plasma call than inside, however, even
for a factor ten higher $\beta^{*}$, a single quadrupole would still
give unacceptable emittance growth. Chromatic errors can either be
corrected with sextupoles, or with achromatic quadrupole lattices
\cite{EA:APOCHROM}. Figure \ref{EA:FIG1}, from \cite{EA:PWFA_LC_ISTAGE},
shows an example of an achromatic interstage lattice, with dispersion
cancelled to first order. The second-order dispersion leads to a 2\%
emittance growth. For a 500 GeV beam, the length of this lattice is
39 m. The required lattice length, keeping the lattice structure and
thus chromaticity suppression constant, scales in general as $\sqrt{E_{\text{m}}}$
for cases where the main beam energy, $E_{m}$, is much higher than
the drive beam energy. I.e. all but the lowest energy stages. The
use of axisymmetric lensing, for example active plasma lenses \cite{EA:BELLA_PL},
can reduce the lattice length by a significant factor \cite{EA:APOCHROM}.
However, in the case of active plasma lenses, too intense beams, including
beams with parameters as proposed in \cite{EA:PWFA_LC}, will generate
strong wakefield in the lenses leading to unacceptable emittance growth
\cite{EA:PL_NONLIN}.

\begin{center}
\includegraphics[width=0.25\textwidth]{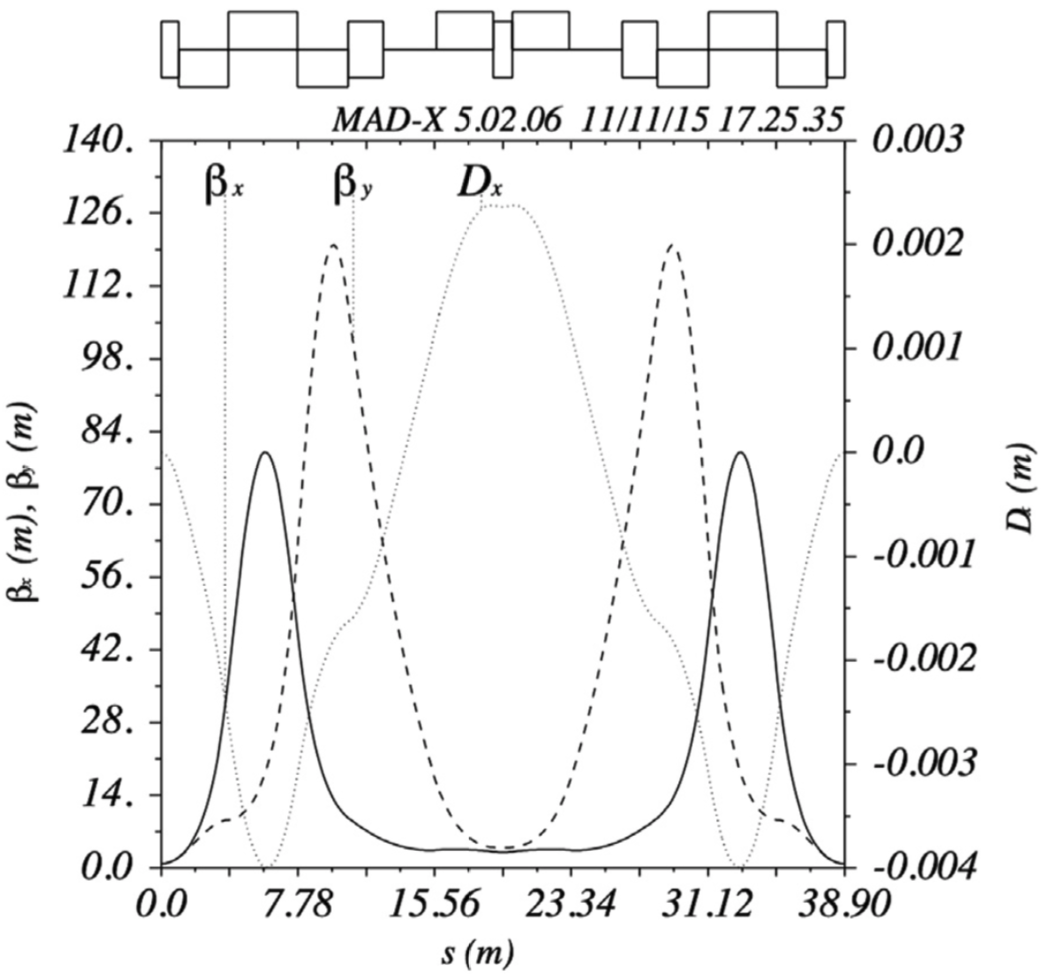}
\par\end{center}

\emph{Figure \label{EA:FIG1} }An example of an achromatic interstage
lattice, with dispersion canceled to first order. The second-order
dispersion leads to a 2\% emittance growth. For a 500 GeV beam, the
length of this lattice is 39 m. From \cite{EA:PWFA_LC_ISTAGE}.

\textbf{CSR considerations}
 The injection and extraction of drive beams could be done using dipole magnets (septum), combined with the fact the energy difference of the main beam and the drive beam is large for all but the first stages.  To close the dispersion in the interstage section, at least four dipole magnets,forming a chicane, are needed.


\section{Integrated system}
\subsection{Instrumentation:}
\subsubsection{Plasma instrumentation}

Accurately (<1\%) measuring plasma density is challenging. %
Therefore, systems that naturally provide the the plasma density through simple control of the source parameters, such as the temperature of the alkali metal vapor source, provide an effective and simple means to insure the plasma density target value is reached for every event. %
Field ionization then turns the gas/vapor density into the same plasma density and thus uniformity. %
One assumes that the gas/vapor/plasma fully recover between acceleration events. %
This has led to a choice of low frequency and continuous(hundreds of kHz) beam time structure. %

The applicability of specific plasma density diagnostics depends upon the type of plasma source in use. Here we will describe several methods that are applicable to one or more of the plasma sources described in this document.

Direct optical observation of the plasma glow with a CCD or CMOS camera is a simple and robust method of assessing the plasma density. The application of a line filter to select only a specific de-excitation line can allow one to selectively observe the plasma density of a particular species in a mixed gas environment. The use of a gated ICCD can reduce the integration time to a few ns, which provides a degree of temporal resolution of the measurement comparable to the plasma recombination time.

The use of an imaging spectrometer can provide longitudinal density information by observing the Stark broadening of the plasma's characteristic spectral lines~\cite{DS:starkBroadening}, which can play a role in the line width broadening. Once again, the use of a few-ns gated ICCD can provide some temporal resolution to the measurement. The resolution of the density measurement is limited by the dispersion of the spectrometer and the pixel density of the detector. Commercially available models can reach a spectral resolution of ~0.1 nm. The spatial resolution of the measurement is determined by the field of view and the pixel density of the detector. One meter of plasma imaged onto a pixel array with a width of 1000 px, for example, can provide roughly 1 mm spatial resolution.

A floating Langmuir probe can be used to measure the plasma temperature. Using a double probe setup, an I-V curve can be produced by sweeping the applied voltage, or in the case of pulsed plasma sources, by combining many current measurements for a given fixed voltage that is swept out over many instances of plasma generation (i.e. firing of the electrodes or ionizing laser). The temperature of the plasma can then be inferred from the I-V curve~\cite{EJ:doubleProbe}, once the floating potential of the plasma has been determined~\cite{SC:tripleProbe}. The single-shot nature of the latter method has clear advantages for pulsed plasma sources, though the uncertainty of the measurement can become sizable if the plasma temperature is too large.

Perhaps the most promising, though most complex plasma density diagnostic relies upon index object reconstruction techniques that utilize phase front distortion of a probe laser pulse. This actually describes a family of techniques with various names such as diffractometry, ptychography, shadowgraphy, phase contrast imaging, and others. In all cases, the fundamental setup includes sending a low energy, ultrashort laser pulse through the plasma source and observing the phase pattern that is induced by the plasma~\cite{ZL:phaseContrastImaging}. In some cases where spectral interferometry is utilized, the pulse may be chirped. Phase retrieval may be accomplished through several means. One example that was used in FACET~\cite{RZ:phaseProbeFACET} relied upon the Gerchberg-Saxton method to take intensity profiles of the laser at multiple object planes and iteratively build a phase map of the laser at each observed plane. From these phase maps, a 2-D or 3-D index profile can be reconstructed, from which a plasma density profile can be inferred. This method provides near instantaneous snapshots of the plasma density profile, and is resolution limited only by the nature of the optical setup and the number of object planes that are simultaneously imaged, which essentially corresponds to the number of cameras that are deployed. Such techniques have even been used to successfully image lightspeed index objects~\cite{ZL:lightSpeedMovies} and could as well be used to image plasma wakes a beam-driven PWFA.

\subsubsection{Beam line instrumentation}
The beam instrumentation requirements for both the Drive and Main beam injector complexes have great similarities with those of the CLIC and the ILC injector complex. They rely on the production of a series of drive beams with an adequate time structure and on the generation and preservation of small emittance for the main beams. 
The technical challenges to measure very small beam size have been already addressed. Single-shot sub-micron beam sizes can be measured using Optical Transition Radiation (OTR) monitors \cite{PK:OTRPSF}\cite{BB:OTRPSF} and Laser Wire Scanners (LWS) \cite{SB:LWS} can provide a non-invasive alternative to OTR, but are more costly and complex to operate. 
PWFA-based colliders are based on the use of shorter bunches than ILC or CLIC, both for the Drive and the Main beams with bunch lengths of 40 and 20um respectively. In the framework of X-ray Free Electron Laser machines, which require even shorter bunches, measurement techniques have been developed and already achieved time resolution in the femtosecond range using either RF deflectors \cite{JM:RFdeflector}, Coherent radiation spectrometry \cite{SW:THzspectrometry}\cite{TM:THzCR} or electro-optical techniques \cite{IW:EOSD}\cite{AC:EOSpatialD}\cite{GB:EOTD}\cite{GB:EOflash}.

The main challenges for PWFA beam instrumentation are all related to the specificity of the acceleration of charged particle in plasma channels, i.e. an excellent spatial overlap and temporal synchronization between the drive and the main bunches. 
The relative transverse alignment between the two beams in the plasma source needs to be better than a fraction of their beam size (i.e. 1/10), which would require both an adequate beam profile and beam position monitoring systems at the entrance of each plasma source. The short distance between the drive and the main bunch of 100-200um (300-500fs) [as mentioned in paragraph xx] puts an additional constrain on the choice of diagnostic technique that would need to measure both beams independently. 

RF cavity beam position monitor have already demonstrated single-shot, sub-micron resolution \cite{SW:CavBPM}, but they have typically a time response in the nanosecond range \cite{FC:CavBPMtimeres} that will not enable to measure both beams separately. In order to provide such a fast time resolution, beam radiation processes like diffraction radiation \cite{PK:ODRbeamsize} or Cherenkov diffraction radiation \cite{RK:Chdrobservation} could be used and would provide beam position measurements with micron resolution in space \cite{RK:DRBPM},\cite{TL:ChDRBPM} and sub-picosecond resolution in time. The measurements of optical photons with such a time resolution was never demonstrated experimentally but X-ray pulses were resolved for example when using detection systems based on fast streak camera. This would need to be studied in the future. 
	
    Measuring beam size in main linac could be done using OTR screens located before and after the plasma sources. One can also consider to install retractable screens inside the plasma source. The tolerance of screens and mechanical interfaces to the high temperature plasma environment should be studied. The use of screens will be anyway limited for long trains due to the thermo-mechanical damage threshold of the screen material, such that non-interceptive diagnostics would be required to overcome this limitation. Recently, techniques based either on the detection betatron radiation emitted in the plasma source. \cite{AC:betatronradiation}, gas ionisation \cite{RT:Gasionization} or Cherenkov diffraction radiation \cite{RK:Chdrobservation} have shown promising results but should be studied in greater detail. Optical Diffraction Radiation (ODR) \cite{PK:ODRbeamsize} can measure a beam size as small as a few tens of micrometres, but its applicability in machine operations would require a dedicated study.
    
A precise synchronization scheme between the drive and main beams is also required all along the CLIC linac \cite{bib:clic}. The requirements are slightly tighter for PWFA with only a 10fs time jitter allowed. However, the fact that both beams are travelling inside the same beam line provides a great advantage as single-shot longitudinal measurements can be performed on the two beams simultaneously and would provide a direct measurement of the relative time overlap between the two beams as recently demonstrated using electro-optical spatial decoding \cite{RP:EOdualbunch}.

After each plasma cell, the drive bunch is extracted from the main linac and dumped, and a fresh drive beam is injected into the next plasma cell to accelerate the main bunch further. This could be done using different techniques, such as a magnetic chicane or RF/THz deflector \cite{LZ:directTHzstreakingM}. Monitoring the cleanliness of the drive beam extraction and its energy and energy spread after deceleration would be easily done by measuring its transverse beam profile at the end of the dump line. This would guarantee that beam losses and activation are minimized along the main linac and would monitor a safe operation of the drive beam scheme, as this is also done in the CLIC decelerator \cite{MO:DBprofile}.


\section{Partners and resources, current activity level. What is needed? What do we support?}
\subsection{Facilities: current activity levels, partners, resource}
Table \ref{tab:PWFAfacilities} summarises the current PWFA facilities world-wide. Whereas AWAKE, CLEAR, FlashForward and INFN SPARCLAB are in operation, FACET-II, CLARA and EuPRAXIA$@$SparcLAB are currently under construction.  MAX IV is in the planning and design phase. 
Depending on the facilities, their main research topic vary and include PWFA, FEL and other applications. 
Apart from AWAKE, which uses protons as drive beam, high energy physics applications are currently not the main focus in the facilities.
FACET-II and INFN-EuPRAXIA$@$SPARCLAB plan to upgrade their facilities at a later stage to provide positron witness beams. 


\begin{table}[h]
\begin{center}

\caption{Overview of PWFA facilities}
\begin{tiny}
\label{tab:PWFAfacilities}
\begin{tabular}{ccccccccc}
\hline\hline
\textbf{}  & \textbf{AWAKE}  & \textbf{CLEAR}  & \textbf{FACET-II} & \textbf{FF$>>$} & \textbf{SparcLAB} & \textbf{EuPR$@$Sparc} & \textbf{CLARA} & \textbf{MAX IV} \\
\hline
operation start  & 2016  & 2017  & 2019 & 2018 & 2017 & 2022 & 2020 & tbd  \\
\multirow{3}{*}{} & & & & & PWFA, LWFA & & & \\
current status & running & running & construction & commissioning & commissioning & CDR ready & construction & design \\
 & & & & &  & & & \\
 
\multirow{5}{*}{} &  & rapid & high energy & MHz rep rate & PWFA with & PWFA with &  & low emittance, \\
unique  &  & access and& peak-current & 100kW average power & COMB beam,  & COMB beam,  & ultrashort &  short pulse,\\
contribution & protons & operation & electrons, & 1\,fs resolution  &  LWFA external & X-band Linac & e$^-$ bunches & high-density \\
 &  & cycle &  positrons & bunch diagn. & injection,  & LWFA ext. inj.  &  & e$^-$ beam  \\
 &  &  &  & FEL gain tests & test FEL & test FEL &  &  \\

& & & & & & & & \\
 
\multirow{3}{*}{} &  & instrumentation & high intensity  & high average power  & PWFA & PWFA, LWFA,  &  & PWFA, \\

research topic & HEP & irradiation &  e$^-$, e$^+$ beam & e$^-$ beam  &  LWFA &  FEL, other  & FEL & Soft  \\
  &  &  AA technology & driven exp. & driven exp. &  FEL &  applications &  & X-FELs \\
  & & & & & & & & \\
user facility & no & yes & yes & no & no & yes & partially & no \\
 & & & & & & & & \\
drive beam & p$^+$ & e$^-$ & e$^-$  & e$^-$  & e$^-$  & e$^-$  & e$^-$  & e$^-$ \\
driver energy & 400\,GeV & 200\,MeV & 10\,GeV & 0.4$-$1.5\,GeV & 150\,MeV & 600\,MeV & 240\,MeV & 3\,GeV \\
ext. inject. & yes & no & no/yes & yes?? & no & no & no & no \\
witness energy & 20\,MeV & na & tb ugraded & 0.4$-$1.5\,GeV & 150\,MeV & 600\,MeV & na & 3\,GeV \\
 & & & & & & & & \\
plasma & Rb vapour & Ar, He capillary & Li oven & H, N, noble gases & H, capillary & H, capillary & He, capillary & H, gases \\
density [cm$^{-3}$] & 1-10E14 & 1E16-1E18 & 1E15-1E18 & 1E15-1E18 & 1E16-1E18 & 1E16-1E18 & 1E16-1E18 & 1E15-1E18 \\ 
length & 10\,m & 5-20\,cm& 10-100\,cm & 1-30\,cm & 3\,cm & $>$\,30\,cm & 10-30\,cm & 10-50cm \\
plasma tapering & yes & na & yes & yes & yes & yes &  & yes \\
 & & & & & & & & \\
 acc. gradient & 1\,GeV/m average & na & 10$+$\,GeV/m peak & 10$+$\,GeV/m peak & $>$1\,GeV/m & $>$1\,GeV/m & na &  10$+$\,GeV/m peak \\
exp. E gain & 1$+$\,GeV & na & $\approx$10\,GeV  & $\approx$1.5\,GeV & 500\,MeV-4\,GeV  & $>$\,500\,MeV & na & 3\,GeV \\ 

 \hline\hline
 
 \end{tabular}
 \end{tiny}
 \end{center}
 \end{table}




\begin{table}[h]
\begin{center}

\caption{PWFA issues addressed by the PWFA facilities}
\begin{tiny}
\label{tab:PWFAissues}
\begin{tabular}{ccccccccc}
\hline\hline
\textbf{}  & \textbf{AWAKE}  & \textbf{CLEAR}  & \textbf{FACET-II} & \textbf{FF$>>$} & \textbf{SparcLAB} & \textbf{EuPR$@$Sparc} & \textbf{CLARA} & \textbf{MAX IV} \\
\hline
emittance preservation & yes & na & yes & yes & yes & yes & yes & yes \\
efficiency & yes & na & yes & yes & na & yes & na & yes \\

 \hline\hline
 
 \end{tabular}
 \end{tiny}
 \end{center}
 \end{table}

\subsection{What is demonstrated, what is not even demonstrated? }
\subsubsection{Demonstrated parameters}
The accelerator must accelerate e$^-$ and e$^+$ with a high gradient, efficiently produce an accelerated bunch with high energy. %
The bunch must have a low relative energy spread (<1\%) and a low emittance. %
It must be stable and parameters must be reproducible. %

Acceleration of electrons at high gradient ($\sim$50\,GeV/m) over a meter scale plasma was demonstrated\cite{bib:blumenfeld}. The energy gain by electrons was 42\,GeV, i.e., on the order of what is expected for a single stage in a high-energy staged approach. %
The acceleration of a witness bunch, though at lower gradient (due to available electron beam parameters) and with GeV energy gain was also demonstrated. %
An energy transfer efficiency from the drive to the witness bunch of 30\% was achieved\cite{bib:Litos}. %
A relative energy spread of a few \% with some evidence of beam loading was measured~\cite{bib:loading}. %
In these experiments the beams normalized emittance was a few tens of mm-mrad. %
At that high level when compared to the emittance foreseen for a future collider, no significant emittance growth was observed. %
No evidence of transverse instability, such as the hose instability~\cite{bib:hose}, was observed. %
Staging of multi-GeV acceleration stages has not been demonstrated. %
Acceleration of positrons was demonstrated at low~\cite{bib:blue} and high~\cite{bib:corde} with corresponding low and high accelerating gradients (See report by WG8). %

Alkali metal vapor sources as plasma sources with meter-scale length and suitable density and length for the required energy gain per stage are routinely used in experiments (See Sect~\ref{accelelstructures}). %
\subsubsection{Next steps}
PWFA research and progress are strongly dependent on available facilities. %
Though the number of facilities delivering bunches suitable to drive wakefields at the GV/m level, in the non-linear regime, over significant plasma length ($\sim$1m) is increasing, their number is still limited to a few. %
In particular, facilities that can deliver independent drive and main bunches do not exist. %

The main topics that these facilities will address are: emittance preservation at the mm-mrad level (emittance available), energy transfer efficiency in the presence of energy depletion of the drive bunch, beam loading for relative energy spread minimization, operation with bunch train rather than single bunch. %

The issue of the plasma source is a major one. %
In particular, active sources (discharges) must be robust enough to reliably and reproducibly operate for long periods of time. %
Passive, alkali metal sources consume the metal and transparent recycling or reloading must be developed and demonstrated. %
The issue of dissipation of the energy not extracted from the wakefields by the main bunch (possibly corresponding to MWs of power) must be studied
Tolerances to density variation in each and between sources must be developed. %

\section*{Acknowledgements}

We wish to thank all attendees and contributors to the WG5 work summarized here.


%
%
%
%
%


%% file: WG6SWFA.tex
%
%
%
%
%
%
%

\section{Introduction}

\begin{figure}[hhhh!!!!!!!!!!!!]
\centering
\includegraphics[width=0.65\textwidth]{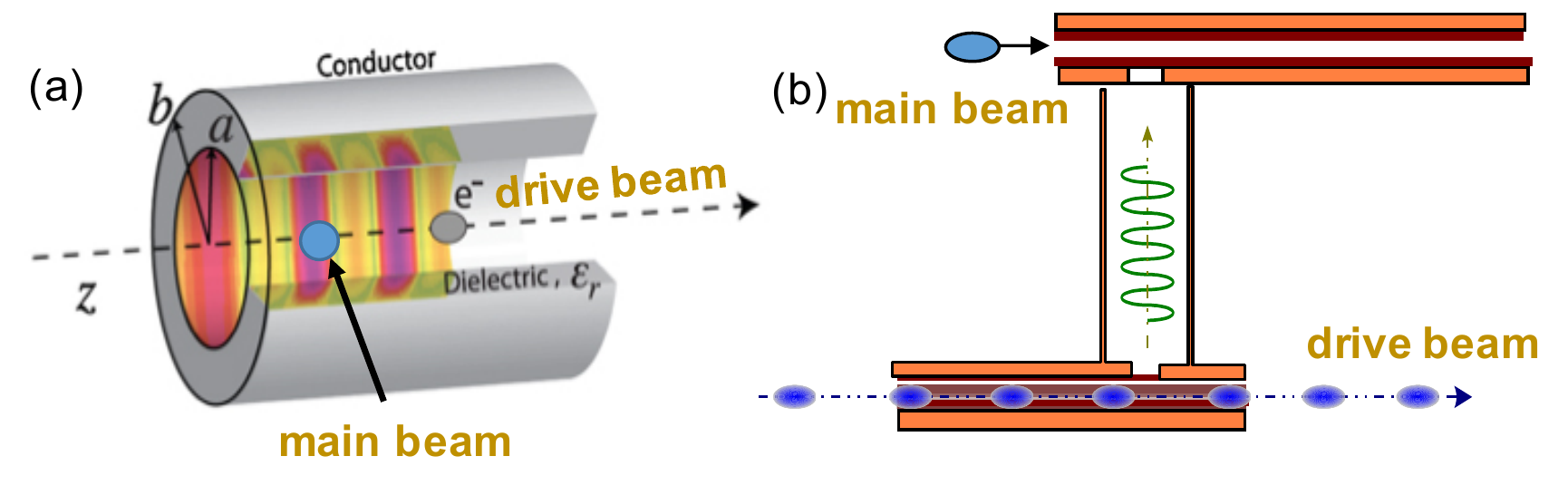}
\caption{Configurations associated with the collinear-wakefield acceleration (CWA) (a) and two-beam acceleration (TBA) (b) schemes.}
\label{fig:cwatba}
\end{figure}

In Structure Wakefield Accelerators (SWFA's) electromagnetic wakefields are excited in a solid-state structure by a charged-particle bunch (the ``drive" bunch). The structures are designed to support a strong axial electric field to accelerate a trailing ``main" bunch. There are two SWFA configurations: collinear-wakefield acceleration (CWA) and two-beam acceleration (TBA); see Fig.~\ref{fig:cwatba}. 

In the CWA configuration, the drive and main bunches follow the same path while in the TBA configuration the fields excited by the drive bunch in one structure are coupled to a separate structure in a parallel beamline to accelerate the main beam.  The CWA has the advantage of requiring only one beamline but the disadvantage of having to stably transport both beams (with very different energy and charge) through the same beamline lattice. It should be noted that some structures have different channels for the beams but are so close together that they still share one lattice so we consider this scheme to fall into the CWA category. An important feature of SWFA for the $e^+/e^-$ linear collider (LC) is its indifference to particle species so that a drive electron bunch will excite the acceleration wakefields for both the electron and positron main bunches. 

The topics discussed during the SWFA working group included a review of electromagnetic structures mediating the wakefield interaction, discussion on auxiliary systems and beam dynamics associated with the drive and main beam, simulation codes, and concepts for staging needed to attain energies required for a LC. 

\section{Machine components}
\subsection{e-/e+ sources, cooling}
At the present stage and given the variety of available structures for SWFA, precise requirements on the transverse beam emittances and bunch duration have not been formulated as it will ultimately depend on the selected structure.
However, one item discussed is the capability to longitudinally shape the main beam to optimize the overall efficiency of the accelerator. For instance in the case of a single-mode structure, it is advantageous to shape the current profile to follow a linearly-ramped shape thereby reducing the main-bunch energy spread by optimally loading the wakefield produced by the main bunch~\cite{vdMeer}.  
%
%
\subsection{Accelerating structures}
A wide of wakefield acceleration structures have been considered for the SWFA scheme. The structure must provide high efficiency and high gradient operation while allowing for control of the beam breakup instability. Over the last two decades, SWFA's have primarily focused on GHz-scale dielectric-loaded waveguides (DLWs)~\cite{gai} in the TBA configuration. More recently, structures utilizing different materials, geometries, and higher frequencies have become an active area of research. High-peak fields have been attained using small-aperture structures which ultimately limits the drive-beam bunch charge and places increased demands on the peak current for a given wakefield amplitude. Average accelerating fields of $\sim150$~MV/m have been achieved~\cite{AWA1} in GHz structures $\sim300$~MV/m have been attained in the THz regime~\cite{thompson,oshea}. Additionally, breakdown tests on metallic and dielectric structures have been performed at several accelerator-test facilities~\cite{dalforno1,antipov3}. Dielectric materials have been considered for a wide array of structures but some possible limitations, along with on-going developments in photonic, have motivated further research on metallic or hybrid (metal-dielectric) structures. Specifically, engineered structures, e.g. using meta-material or plasmonic effects, have opened the path to more-precise control over the properties (spatial distribution, tapering,...) of the excited electromagnetic field to improve the structure accelerating performances. Improvements sought include increase of efficiency ($R/Q$), suppression of harmful electromagnetic modes (e.g. with strong transverse fields), or the mitigation of electromagnetic breakdown. 

An important development has been an increased understanding of the effect of the material properties on the structure performance. The structures boundaries are expected to experience transverse electric field well above ${\cal O}\mbox{(GV/m)}$. The presence of these strong fields can alter the material-properties~\cite{oshea} or induce surface phenomenon such as multipacting~\cite{jingAPL16}. These observations emphasize the need to incorporate material-science considerations in numerical simulations. 

Some of the more advanced structures discussed included photonic band gap (PBG) structures~\cite{simakov}, metallic meta-materials (MTM's),  ``dielectric-assist" Accelerators (DAAs)~\cite{satoh}, photonic topological insulators (PTI's)~\cite{pti}, bragg-reflector dielectric structure~\cite{andonianbragg}, coaxial dielectric wakefield accelerator~\cite{coax}, and conductive-dielectric layered waveguide~\cite{vasili}. Most of these structures have either been tested with beam and their basic accelerating properties verified or will be tested in the near future. Finally, the use of electromagnetic-wakefield amplification by an active medium was also presented~\cite{levi2}.

\subsection{Coupling/transport components between stages}
There is no fundamental physics limitation regarding staging in SWFA. However, it should be pointed out that to date only a simplified version of staging has been demonstrated for TBA in the GHz-regime~\cite{AWA1}, In this experiment the drive beam was sent through both stages and a high-brightness, 1-nC, main beam was accelerated to $\sim 100$~MV/m in two successive stages. 

Staging for CWA scheme is technically more challenging. For instance, the drive-beam transverse jitter will excite transverse wakefields that could cause emittance degradation or loss of the main beam. Tolerance studies of the impact of drive-beam position stability on the main beam dynamics need to be investigated. Likewise, CWA staging at THz frequencies will require sub-picosecond control of the arrival time jitter between the drive and main beams.  These technical challenges associated with CWA staging should eventually be experimentally explored at available facilities. Likewise, CWA-staging experiments will also enable the exploration of main-beam-quality preservation during the staging. 

\subsection{Drivers}
In SWFA, the accelerating wakefield is produced by a relativistic drive electron bunch. One important topic associated with the drive beam formation is the need to tailor its transverse and longitudinal distributions (a need shared with PWFA).

The formation of drive beams with tailored longitudinal distributions~\cite{bane} is critical to the performance of the CWA scheme as it controls the transformer ratio. Longitudinal bunch shaping (LBS) with precision at the picosecond scale has  developed over the last decade owing to advances in phase-space manipulation. A flurry of LBS techniques have been demonstrated including: photocathode-laser pulse shaping~\cite{lsrshaping,lsrshaping2}, transverse-to-longitudinal phase-space exchanger~\cite{eex3}, nonlinear correlations introduced in the longitudinal phase space combined with aberration-controlled beamlines~\cite{england}, achieved using multifrequency linacs~\cite{piotPRL12}, and longitudinal wakefields~\cite{andonianshaping}.  Overall, LBS for the drive has benefited from strong interest and critical proof-of-principle experiments that have been performed at available facilities.

In addition to longitudinal bunch shaping, several of the CWA schemes require transverse bunch shaping. One class of CWA structures incorporate multiple beam channels for one (or more) drive beam(s) and the main beam (e.g. a coaxial structure requires a annular-shaped drive beam with a axial-shaped main beam). Likewise, planar structures,would require the production of drive beams with large transverse-emittance ratios. Transverse shaping has been demonstrated at several facilities using photoemission electron source with parameters comparable to the ones foreseen for the drive-beam electron source of a future LC's. Precise shaping can be obtained by tailoring the photocathode laser-pulse transverse distributions~\cite{marwan,aleksei,ericawa} while large transverse-emittance ratios "flat" beams have been obtained using linear manipulations of magnetized beam~\cite{FBpiot}. 

The formation of drive beams with a transverse emittance partition consistent with THz-scale structures could be challenging. However drive-beam generation benefits from on-going electron-source development in accelerator based light sources. 

\subsection{BDS: IP components and detectors }

The beam delivery system and IP were not discussed in our working group. One question regarding the IP is whether a multiple beam collision scheme would be acceptable to the detector. If so, this could enable the generation of many lower-charge main bunches that could be collided. This would enable lower emittance, and therefore high luminosity, but would also require a detector capable of covering $\sim 4\pi$ solid angle.  Another questions was whether the longitudinally shaped main bunches required to maximize beam loading would degrade the luminosity. 

\section{Integrated system} 

To date, only a straw-man design for a TBA option of a SWFA-based LC has been devised~\cite{AWAflexLC}. In the case of the CWA scheme, subsystems have been investigated separately and a design for a single-stage acceleration module was presented in the context of a short-wavelength free-electron-laser~\cite{ZholentsFEL,LANL_xfel}. 

\subsection{Tolerances}
Small beam misalignment in the SWFA results in beam disruption due to the beam-break-up instability~\cite{bbu} arising from the excitation of transverse wakefields. Such an effect affects both the drive and main beams. 

In the case of the high-charge drive beam, the transverse force leads to a time-dependent kick within the bunch which can lead to the single-bunch beam-break-up (SBBU) instability.  This instability eventually results in beam losses. High-frequency (THz) SWFA's are especially prone to SBBU and mitigation techniques devised for conventional linacs~\cite{bns} are being adapted to SWFA's~\cite{stas1}. The standard SBBU mitigation technique (known as BNS damping) limits the maximum accelerating field for a given SWFA operating frequency~\cite{li_bbu,baturin2}.

The effect of misalignment on the main beam is two-fold. First, a misaligned drive bunch can excite transverse fields which yield to emittance dilution of the main beam due to multi-bunch beam-break up (MBBU) instability. Second, the main beam misalignment can accumulate a time-dependent kick and be subject to the same SBBU instability as discussed above.

\subsection{Instrumentation}
The main challenge with beam diagnostics regards the need to independently resolve parameters associated with the main and drive bunches. When considering the CWA scheme in the THz regime the separation between these two bunches could be a picosecond. Work on this aspect has been sparse to date. During the workshop, experimental results pertaining to a cross-correlation method capable of resolving the temporal separation between the drive and main beams was discussed~\cite{Xdrivemain}. It is conceivable that given the short time scale involve, non-interceptive diagnostics will include frequency-based analysis or radiated field, e.g. in a multi-mode structure, or could possibly include time-resolved diagnostics based on electro-optical sampling. Likewise, the development of ultrafast deflector, e.g., based on the coupling of THz pulse or passive deflection will provide high-resolution interceptive diagnostics.
\subsection{Simulation}
The simulation of an integrated linear accelerator based on SWFA is generally performed piecewise using various degrees of approximation. 

First, the 3D electromagnetic properties of the SWFA structure need to be simulated and this is generally accomplished with electromagnetic eigenmode solvers which are (only) commercially available. An alternative approach is to perform a modal analysis of time-dependent simulation performed with an FDTD electromagnetic program over a sufficiently large number of time step to resolve the frequency associated with the excited electromagnetic modes using electromagnetic solvers available in, e.g., {\sc meep}~\cite{meep} or {\sc warp}~\cite{warp}. The generated Green's functions can then be used in tracking simulation to understand the beam dynamics and guide the design of the SWFA linac. Such an investigation can be performed with open-source programs such as e.g., {\sc elegant}~\cite{elegant}. In fact, very often a simple 1D-1V model considering the beam dynamics in the longitudinal phase space is considered to devise drive-beam shape and optimize the efficiency of the scheme. 

Ultimately, full-electromagnetic simulations using a particle-in-cell framework is performed, e.g. for a single stage of the acceleration, to verify the performance of the scheme.  Over the year open-source tracking codes capable of handling collective effect have been altered to include a set of field ``library" associated with some SWFA structure~\cite{impact1,impact2,astra}. This type of codes has been used to simulate and guide experiments~\cite{yuriclara1}. Likewise, several open-source PIC codes have been available with some capability to model SWFAs~\cite{warp,piconGPU}. Finally electromagnetic simulation can also be performed using wakefield code available within the {\sc echo}~\cite{echo} and {\sc ace3P}~\cite{ace3p} suite of codes. 

One main issue with the modeling of SWFA regards the electromagnetic simulations needed to optimized the performance of a structure. First, very few open-source programs are available thereby preventing the customization of the Physics available in the code. This is especially troublesome as the models associated with some of the physical properties of the SWFA are very limited (for instance the frequency dependence of the dielectric permittivity cannot be included in the programs mentioned above). Likewise, the material-science aspects of the SWFA and the fact that its properties can be altered in strong-field regime are not captured by any of the program currently available.

\section{Partners and resources, current activity level. what is needed? What do we support?}

\begin{figure}[hhhh!!!!!!!!!!!!]
\centering
\includegraphics[width=0.85\textwidth]{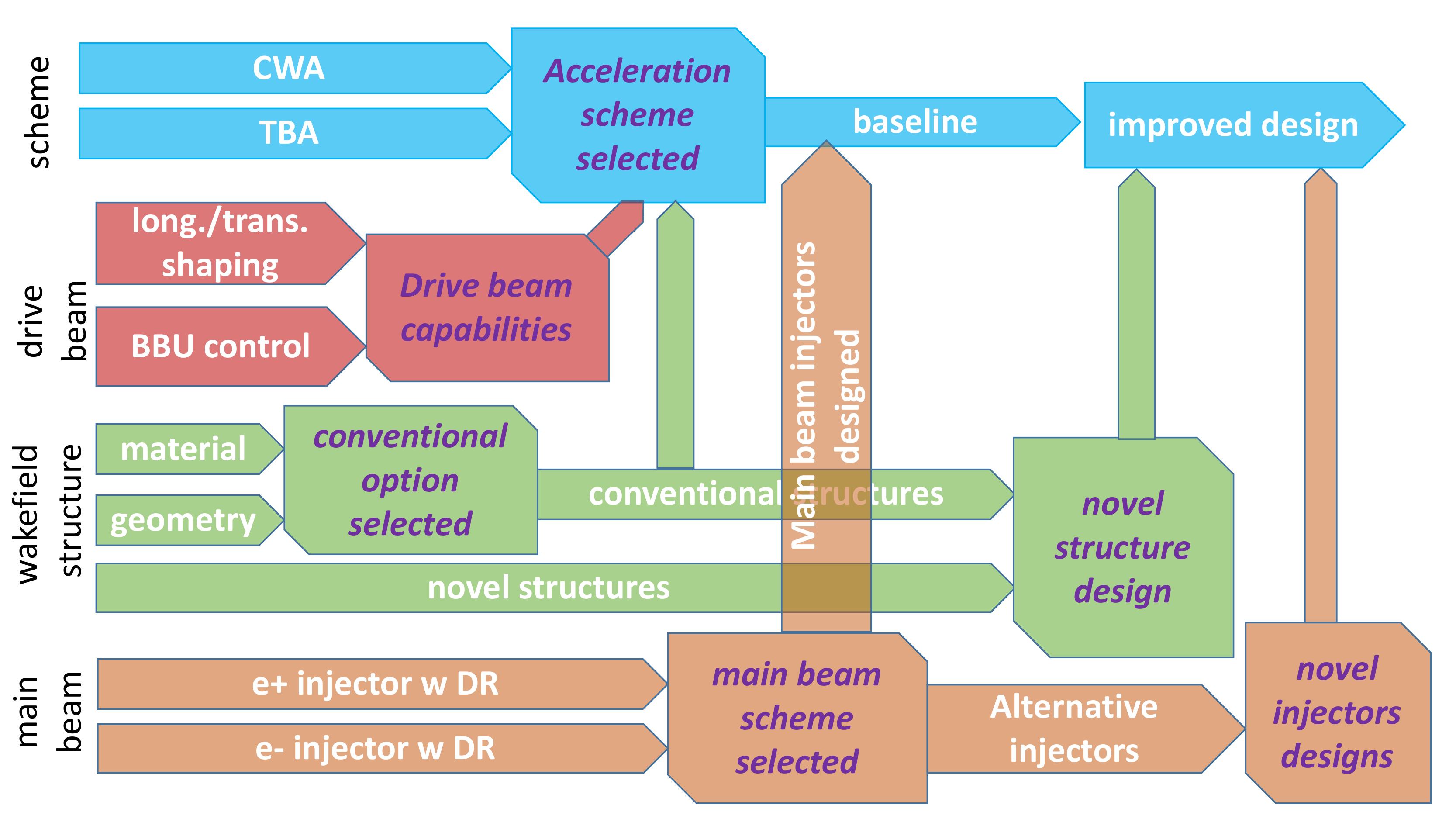}
\caption{Road map toward the development of an SWFA-based linear collider.}
\label{fig:roadmap}
\end{figure}

There are several SWFA options (CWA vs TBA, structure geometry and material) capable of supporting an LC facility. An important near-term goal is to explore and compare the CWA and TBA options and elaborate an algorithm to compare the integrated performances (e.g. gradient, efficiency, impact on beam dynamics) of the various wakefield structure being explored by different groups for a similar straw-man design. The flowchart diagrammed in Fig.~\ref{fig:roadmap}  attempts to summarize a possible strategy toward the elaboration of a baseline design for a LC.  In the near term, we expect a possible CWA straw-man design to be formulated and compared with the proposed TBA. In parallel, a mid-term research effort on the structure should identify structures geometry and material to improve the efficiency and mitigate some of the effects reported above. Once a structure and a scheme is identified the drive and main beams parameters can be specified. 

On the experimental side, there is a number of research activities related to SWFA. A few accelerator test facilities are currently involved in research related to SWFA while other have expressed interest in hosting experiments on the topics. It should be pointed out, however, that most of the on-going experimental work on SWFA is not directly related to linear-collider R\&D but rather aimed at understanding basic properties of a given structure or focus on alternate applications, e.g. to manipulate and diagnose charged-particle beams.

\section{Participants}
The following individual participated to the working group: Alexei Lyapin, Francois Lemery, Yuri Saveliev, Levi Schachter, Sergey Shchelkunov, Vasili Tsakanov, Dan Wang, Igor Zagorodnov, Christina Swanson , Erik Adli. 


%% file: WG7_DLA.tex
\section{Introduction}
\label{sec:intro}

Modern state-of-the-art particle accelerators operate at microwave frequencies and use metallic cavities to confine electromagnetic modes with axial acceleration forces.  The achievable fields in these devices are ultimately limited by the electrical breakdown of the metallic surfaces to accelerating fields of order 10 to 50 MV/m.  However, the basic technologies employed in modern accelerators (metal cavities powered by microwave klystrons) are over 50 years old.  Even particle accelerators with modest particle energies of a few hundred MeV are large and expensive devices accessible mainly to government laboratories.  The largest accelerators used for high energy physics have construction costs in the billions of dollars and occupy many kilometers of real estate.  Constraints on the size and cost of accelerators have inspired a variety of advanced acceleration concepts for making smaller and more affordable particle accelerators.

The use of lasers as an acceleration mechanism is particularly attractive in this regard, due to the intense electric fields they can generate combined with the fact that the solid state laser market has been driven by extensive industrial and university use toward higher power and lower cost over the last 20 years.  Metallic surfaces have high ohmic loss and low breakdown limits at optical frequencies, making them generally undesirable as confining structures for laser-powered acceleration.  Dielectrics and semiconductor materials, on the other hand, have damage limits corresponding to acceleration fields in the 1 to 10 GV/m range, which is orders of magnitude larger than conventional accelerators.  Such materials are also amenable to rapid and inexpensive CMOS and MEMS fabrication methods developed by the integrated circuit industry.  These technological developments over the last two decades, combined with new concepts for efficient field confinement using optical waveguides and photonic crystals, and the first demonstration experiments of near-field structure-based laser acceleration conducted within the last few years, have set the stage for making integrated laser-driven micro-accelerators or ``dielectric laser accelerators" (DLA) for a variety of real-world applications.  

\begin{figure}
\begin{center}
 \includegraphics[height=.35\textheight]{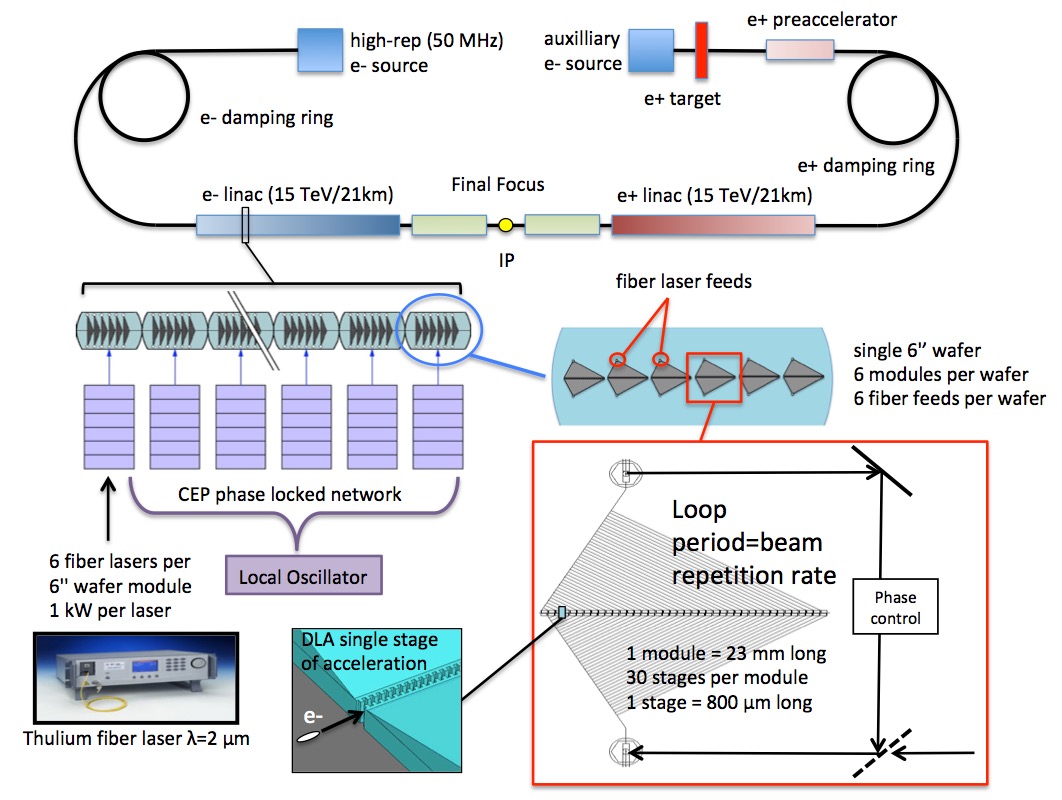}
\caption{Conceptual schematic of a 30 TeV DLA e+ e- collider driven by a carrier envelope phase locked network of energy-efficient solid-state fiber lasers at 20 MHz repetition rate. Laser power is distributed by photonic waveguides to a sequence of dielectric accelerating, focusing, and steering elements co-fabricated on 6-inch wafers which are aligned and stabilized using mechanical and thermal active feedback systems.}
\label{collider}
\end{center}
 \end{figure}

A future DLA-based linear collider, schematically illustrated in Fig. \ref{collider}, will require the development of high-gradient accelerator structures as well as suitable diagnostics and beam manipulation techniques, including compatible small-footprint deflectors, focusing elements, and beam position monitors (BPMs).  Key developments in these areas have been made within the last 5 years, including the demonstration of high average gradients (300--850 MeV/m) with speed-of-light synchronous acceleration in laser-driven dielectric microstructures \cite{peralta_demonstration_2013,wootton_demonstration_2016,cesar_nonlinear_2018}, non-relativistic acceleration with gradients up to 350 MV/m \cite{breuer_laser-based_2013,leedle_dielectric_2015}, and development of preliminary design concepts for compatible photonic components and power distribution networks \cite{hughes:chip:2018,mcneur_elements_2018}. The power distribution scheme is then envisioned as a fiber-to-chip coupler that brings a pulse from an external fiber laser onto the integrated chip, distributes it between multiple structures via on-chip waveguide power splitters, and then recombines the spent laser pulse and extracts it from the chip via a mirror-image fiber output coupler \cite{colby:2011}, after which the power is either dumped, or for optimal efficiency, recycled \cite{siemann:2004}.  Maintaining phase synchronicity of the laser pulse and the accelerated electrons between many separately fed structures could be accomplished by fabricating the requisite phase delays into the lengths of the waveguide feeds and employing the use of active feedback systems.  

The DLA mechanism is also equally suitable for accelerating both electrons and positrons. Example machine parameters for a DLA collider have been outlined in the Snowmass 2013 report and several other references \cite{colby:2011,dla:2011,snowmass:2013,england:rmp2014}.  In these example parameter studies, DLA meets desired luminosities with reasonable power consumption and with low beamstrahlung energy loss.  

\section{Summary of the ANAR 2017 Report}
\label{sec:anar}
\begin{figure}
\begin{center}
 \includegraphics[height=.35\textheight]{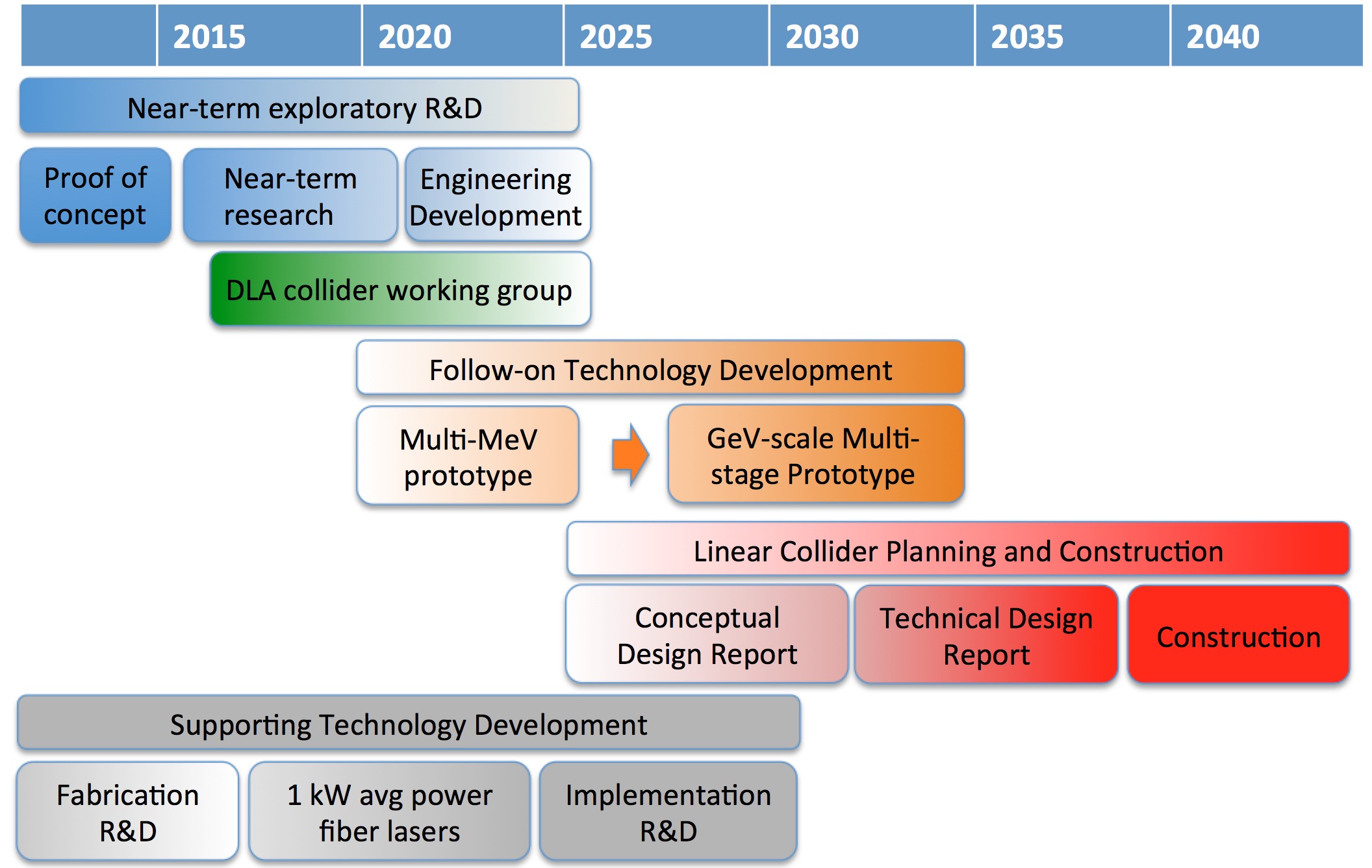}
\caption{Thirty-year roadmap for a DLA collider, reproduced from the ANAR 2017 Working Group 4 Report [Ref].}
\label{30Year}
\end{center}
 \end{figure}

The Advanced and Novel Accelerators for High Energy Physics Roadmap Workshop (ANAR) was held at CERN in June 2017, with the goal of identifying promising advanced accelerator technologies and establishing an international scientific and strategic roadmap toward a future high energy physics collider \cite{anar:2017}.  Four concepts were considered:  laser-driven plasma wake field acceleration (LWFA), beam-driven plasma wake field acceleration (PWFA), structure-based wake field acceleration (SWFA), and dielectric laser acceleration (DLA). Dedicated working groups were convened to study each concept.  The working group on DLA, co-chaired by R.~J.~England (SLAC), J.~McNeur (FAU-Erlangen), and B.~Carlsten (LANL), produced a roadmap to a DLA based collider on a 30 year time scale, as shown in Fig.~\ref{30Year}.  Compact multi-MeV DLA systems for industrial and scientific use are expected on a 10 year time scale.  A dedicated GeV-scale multi-stage prototype system is recommended within 20 years to demonstrate energy scaling over many meters with efficiency and beam quality suitable for HEP applications.  A conceptual design report is projected to commence in 2025, followed by a technical design report, with linear collider construction commencing by 2040.

The DLA working group evaluated current state of the art in the field and identified both significant advantages as well as technical challenges of the DLA approach as a future collider technology. Key advantages include the fact that the acceleration occurs in vacuum within a fixed electromagnetic device, that the acceleration mechanism works equally well for both electrons and positrons, and that the approach is readily amenable to nanometrically precise alignment and optical stabilization of multiple stages.  Complex integrated photonic systems have been shown to provide phase-stable operation for time periods of order days \cite{hulme:2014,xiang:2016}, and nanometric alignment of optical components over kilometer-scale distances has been well established by the LIGO project with stability of 0.1 nm Hz$^{-1/2}$ \cite{ligo:stability}.  In addition, the low-charge and high-repetition-rate particle bunch format inherent to the DLA scheme would provide a very clean crossing at the interaction point of a multi-TeV collider, with estimated beamstrahlung losses in the single percent range, as compared with 10s of percents for more conventional accelerators. 

\begin{figure}
\begin{center}
 \includegraphics[height=.35\textheight]{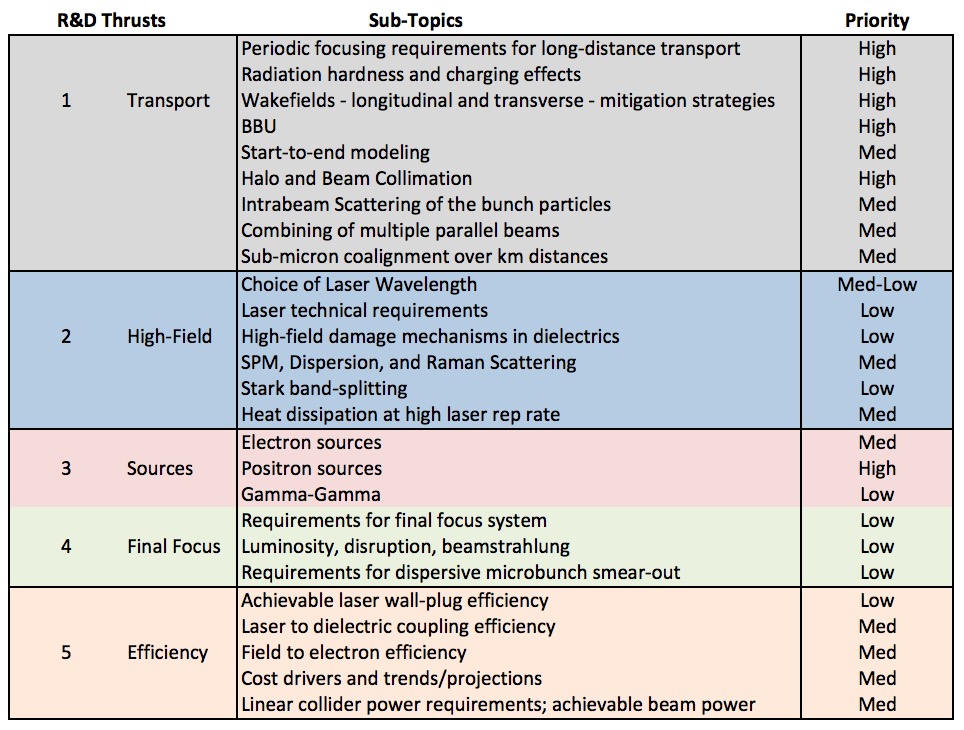}
\caption{Prioritization of technical challenges for a DLA based collider.}
\label{Topics}
\end{center}
 \end{figure}

Technical challenges were prioritized from High to Low, as shown in Fig.~\ref{Topics}, with higher priority items being addressed earlier in the timeline.  Supporting technologies, including high average power solid state lasers and precise nanofabrication methods, were deemed low priority, since the current state of the art in these areas is already at or near required specifications. Detailed considerations of the final focus design and beam collimation were also deemed lower priority:  Due to the very low charge and low emittance beams that a DLA accelerator would intrinsically provide, existing approaches for more conventional accelerators would already be over-engineered for the DLA scenario and could thus be directly applied or perhaps even made more compact. The highest priority challenges identified by the ANAR working group largely pertain to the transport of high average beam currents in the relatively narrow (micron-scale) apertures of nanostructured devices.  These include effects such as beam breakup instability, charging, radiation damage, and beam halo formation, which may be less relevant at low beam powers and beam energies, but can be highly detrimental in a collider scenario.  To these ends, it was recommended to establish a core working group on DLA that would oversee the strawman collider design and motivate these feasibility studies.  It is our intention that the working group formed under the ALEGRO workshop will serve in this capacity. %
This document was submitted as an Addendum to the main input document to the European Strategy for Particle Physics. %

\section{ Facilities using advanced acceleration}
\label{sec:facilities}
{\hskip 0.13in}
Dielectric laser accelerators have a lot of ground to cover to be competitive with other advanced accelerator techniques which have already demonstrated gradients in excess of 50 GV/m and multi-GeV energy gains. The current state-of-the-art in DLAs is accelerating gradients approaching 1 GV/m \cite{wootton_demonstration_2016,cesar_nonlinear_2017} and energy gains on the order of 300 keV \cite{peralta:2013,cesar_pft_2018}.  DLA has the potential to provide high efficiency accelerators operating at very high repetition rates, with bunch formats (charge, beam sizes, emittances, time duration and temporal separation) significantly different than what is commonly available in conventional accelerator facilities. 

For these reasons, it is important for a beam test facility to match the unique characteristics of the DLA accelerators. For example, very high repetition rate electron sources combined with compact, efficient, and relatively low energy ($\mu$J to mJ class) lasers will enable testing of linear-collider relevant concepts such as beam loading and wall-plug efficiency. In this regard we note a general trend in DLA research towards longer wavelengths.  This is motivated in part by a desire to increase the phase space acceptance of the accelerator, which scales with the wavelength both in longitudinal and transverse dimensions. Therefore availability of suitable laser driver pulses in the mid-infrared will be needed. 

Compatibility with DLA's unique features requires high brightness beamlines equipped with diagnostics suitable for the measurement of ultralow (sub-pC to few-fC) bunch charges and ultralow (< 1 nm) normalized emittance. Current relativistic DLA experiments have mostly taken place at low repetition rate facilities such as the next linear collider test accelerator (NLCTA) at SLAC and the Pegasus facility at UCLA \cite{peralta:2013,wootton_demonstration_2016,cesar_nonlinear_2017}. Such facilities are sufficient for initial proof-of-principle experiments, but should in future be coupled with the novel electron sources being developed for high repetition rate free-electron lasers such as superconducting and very high-frequency RF guns. 

\begin{figure}
\begin{center}
 \includegraphics[height=.2\textheight]{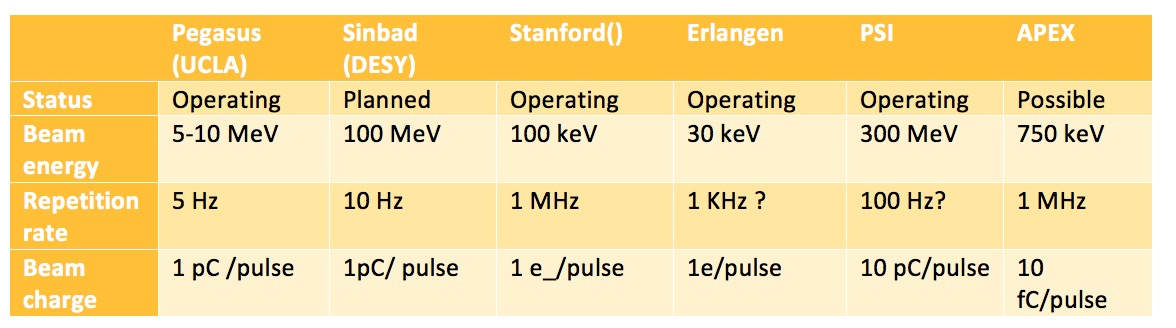}
\caption{Table of DLA Test Facilities}
\label{facilities}
\end{center}
 \end{figure}

A significant body of exploratory research has also been carried out using refurbished electron microscope columns \cite{breuer:2013,leedle:2015,leedle_dielectric_2015,mcneur_elements_2018}.  However, it should be noted that the brightness of these sources is not sufficient to efficiently couple the beam to the small phase space acceptances typical of DLAs. Furthermore, an analysis of coupling efficiency shows that at least mildly relativistic electron energies are strongly favored to maximize the accelerating gradients in the first stages of acceleration.  Another very important characteristic will be the availability of a complete suite of optical diagnostics which should be employed whenever possible to monitor performance and provide active feedback of the laser illumination of the DLA structures.  With all these elements available, the logical progression of DLA experiments to demonstrate suitability for high energy physics applications includes the demonstration of bunching, beam control over longer distances, staging, wakefield mitigation, beam halo, emittance preservation, and efficient energy transfer.

We list in Fig.~\ref{facilities} the major current and future facilities for DLA experiments and the most relevant operating parameters.  To provide an estimate of projected facilities costs, the current ACHIP program includes roughly \$1.3M/year of combined in-kind support from three national laboratories (SLAC, DESY, and PSI) in the form of access to personnel, resources, and beam time. However, for demonstrating many-staged DLA accelerators, ultra-low emittance particle sources need to be developed and combined with DLA devices to make compact injectors. Ongoing development of DLA prototype integrated systems will provide a pathway for scaling of this technology to high energy (MeV to GeV) and to beam brightnesses of interest both for high energy physics and for a host of other applications, as discussed in Ref. \cite{england:review:2016}. 

It is likely that the path to a multi-TeV linear collider will include a multi-stage GeV-scale prototype to demonstrate the feasibility of the candidate collider technology or technologies to confirm gradient, emittance control, and wall-plug to beam efficiency, and to validate the fabrication cost model \cite{P5:2014}. Prior to building a large facility based upon a cutting-edge concept, a demonstration system of intermediate scale is well advised.  The primary purpose of such a demonstration system would be to incorporate interrelated technologies developed under a prior sequence of R\&D steps in order to identify and address new challenges arising from the integration of these components.  Ideally, such a demonstration system would combine all or most of the technological sub-units needed to build a larger-scale system, and would simultaneously possess utility in its own right as a compelling scientific tool.  We envision such a demonstration system based upon the DLA concept to consist of a sequence of wafer-scale modules, which would each incorporate of order tens of single-stage acceleration sections individually driven by on-chip fiber or SOI type guided wave systems for directing laser light and phasing them in sync with the passing speed-of-light particle beam, as described in Refs. \cite{england:rmp2014,colby:2011,dla:2011,snowmass:2013}.  Such a system would illustrate:  (1) integration of the DLA concept with compatible MeV particle sources with nanometric beam emittance and attosecond particle bunch durations, (2) implementation of an accelerator architecture with a pathway to TeV beam energies, (3) carrier envelope phase-lock synchronization of multiple lasers and correct phasing and delivery of laser light over multiple acceleration stages, (4) beam alignment and steering between wafer-scale modules using interferometric techniques combined with feedback, and (5) efficient power handling and heat dissipation.  Due to the high potential cost of such a multi-stage prototype, cost sharing may be possible if the prototype also serves a secondary purpose, such as driving a future light source.
 
\section{Physics considerations that motivate collider energy and luminosity}
\label{sec:physics}
\begin{figure}
\begin{center}
 \includegraphics[height=.25\textheight]{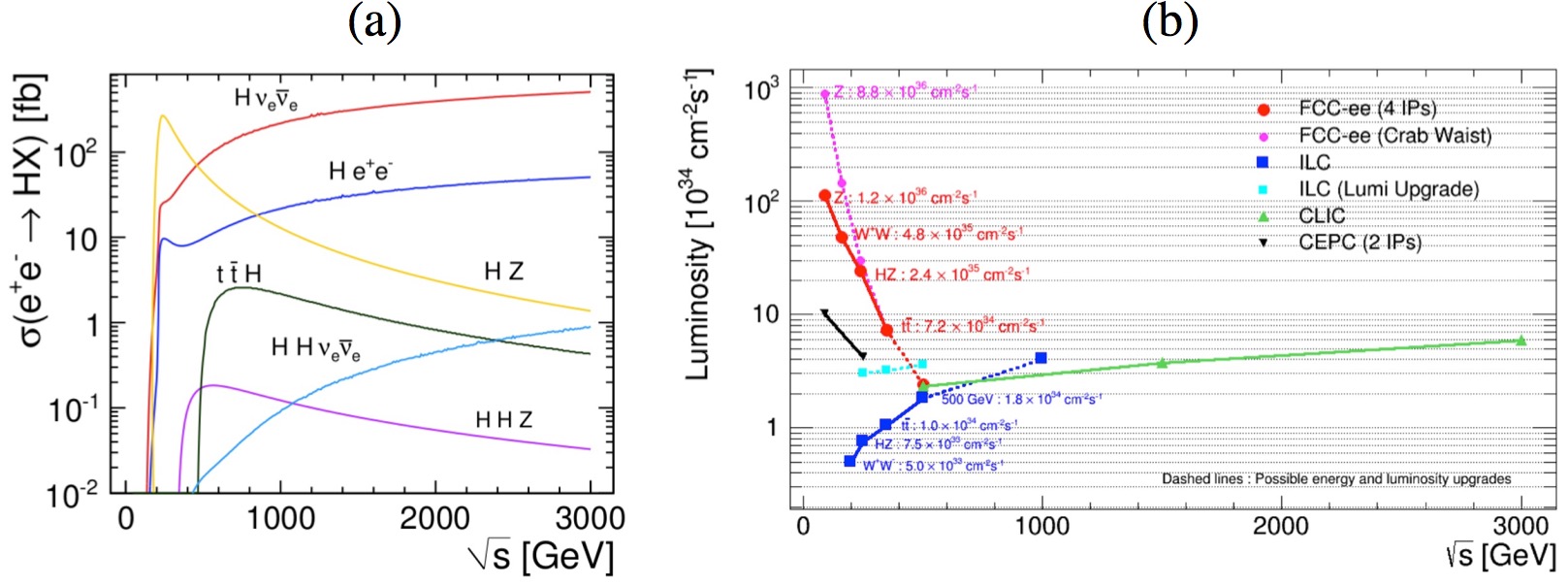}
\caption{Plots of (a) cross section of Higgs production as a function fo center-of-mass energy for a Higgs mass of 120 GeV from Ref. \cite{blanco:thesis:2015}; and (b) luminosities of proposed future e+e- colliders from Ref. \cite{zhu:2015}}
\label{physics_case}
\end{center}
 \end{figure}

{\hskip 0.13in}
The international high-energy physics community recognizes the potential of a multi-TeV e+e- linear collider to make high precision measurements of the Higgs boson's properties, including direct measurements of couplings to top quarks and self-couplings and to study the Higgs contributions to vector boson scattering \cite{P5:2014}. Comparing measured cross sections to predictions (Fig. \ref{physics_case}(a)) will differentiate between the Standard Model and Supersymmetry with an extended Higgs sector. Current plans are to construct the International Linear Collider (ILC) at 500 GeV (upgradable to 1 TeV) to complement CERN's Large Hadron Collider (LHC) upgrade to higher luminosity to explore the TeV energy scale where Higgs superpartners predicted by the Supersymmetry can be observed and to study the interactions of the top quark with the Higgs boson. In addition to a multi-TeV e+e- collider, future high-energy physics machines include a 100-TeV class hadron collider to discover and study new particles and their interactions up to energies scales of about 50 TeV to clarify the Electroweak Symmetry Breaking mechanism. Importantly, a multi-TeV e+e- collider complements this very high energy future hadron collider by providing a cleaner environment to facilitate the detection of new weakly interacting particles more precise measurements of particles \cite{P5:2014}. Particularly, in a linear collider, two elementary particles with known kinematics and spin define the initial state and high-resolution detectors allow high-precision measurements of the final state. In the US, physics using a 100-TeV class hadron collider and a multi-TeV e+e- collider are expected to start in the 2035--2050 time period, first with the hadron collider and then followed by the multi-TeV e+e- collider. In Europe, plans are already underway for a 0.5-TeV CLIC upgradable to 3 TeV.

Practically, a future multi-TeV collider is constrained by construction cost, required experimental luminosity, and power consumption. Ideally, advanced accelerator concepts like DLA will bring construction costs to an acceptable level (of order a few \$B). Experimental precision is aided by having a large collider luminosity which scales as the repetition rate times the number of particles in each colliding bunch, divided by the collision area. Target luminosities of a future multi-TeV e+e- collider will be of the order of 10$^{34}$ to 10$^{35}$ cm$^{-2}$ sec$^{-1}$ as shown in Fig. \ref{physics_case}(b). In rough numbers, a 60 mA average current at 3 TeV center-of-mass will lead to a luminosity of $6 \times 10^{34}$ cm$^{-2}$ sec$^{-1}$ (these numbers correspond to the CLIC value at 3 TeV in Fig. \ref{physics_case}(b), assuming an interaction cross section of 40 nm$^2$), with a total beam power of about 180 MW. It is likely that the total linear collider site power consumption must not exceed 500 MW for such a discovery tool to be practical. Thus, a high wall-plug to beam efficiency will be critical.

\section{Our long-term goal: the Advanced Linear Collider (ALIC)}
\label{sec:parameters}
{\hskip 0.13in}
To reach 30 TeV center-of-mass energies, a next generation lepton collider based on traditional RF microwave technology would need to be over 100 km in length and would likely cost tens of billions of dollars to build.  Due to the inverse scaling of the interaction cross section with energy, the required luminosity for such a machine would be as much as $100\times$ higher than proposed 1 to 3 TeV machines (ILC and CLIC), producing a luminosity goal of order $10^{36}$ cm$^{-2}$s$^{-1}$.  In attempting to meet these requirements in a smaller cost/size footprint using advanced acceleration schemes, the increased beam energy spread from radiative loss during beam-beam interaction (beamstrahlung) at the interaction point becomes a pressing concern.  Since the beamstrahlung parameter is proportional to bunch charge, a straightforward approach to reducing it is to use small bunch charges, with the resulting quadratic decrease in luminosity compensated by higher repetition rates.  This is the natural operating regime of the DLA scheme, with the requisite average laser power (>100 MW) and high (>10 MHz) repetition rates to be provided by modern fiber lasers.

\begin{figure}
\begin{center}
 \includegraphics[height=.35\textheight]{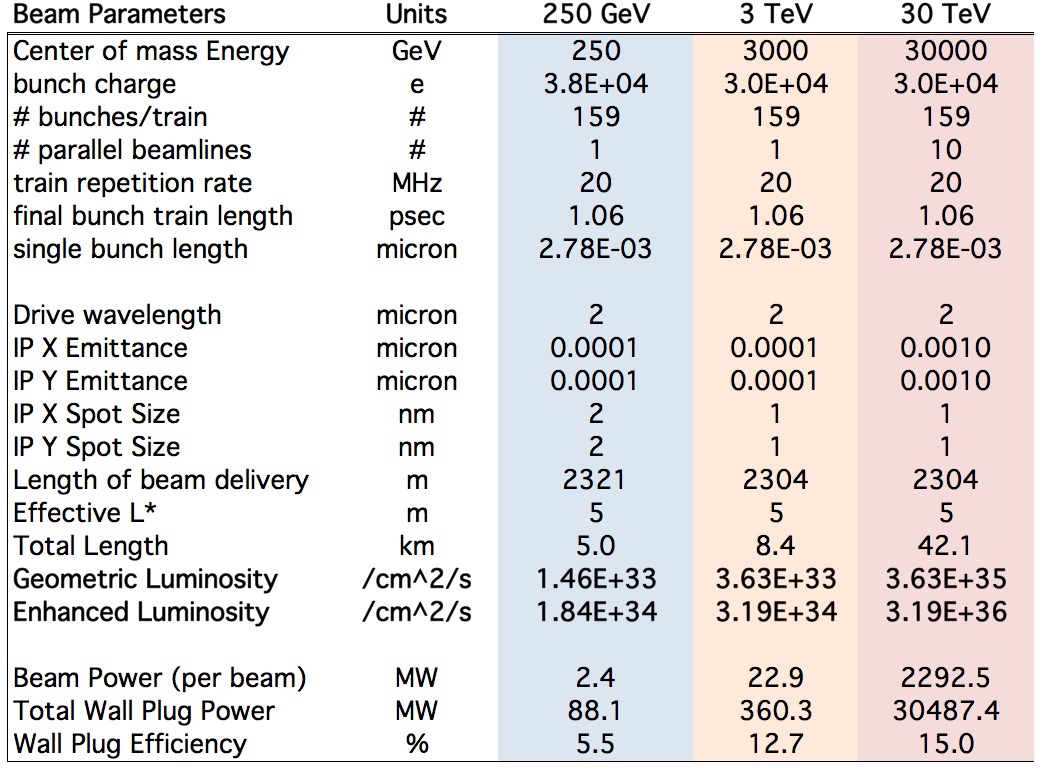}
\caption{Strawman parameters for a DLA based collider at 250 GeV, 3 TeV, and 30 TeV center-of-mass energies.}
\label{parameters}
\end{center}
 \end{figure}

Strawman parameters for the 250 GeV and 3 TeV cases have been previously reported \cite{england:rmp2014,rast:2016}, and these are reproduced in Fig.~\ref{parameters}.  To scale this scenario to 30 TeV we note that the total wall-plug power is proportional to the beam power $P_\text{wall} = P_\text{beam}/\eta$, where $\eta$ is the wall-plug efficiency and the beam power (of both beams together) is $P_\text{beam} = E_\text{cm} n N f_\text{rep}$, where $n$ is the number of bunches per train, $N$ the number of electrons per optical microbunch and $E_\text{cm}$ the center-of-mass energy.  The geometric luminosity scales as
\begin{equation}
\mathcal{L} = \frac { (n N)^2 f_\text{rep}} {4 \pi \sigma_x \sigma_y} = \chi E_\text{cm}^2 ,
\label{eq:luminosity}
\end{equation}
where $\chi = 2.3 \times 10^{33} \text{cm}^{-2} \text{s}^{-1} \text{TeV}^{-2}$ is a scaling constant \cite{king:2000}. We note that the luminosty here scales as $(n N)^2$ rather than $n N^2$ because it is assumed that entire bunch trains (each a single laser pulse in duration) collide at the IP.  For purposes of calculating the disruption parameter, luminosity enhancement, and beamstrahlung energy loss, it is further assumed that the microbunch structure is smeared out prior to the IP, giving a luminosity enhancement of approximatey 10.  Hence the required particle flux scales as 
\begin{equation}
n N f_\text{rep} = E_\text{cm} \sqrt{4 \pi \sigma_x \sigma_y \chi f_\text{rep}} .
\label{eq:flux}
\end{equation}
Combining these relations we obtain the following scaling for wall-plug power with center-of-mass energy:
\begin{equation}
P_\text{wall} = \eta^{-1} E_\text{cm}^2 \sqrt{ 4 \pi \sigma_x \sigma_y \chi f_\text{rep}} .
\label{eq:wallplug}
\end{equation}
Consequently, if the center of mass energy is increased by a factor of 10 (from 3 TeV to 30 TeV), then for similar repetition rate and efficiency, the wall-plug power will increase by a factor of 100, as reflected in Fig.~\ref{parameters}.  At the same time, Eq.~\eqref{eq:flux} requires a 10 times increase in average beam current.  Due to the scaling in Eq.~\eqref{eq:flux} with $f_\text{rep}$ there is a tradeoff between charge and repetition rate. However, since the bunch charge $N$ and laser pulse duration are constrained by the efficiency, gradient, and space charge arguments of Section \ref{structures}, a potential solution is to incorporate 10 parallel DLA beamlines in a matrixed configuration such as that of the 2D honeycomb DLA of Fig.~\ref{structures}(b). The physics of the beam recombination mechanism at the IP requires further study, but for the purposes of Fig.~\ref{parameters} we assume a linear emittance scaling with number of parallel beamlines. In these examples, DLA meets the desired luminosity, and with a small (few percent) beamstrahlung energy loss.  Although the numbers in Fig.~\ref{parameters} are merely projections used for illustrative purposes, they highlight the fact that due to its unique operating regime, DLA is poised as a promising technology for future collider applications.

\section{ALIC Machine components}
\label{sec:machine}

\subsection{e-/e+ sources, cooling}
\label{sec:sources}
{\hskip 0.13in}
Considerable effort at the university level has been directed towards development of DLA-compatible compact electron sources based on laser-assisted field emission from nanotips, which can produce electron beams of unprecedented brightness.  However, comparable techniques for generating high-brightness positron beams in a compact footprint have not been identified.  While miniaturized particle sources are well adapted to the DLA approach and are highly desirable for a variety of applications, they are less critical at the size and cost scales of a collider facility.  Consequently, a possible solution would be to use a high repetition rate cryogenic RF photoinjector as an electron source, with a separate DLA linac for positron production at a target followed by a damping ring.  A feasibility study is needed to understand how to adapt such a source to meet linear collider luminosity requirements with a DLA bunch format.

However, high gradient, high energy superconducting radio-frequency (SRF) guns operated in continuous wave (CW) mode are promising candidates for delivering relevant beams for DLA-based linear colliders. SRF guns have demonstrated reliable operation at 10 MV/m gradient, and there are active R\&D efforts to improve the gradient to >20 MV/m and even 40 MV/m. The CW operation allows in principle that each RF bucket be filled with a photoelectron bunch up to the resonant frequency of the cavity, i.e. typically 1.3 GHz for elliptical geometry guns and 100-200 MHz for quarter-wave resonator type guns \cite{arnold:2011}. The clean vacuum environment inside SRF guns also potentially allow advanced photocathodes to be used, while care must be taken to avoid contamination of the SRF cavity surface by nanoparticles from the cathode. Preliminary simulations show that it is possible to deliver 10 fC, 1 ps, 1.0 nm-rad emittance, 2 MeV electron beams from a 2$\mu$m RMS laser spot on the cathode with 0.2 mm-mrad/mm RMS intrinsic emittance in a 20 MV/m, 200 MHz quarter-wave resonator type SRF gun. 

An alternative approach outlined in Ref.~\cite{schachter:aac14} combines static and laser fields to extract electrons from carbon nano-tubes via field emission. A few thousands of electrons per micro-bunch are predicted theoretically. Immediately after the anode of the injector the bunches are trapped by the field in a tapered booster and their energy is elevated from 40-70 keV to 10 MeV where the relativistic effect becomes substantial, as seen in Ref. \cite{hanuka:trapping:2017}. 

\subsection{Accelerating structures}
\label{sec:structures}
{\hskip 0.13in}
Use of lasers to accelerate charged particles in material structures has been a topic of considerable interest since shortly after the optical laser was invented in the early 1960s.  Early concepts proposed using lasers to accelerate particles by operating known radiative processes in reverse, including the inverse Cherenkov accelerator \cite{shimoda:1962} and the inverse Smith-Purcell accelerator \cite{takeda:1968,palmer:1980}.  Energy modulation of relativistic electrons has also been observed in a laser field truncated by a thin downstream metallic film \cite{leap:2005,sears:2008}.  In these direct optical-scale interactions, an electron bunch longer than the operating wavelength of the accelerator will sample all phases of the accelerating field and will therefore experience an energy modulation.  In order to produce a net acceleration of the electrons using the DLA concept, the bunch must therefore be microbunched with a periodicity equal to the laser wavelength.  Techniques for accomplishing this at the optical period of a laser have been previously demonstrated at wavelengths of 10 $\mu$m and 800 nm \cite{kimura:2001,sears:atto2008}. 

\begin{figure}
\begin{center}
 \includegraphics[height=.18\textheight]{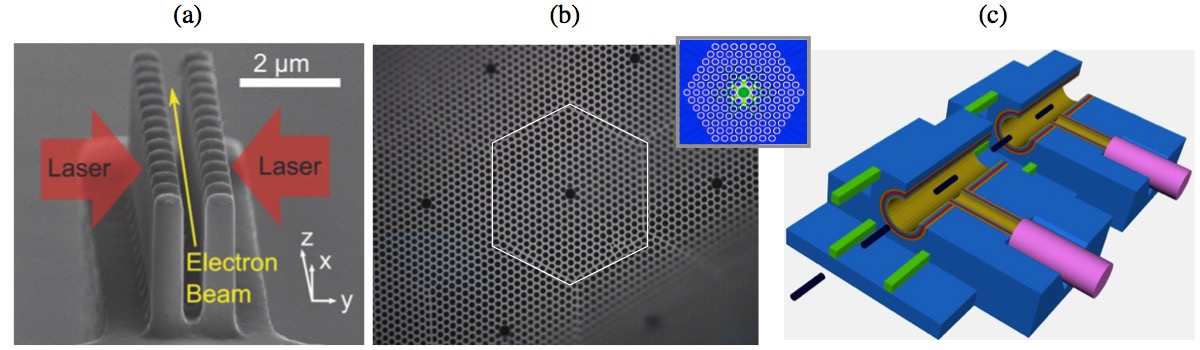}
\caption{Various DLA structures: (a) the planar-symmetric dual-pillar collonade geometry \cite{leedle:2018}; (b) a hexagonal hollow-core photonic crystal fiber geometry with TM$_{01}$ like mode (inset) \cite{noble:2011}; and (c) a proposed cylindrically symmetric Bragg geometry with cutaway revealing the interior \cite{dla:2011}.}
\label{structures}
\end{center}
 \end{figure}

These initial experiments laid the ground work for efficient phased laser acceleration.  However the interaction mechanism used to accelerate the particles in the experiments of Refs. \cite{takeda:1968,palmer:1980,leap:2005,sears:2008} is a relatively weak effect, requiring laser operation at fluences above the damage limit of the metal surface.  This points to the need to use materials with high damage limits combined with acceleration mechanisms that are more efficient.  Due to these considerations combined with the fact that metallic surfaces suffer high ohmic losses and low damage threshold limits at optical wavelengths, a shift in focus has occurred towards photonic structures made of dielectric materials and incorporating new technologies such as photonic crystals and meta-surfaces \cite{rosing:1990,lin:2001,cowan:2003,mizrahi:2004,schachter:2004,naranjo:2012,scheuer:2014}.  The slab-symmetric or planar 1D type of geometry illustrated in Fig. \ref{structures}(a) is simpler to fabricate and its wide aspect ratio helps in improving charge transmission, making demonstration experiments simpler.  This has led to structures of type (a) being the first to be successfully fabricated and undergo demonstration experiments \cite{peralta:2013,mcneur:2012}.  Due to the stringent requirements of a linear collider on beam quality and luminosity, we outline below various key constraints on DLA structure design and beam operation which we will then observe in developing a strawman collider scenario in Section \ref{sec:parameters}.  

\textbf{Loaded Gradient and Efficiency.} For the case of a 30 TeV collider, the average power carried by the electron beam is of order 0.5 GW. In order to get this power into the electron beam, we need twice this power (assuming 50\% laser-to-electron coupling efficiency) in the laser beam. It has been shown \cite{hanuka:single:2018} that maximum efficiency does not occur for the same parameters as maximum loaded gradient. Therefore we must either operate at maximum efficiency and compromise the gradient (increasing the required length of the accelerator) or operate at the maximum loaded gradient and compromise the efficiency and thus running into severe problems of wall-plug power consumption. Several possible regimes of operation have been analyzed in Ref. \cite{hanuka:regimes:2018}.  For our considerations in Section \ref{sec:parameters}, we take as a conservative number a loaded gradient of 1 GeV/m, and a multi-bunch laser-electron coupling efficiency of 40\%. By microbunching the beam, the coupling efficiency of the axial laser field to the particles in a DLA can in principle be as high as 60\% \cite{siemann:2004,na:2005}.  Combined with recent advances in power efficiencies of solid state lasers, which now exceed 30\% \cite{moulton:2009} and designs for near 100\% power coupling of laser power into a DLA structure \cite{wu:2014}, estimates of wall-plug power efficiency for a DLA based system are in the range of 10--12\%, which is comparable to more conventional approaches \cite{england:rmp2014}.  

\textbf{Single Mode Operation.}  The 2D (fiber) type geometry of Fig.~\ref{structures}(b) is similar conceptually to more conventional RF accelerators in that the confined accelerating mode is close to azimuthally symmetric, mimicking the transverse magnetic TM$_{01}$ mode of a conventional accelerating cavity.  However, the preferred fabrication technique (telecom fiber drawing) is less amenable to a fully on-chip approach. A lithographically produced variant of such an azimuthal structure, shown in Fig.~\ref{structures}(c), would allow confinement of a pure TM$_{01}$ accelerating mode, which is beneficial for stable beam transport over multi-meter distances. For such a structure, the radius of the vacuum channel is roughly $R = \lambda / 2$ whereas the bunch radius should satisfy $\sigma_r < 0.1 \lambda$, where $\lambda$ is the laser wavelength. This leads to two important aspects that should be emphasized. One is that due to constructive interference of a train of microbunches, the projection of the total wake on the fundamental mode is magnified. The other is that dielectric materials are virtually transparent over a large range of wavelengths. Consequently, whereas tens of thousands of modes are used for wake calculations in metallic RF cavities, only a few hundred modes contribute to long-range wake effects in a DLA. 



\textbf{Bunch Format.} The luminosity requirement for a 30 TeV collider necessitates of order of 10$^{14}$ electrons per second at the interaction point. Current laser technology permits repetition rates as high as 100 MHz with pulse durations on the order of picoseconds. Consequently, each such laser pulse may contain an electron train of 100 to 1000 microbunches.  The bunch structure and the aperture of the acceleration structure determine the number of electrons to be contained in one microbunch. The effective radial field $E_\perp = (e n \sigma_r / 2 \epsilon_0 \gamma^3)$ generated by a pencil beam is suppressed by the relativistic Lorentz factor $\gamma$, but becomes very significant at low energies. For example, in the non-relativistic case $10^4$ electrons confined in a sphere of radius 0.1 $\mu$m  is of the order of 1 GV/m, which is comparable with the laser field. Consequently, one micro-bunch can not contain more than on the order of a few $10^4$ electrons.  In the example parameters of Fig. \ref{parameters} we assume a microbunch charge of $3 \times 10^4$ electrons/positrons, which is also consistent with efficient multi-bunch operation.  

\textbf{Beam Break-Up (BBU)}.  According to Panofsky \cite{panofsky:BBU:1968}, BBU was first observed in 1966 as the pulse length of the transmitted beam appeared to shorten, provided the beam current exceeds a threshold value at a given distance along the accelerator; the greater the distance, the lower the threshold. Its essentials were found to be transverse fields generated by the beam \cite{chao:1980}. During the years BBU attracted attention every time a new acceleration paradigm came to serious consideration: this was the case for NLC \cite{dehler:1998} and CLIC \cite{braun:2008}, where it has been suggested to damp and detune the structure (DDS) in order to suppress hybrid high order modes (HOM) that leads to BBU instability. Later when the energy recovery linac (ERL) was in focus BBU was investigated in this configuration \cite{hoffstaetter:2004} and further for superconducting RF gun \cite{volkov:2011}. Without exception, the acceleration structure is initially azimuthally symmetric and the hybrid modes are excited due to transverse offset or asymmetry of the beam or to the coupling of input or output arms. In the case of dielectric structures, with the exception of Bragg waveguide \cite{mizrahi:2004}, the acceleration modes are quasi-symmetric since the TM$_{01}$ mode is actually a hybrid mode with a transverse electric component, in addition to other possible hybrid modes. Consequently, for a linear collider, it may be desirable to use an azimuthally symmetric structure, such as the Bragg waveguide of Fig. \ref{structures}(c), or to suppress dipole modes by careful structure design as discussed in Ref. \cite{lin:2001} for the honeycomb photonic crystal geometry of Fig. \ref{structures}(b).

\textbf{Emittance.} As a figure of merit we keep in mind that to remove the beam sufficiently from the structure's wall we assumed that $\sigma_\text{r} \leq \lambda/10$. For a heuristic estimate of the geometric emittance we note that the transverse velocity should be smaller than the velocity required for an electron on axis at the input to hit the wall $r = R$ at the exit of one acceleration stage $z = L$. Therefore, the transverse emittance must be at least of the order $\epsilon_\text{rms} = R \sigma_\text{r} / L \simeq 0.5$ nm.  For the parameter tables of Section \ref{sec:parameters} we assume a normalized emittance of $\epsilon_N = \gamma \epsilon_\text{rms}$ = 0.1 nm at $\lambda$ = 2 $\mu$m. Since $\epsilon_N$ is preserved as $\gamma$ increases, we require that $\epsilon_N$ be matched at low energy ($\gamma \simeq 1$).


\textbf{Laser Power and Heat Dissipation}.  Assuming an average loaded gradient of 1 GV/m, the active length of each arm of the collider is 1.5 km yielding approximately 0.1 MW/m of average laser beam power per accelerating channel, with 33 lasers per meter of active accelerator length and 1 kW of average power required per laser. While CW lasers exceed 50\% efficiency at slightly lower power levels, high-repetition rate lasers currently cannot deliver the necessary average power.  However, laser experts predict that modern fiber lasers will reach the requisite average power levels within 5--10 years \cite{dla:2011}.  Close concentration of such laser energy in a dielectric substrate raises concern about heat dissipation. Compared with metallic surfaces at RF frequencies, the absorption coefficients for dielectrics at optical wavelengths are relatively low. It is found in Ref. \cite{karagodsky:2010} that, ignoring wake field effects, the heat dissipated by the fundamental mode in a Bragg waveguide is at least three orders of magnitude below the practical limit for thermal heat dissipation from planar surfaces (1500 W/cm$^2$). 


\textbf{Guiding and Focusing.} For long-distance particle transport there exists a serious need for a focusing system. Techniques for DLA have been proposed that utilize the laser field itself to produce a ponderomotive focusing force either by excitation of additional harmonic modes or by introducing drifts that alternate the laser field between accelerating and focusing phases to simultaneously provide acceleration as well as longitudinal and transverse confinement \cite{naranjo:2012,niedermayer:focusing:2018}. In simulation, such focusing techniques can adequately confine a particle beam to a narrow channel and overcome the resonant defocusing of the accelerating field. New structure designs and experiments are currently underway to test these approaches. Requiring that the confining force of a focusing lattice is stronger than the repelling force of the charged particles sets a limit on the total number of electrons in a bunch. The latter is determined by the momentum of the electrons and the energy density of the focusing system. In Ref. \cite{hanuka:regimes:2018} the maximum charge is investigated that could be transported in four types of focusing lattices: Einzel lens, Solenoid, Electric or Magnetic quadrupole. While the Electric Quadrupole would facilitate the highest amount of charge, the applied voltage will be limited by the distance between two adjacent electrodes, such that breakdown is avoided. Therefore, it seems inevitable to split the bunch into a train of bunches in order to weaken the space-charge. This space-charge reduction comes at the expense of the maximum efficiency, which is at least 20\% lower than the single bunch configuration \cite{hanuka:multi:2018}.  

\subsection{Coupling/transport components between stages} 
\label{sec:coupling}


A major challenge of DLA is scaling up the interaction length between the driving laser and the electron beam, which is limited by both the beam dynamics and the laser delivery system. A promising solution is to use integrated optics platforms, built with precise nanofabrication, to provide controlled laser power delivery to the DLA, which would further eliminate many free-space optical components, which are bulky, expensive, and sensitive to alignment.  The laser control mechanisms may additionally be implemented on-chip, which will add to the compactness and robustness of the device and allow for precise implementation of laser-driven focusing schemes.

\begin{figure}
\begin{center}
 \includegraphics[height=.2\textheight]{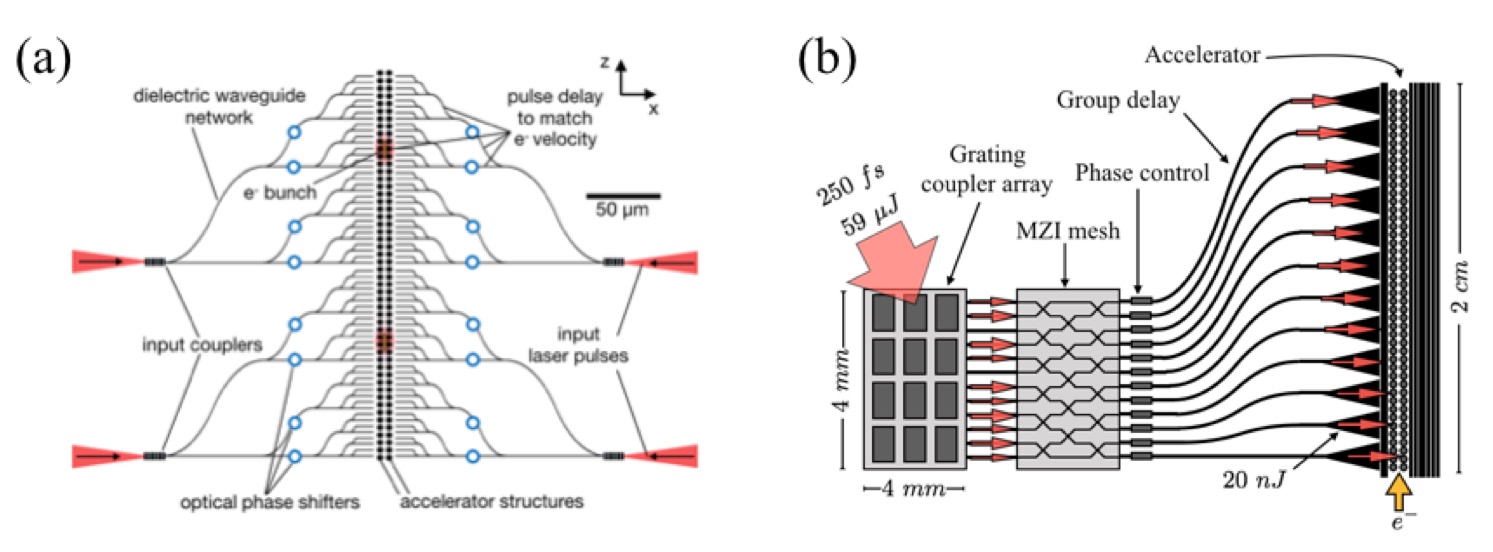}
\caption{(a) The DLA power distribution network concept proposed in \cite{hughes:chip:2018} using dielectric waveguides to split and delay a single input pulse to the accelerator structure, (b) A schematic of a proposed waveguide-fed DLA designed for long interaction length (not to scale).}
\label{treebranch}
\end{center}
 \end{figure}

A system for laser coupling to DLA was recently proposed in Ref. \cite{hughes:chip:2018}, in which the laser beam is first coupled into a single dielectric waveguide on the chip and then split several times to spread over the accelerator structure.  Here, waveguide bends are designed to implement an on-chip pulse-front tilt, which delays the incident laser energy to arrive at the accelerator structure at the same time as the moving electron beam \cite{plettner_proposed_2006,wei:2017,cesar_pft_2018}.  While this work provides a way to achieving interaction lengths on the 100 $\mu$m to 1 mm scale, the power splitting approach has the disadvantage of concentrating the optical power at a single input facet.  For longer length structures, requiring more splits, the input facet becomes a bottleneck for damage and nonlinear effects, and future versions of DLA using integrated optical power delivery systems would ideally have ``one-to-many'' coupling mechanisms where a single laser beam is directly coupled into several waveguides, eliminating this bottleneck. One approach to this kind of coupling uses a large array of grating couplers on the surface of the chip, each supplying power to an individual waveguide, as shown in Fig.~\ref{treebranch}(b).  Theoretical studies of grating couplers, combined with inverse design optimization have shown that coupling efficiencies close to 100\% may be possible \cite{su:2018}.  With areas of several $\mu$m$^2$ having been demonstrated for grating couplers, several thousand may fit on a mm$^2$ area, which may easily be aligned with a free-space laser source.

To increase the robustness of the DLA coupling, an integrated mesh of Mach-Zehnder Interferometers (MZIs) could be fabricated onto the chip.  These MZIs have been experimentally demonstrated and act as tunable beamsplitters that may share power between waveguides as controlled by integrated optical phase shifters  \cite{miller:2015}.  Initial simulations suggest that for a 250 fs pulse, up to 10 MZIs can be accommodated, which would roughly correspond to being able to share power between 10 adjacent waveguides and should be sufficient for the purposes of DLA.  Phase control control may be accomplished by thermal or electro-optic phase shifters integrated on final waveguide sections.  Either the electron beam signal or the light scattered out of plane may be used as a diagnostic tool for sequentially optimizing the phase shifters.  These phase shifters may also be used to implement laser-driven focusing schemes, such as ponderomotive focusing \cite{naranjo:2012} or alternating phase focusing \cite{swenson:1976}, which have shown significant promise for DLA in recent simulation studies \cite{niedermayer:2017}.  Thus, integrated optical phase control gives a path forward for combined acceleration and focusing of the electron beam.

The group delay necessary for matching the arrival of each pulse to the moving electron bunch can be implemented by designing the fixed waveguide geometry, such as the bends described in \cite{hughes:chip:2018}, in combination with subwavelength gratings \cite{wang:2015} embedded the waveguides.  With bend radii as low as 50$\mu$m, it is possible to get close to 100\% transmission through the bends \cite{hughes:chip:2018}.  Additional stages may be necessary for compensating dispersion encountered in the waveguides.  This will be especially important for longer structures.  Previous simulations \cite{hughes:chip:2018} showed that these effects will occur at around 1 cm waveguide lengths when using weakly-guided SiN waveguides.  An attractive option is to engineer this dispersion to avoid damage and nonlinearities, by sending in an initially chirped and broadened pulse, providing recompression closer to the accelerator.

The coupling from waveguide to several DLA periods may be accomplished with an inverse-taper on the waveguide.  Alternatively, the DLA structures may be etched directly into the waveguide, such as in a buried grating \cite{chang4}.  This method is currently being tested experimentally.  It was shown in Ref. \cite{hughes:chip:2018} that a moderate amount of resonance may be beneficial for enhancing the electric fields in the accelerator gap and avoiding the damage and nonlinear constraints in the waveguides.  Thus, a quality factor of about 10 may be useful to design into the DLA structures either by inverse design using the adjoint method \cite{hughes:avm:2017} or by defining dielectric mirrors surrounding the DLA structures.  The entire structure may either be driven symmetrically on each side, or, alternatively, a dielectric mirror may be used to reflect the incoming light from one side of the device.

\subsection{Drivers}  
\label{sec:lasers}
{\hskip 0.13in}
The drive laser requirements for a DLA based accelerator reflect the power and efficiency requirements for future real-world applications as well as the unusual pulse format of the electron beam:  namely pulse energies in the range of 1 to 10$\mu$J, 1 ps pulse duration, 10 to 100 MHz repetition rates, and high (> 30\%) wall-plug efficiencies. In addition, the optical phase of the base carrier wave needs to be locked to the phase of the accelerating electron beam.  The nominal laser type will probably be a fiber laser because of its efficiency and robust, low maintenance operation.  Fiber lasers at 1 micron wavelengths and hundreds of Watts of average power have already been demonstrated to be capable of meeting most of these parameter requirements, and higher power (>1kW) mode-locked systems at longer wavelengths (e.g. 2$\mu$m Thulium-doped lasers) are now commercially available. Consequently, the current state of the art in laser systems is not far from what will eventually be required for large-scale accelerators based upon DLA.

\begin{figure}
\begin{center}
 \includegraphics[height=.25\textheight]{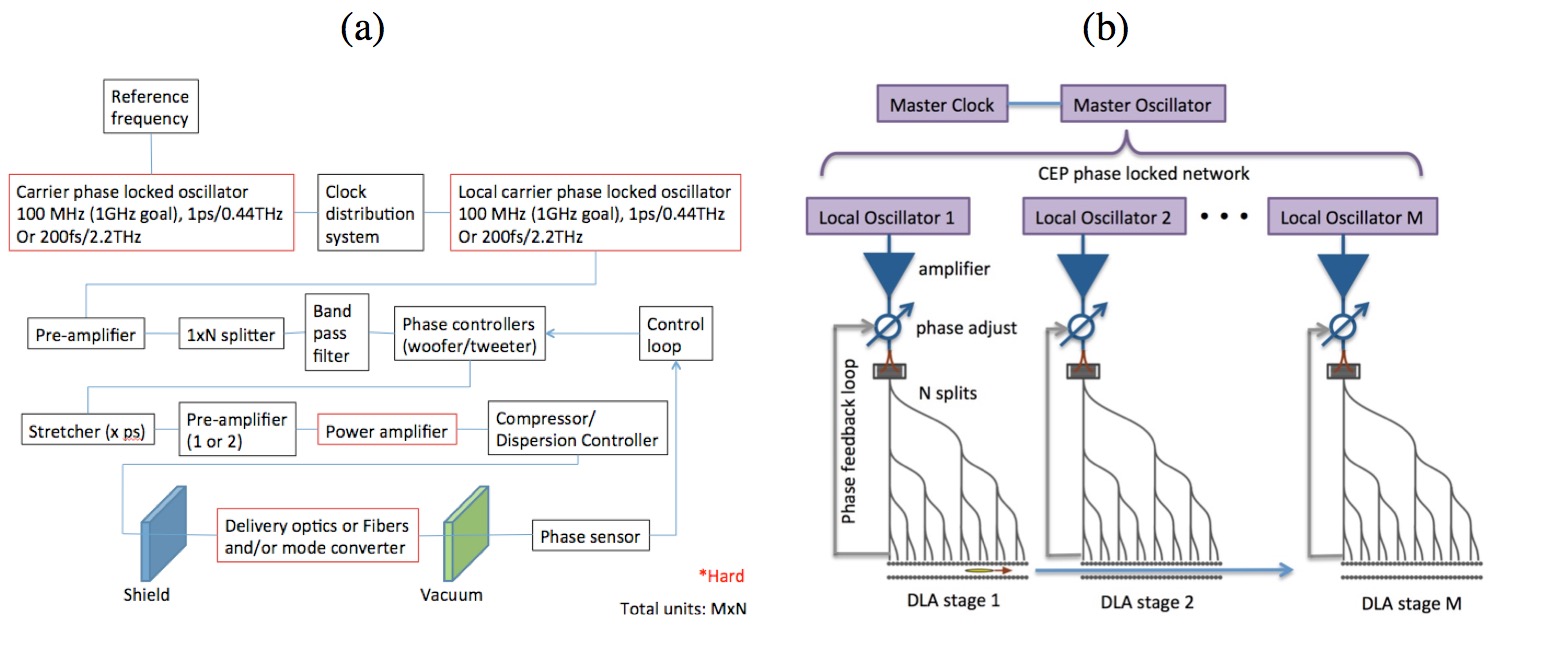}
\caption{Conceptual laser system baseline design for a multi-stage DLA accelerator with $M$ local oscillators, showing (a) block diagram as outlined in Ref. \cite{dla:2011} and (b) a simplified component-level schematic.}
\label{laser_schematic}
\end{center}
 \end{figure}

A baseline conceptual design for a laser system for a multi-stage DLA is shown in Fig.~\ref{laser_schematic}, adapted from \cite{dla:2011}. The design is modular to enable scaling to higher energies with more stages, with timing across a long accelerator as one of the significant technical challenges. The baseline design begins by producing a carrier envelope phase (CEP)-locked oscillator with its repetition rate matched to a stable RF reference frequency source in the range of 100 MHz to 1 GHz, with 1 GHz being the target. This oscillator will serve as the clock for the accelerator. The global oscillator or clock will be distributed via optical fiber to local oscillators, which are phase-locked to the global oscillator. Each structure will require a phase control loop to allow for acceleration through successive structures. Both fast and slow control of the phase will be necessary. By monitoring the energy linewidth as well as the timing of the electron bunches, successful acceleration through the structures may be confirmed.

The baseline conceptual design looks to be a manageable system, with the toughest challenges coming from the requirements for the oscillators, the power amplifier, and the delivery optics. In addition, it will be necessary to repeat the local system multiple times, with each local system phase-locked to the global oscillator. We emphasize that the conceptual illustration of Fig.~\ref{laser_schematic} is our envisioning of a fully scalable laser drive network for a multi-stage accelerator. The main areas of development needed for the laser technology development are timing accuracy and distribution (combined with phase sensing and feedback at the point light is coupled to the electron beam), and beam transport and coupling of the laser to the accelerator structure. Consequently, the key areas of technology development required are: (1) development and demonstration of scalable techniques for sub-cycle phase-locking of multiple fiber lasers, and (2) numerical design, fabrication, and benchtop testing of ultracompact delivery optics. 

As the acceleration process of DLA is linear with the electric field, the optical phase must be well controlled. Poor synchronization would result in either a decrease of efficiency or an electron energy spreading or even defocusing. Frequency comb technologies can detect and control both the repetition rate of the delivered pulses and the carrier to envelop phase (CEP). Frequency comb techniques can be used to precisely measure and fine-tune repetition rate and CEP of delivered pulses. Because the electrons in a dielectric laser accelerator will be optically compressed to form microbunches less than one optical cycle in duration and separated by a single laser wavelength, this requires sub-cycle stabilization of the absolute frequency and relative CEP of each pulse. 

\subsection{Beam Delivery System }

Beam delivery for a DLA based linac could take advantage of the same techniques developed for modern colliders, including conventional collimation systems and magnetic optics at the final focus. The primary purpose of the collimation system is to protect the detector from background due to the particle spray from beam halo that intercept the apertures of the accelerator. However, due to the extremely low-charge, low-emittance bunches that would be produced and the narrow beam apertures, a DLA electron beam would already be inherently collimated to less than a micron in size. Consequently the need for an additional long collimation region prior to the IP may be significantly mitigated. In the parameter table of Fig. \ref{parameters} we assume a collimation length of 1.92 km prior to the final focus section. This section could simultaneously employ dispersive elements to wash out the microbunches so that the collided beams at the IP correspond to ps duration bunches with the combined charge of the bunch train. This allows for a luminosity enhancement of approximately 10 due to the pinching effect at the IP from the beam-beam interaction. Since the gross transverse dimensions of a DLA accelerator would be extremely small (of order 1 mm), particle spray from halo could also potentially be intercepted or angularly dispersed by shielding around the linac itself. The drift length $L^*$ to the detector is difficult to estimate absent details of the detector design. In Fig. \ref{parameters} we assume a drift $L^*$ = 5 m, similar to ILC values.

\section{Integrated system}

\subsection{Tolerances}
The Dielectric Laser Accelerator Workshop held at SLAC National Accelerator Laboratory in 2011 examined required tolerances for a DLA based collider \cite{dla:2011}. Since the emittance must be preserved through several kilometers of acceleration, misalignments must be small enough that they do not result in significant emittance growth. Conventional magnetic focusing would require tolerances of order 1 $\mu$m in quadrupole magnet positioning, 100nm in the accelerator structure alignment, and quadrupole jitter of less than 0.1 nm. This was based on requiring a maximum centroid motion of 10\% of the beam size from magnetic center vibration, assuming 1000 quads and a normalized transverse emittance of 0.1 nm. However, proposed electromagnetic focusing schemes which are now being incorporated into structure designs and experiments, such as alternating phase focusing and nonresonant harmonic focusing \cite{naranjo_stable_2012,niedermayer:focusing:2018}, can be built into the DLA structure design with nanometric precision that should well exceed such tolerances. A preliminary study of beam breakup instability (BBU) using a simple two-particle model found that a 30 nm average misalignment resulted in a transverse normalized emittance growth of 2.2 nm from a cold beam over 500 GeV of acceleration in 1 kilometer. A scan of emittance growth vs. bunch charge was conducted, and it was found that accelerating sufficient charge with tolerable beam degradation for high-energy physics applications requires about 50 nm alignment. Beam stability may be improved by using a shorter focusing period or by use of Balakin-Novokhatsky-Smirnov (BNS) damping. While achieving such tolerances over several kilometers is challenging, the high repetition rate of a DLA collider provides imformation at MHz frequencies, which can be used for feedback stabilization. Stabilization of optical components to better than 1 nm Hz$^{-1/2}$ has already been demonstrated over similar length scales at the LIGO facility \cite{ligo:stability}. Furthermore, since the acceleration process of DLA is linear with the electric field, the optical phase must be well controlled. Frequency comb technologies can detect and control both the repetition rate of the delivered pulses and the carrier to envelope phase (CEP). The technology used to generate frequency combs in ultra-high finesse Fabry Perot cavities is able to control phase noise in the range of 0.01 Hz to 100 kHz. Further stabilization will necessitate control systems operating above 100 kHz and requires important efforts in feedback loop electronics and ultrafast low-noise detectors. 

\subsection{Instrumentation}

A future DLA-based linear collider will require the development suitable diagnostics and beam manipulation techniques, including compatible small-footprint deflectors, focusing elements, and beam position monitors (BPMs), concepts for which have been proposed \cite{plettner:2008,plettner:2011,soong:2012,naranjo:2012,soong:2012b}.  A dielectric laser-driven element that produces transverse deflection forces in both transverse dimensions as well as a longitudinal accelerating force was proposed by Plettner and Byer \cite{plettner:2008}. The concept is similar to the planar symmetric grating accelerator but with the gratings tilted at an angle relative to the particle beam axis. By exciting a superposition of TE and TM modes in such a deflector with a tilt angle of 45$^\circ$, it has been recently shown that a pure deflection mode can be excited (all other force components cancel) which could be used to make nanoscale orbit corrections, to make a laser driven undulator, or to steer the electron beam between successive chips or wafers \cite{englandAAC:2018}. 

A proof-of-principle demonstration of the BPM concept was recently conducted  \cite{soong:2014}.  The concept uses a dual-grating with a tapered grating period to produce a linear variation in operating wavelength along the dimension transverse to the electron beam axis.  Light emitted by wakefield excitation of the electron beam (via the inverse of the acceleration process) has a different center wavelength depending on transverse position of the electrons, permitting a measurement of beam position from the power spectrum of emitted light.  Similarly the spectral width of the emitted radiation is a measure of the beam size, permitting a simultaneous determination of beam position and transverse size.  The direction of the variation is chosen to correspond to either horizontal or vertical offset of the particle beam. When a particle beam traverses this BPM structure, it generates radiation at a wavelength corresponding to the grating period at the beam location. Combined with a DLA-generated sub-micron sized electron beam and a typical 0.1 nm resolution spectrograph, it is estimated that this technique could be employed to resolve beam position in a DLA with 0.75 nm precision. 

\subsection{Simulation}
\label{sec:simulation}
{\hskip 0.13in}
The DLA scheme is analogous in many respects to a frequency-scaled version of a conventional accelerator.  Consequently, modelling many aspects of the particle dynamics and transport in a DLA collider can rely largely upon well established accelerator codes and computational methods.  This reduces the need for new code developments or large-scale computing infrastructure to a subset of DLA simulation tasks.  For simulation purposes, a linear collider can be broken into three regimes which require different kinds of modeling:  (1) Injector (source, emittance preparation, and acceleration up to 1 GeV); (2) Accelerator (1 GeV to 30 TeV) and (3) beam delivery, including final focus, crab kissing, IP design, and beamstrahlung.  The proposed injection scheme outlined in Section \ref{sec:parameters} is based on superconducting RF technology, and thus is amenable to the same computational methods used to model such systems and will therefore not be addressed here. The main accelerator needs longitudinally coupled structures (traveling wave structures) for energy efficiency and mode stability reasons as described in Section \ref{sec:structures}. The main accelerator can then be split hierarchically into separately addressed sections corresponding to their respective length scales:
\begin{enumerate}
\item One Optical Period (2$\mu$m)
\item Power coupling cell (10 cm)
\item Focusing cell (20 m at 3 TeV)
\item Betatron period (100 m at 3 TeV)
\item Bunch compression stage (1 km) 
\item Full linac (4.2 km each side at 3 TeV)
\end{enumerate}

An individual DLA cell (one optical period in length) can be straightforwardly simulated by any of a variety of FDTD, FDFD, and FEFD codes combined with various optimization techniques \cite{shin:2013,egenolf:2017,hughes:avm:2017}. The resulting single-cell Fourier coefficients, combined with the corresponding phase and group velocities, can then be combined with 6D tracking codes to obtain long-distance phase space evolutions by symplectic one-kick-per-cell tracking \cite{niedermayer:2017}.  Advanced codes and large-scale computing become necessary for the integration of particle trajectories through longer structure segments.  Compared to conventional accelerators, DLAs have much smaller feature sizes, which must be resolved in simulations, leading to much more demanding computations.  Particle-in-cell (PIC) simulations can be used to model full particle and field dynamics up to a few thousand or tens of thousands of structure periods but becomes computationally prohibitive at longer length scales.  At optical-scale frequencies, a 10 cm interaction distance would be near the limit of what is possible with a modern supercomputer.  A process of using simplified models to construct larger building blocks can be applied to successive levels of the design, using transfer maps to represent larger-scale components such as an entire power coupling cell or focusing cell. 

Moreover, wake functions and beam coupling impedances must be included in the various structures, which can be done both in full 3D or by simplified 2D models in the frequency domain. The most crucial issues here are to find the beam loading and beam break up limits using the longitudinal and transverse wakes. For the extremely short sub-optical bunches employed in a DLA scenario (with bunch length small compared to transverse size) the theory of beam instabilities might require extension by nonlinear parts of the transverse wakes and transverse position dependent parts of the longitudinal wakes. Such theories have to be validated by extensive PIC simulations. Moreover, the consequences of slight steering errors in the 10-nanometer range, leading to severe average beam power deposition in the structures needs to be studied. Already available radiation damage codes such as Fluka can be applied here.  While much of the above simulation work can be accomplished by use of existing codes and/or well-established computational techniques, the beam dynamics in the final focus of a DLA collider needs to be completely redeveloped for DLA-type beams, unless, as assumed in Section \ref{parameters}, the microbunching is washed out prior to interaction at the IP.  Direct collision of trains of extremely short, low intensity bunches with high repetition rate would behave very differently than conventional bunch crossings when it comes to beamstrahlung or crab kissing.  Fast beam-beam interaction codes are available, but have yet to be adapted to attosecond bunches.

\section{Partners and resources, current activity level. what is needed? What do we support?}
\label{sec:partners}

Progress towards an energy scalable architecture based upon laser acceleration in dielectric materials requires an R\&D focus on fabrication and structure evaluation to optimize existing and proposed concepts, and development of low-charge high-repetition-rate particle sources that can be used to demonstrate performance over many stages of acceleration. To tackle these challenges, a concerted effort is required that leverages industrial fabrication capabilities and that draws upon world-class expertise in multiple areas. A number of university, national laboratory, and industrial institutions and collaborations are now actively conducting research in this area, including the multi-institutional Accelerator on a Chip International Program (ACHIP), which includes 6 universities, 1 company, and 3 national laboratories, as well as Los Alamos National Laboratory, University of Tokyo, Tel-Aviv University, The Technion, and University of Liverpool.  The DLA effort could be made technology-limited through appropriate leveraging of the semiconductor and laser R\&D industries and appropriate growth of the DLA research community.  Current collaborative efforts in the U.S. and Europe are aimed at developing a first R\&D demonstration system incorporating multiple stages of acceleration, efficient guided wave systems, high repetition rate solid state laser systems, and component integration by the year 2020. 

Testing of speed-of-light prototype devices and initial staging experiments will require suitable test facilities equipped with relativistic beams. Existing conventional RF facilities are suitable for near-term tests over the next few years.  To provide an estimate of projected facilities costs, the current ACHIP program includes roughly \$1.3M/year of combined in-kind support from three national laboratories (SLAC, DESY, and PSI) in the form of access to personnel, resources, and beam time. However, for demonstrating many-staged DLA accelerators, ultra-low emittance particle sources need to be developed and combined with DLA devices to make compatible injectors. Ongoing development of DLA prototype integrated systems will provide a pathway for scaling of this technology to MeV, GeV, and then TeV energies and to beam brightnesses of interest both for high energy physics and for a host of other applications, as discussed in Ref. \cite{england:review:2016}.



\bibliographystyle{unsrt}
\bibliography{alegro}


%% file: WG8Positrons.tex
%
%
%
%
%
%
\section{Introduction}
Laser Wakefield Acceleration (LWFA) and Plasma Wakefield Acceleration (PWFA) are attractive methods for accelerating particle beams to the highest possible energies because of the large accelerating gradients that are achieved in the plasma. This has led to a number of proposals for Linear Colliders using plasma technology, where the colliding beams are composed of electrons and positrons. However, there has been relatively little research on how to accelerate positron beams in plasma, in part because there is only one facility world-wide that provides positron beams for PWFA research. In addition, the acceleration of positrons in a plasma wakefield is fundamentally different than that of electrons. This is because the plasma is composed of light, mobile electrons, and heavy, immobile ions, leading to an asymmetric response of the plasma to beams of opposite charge. Despite some effort, no one has proposed a self-consistent scheme for accelerating positrons in plasma while preserving the quality of the beam.

Here, we review the existing approaches for accelerating positrons in plasma that are currently under theoretical and experimental investigation. We study proposals for novel sources of positron beams. Finally, we identify future facilities that should provide positron beams for plasma acceleration research.

\section{Proposed Paths}
There are four regimes that are currently being investigated for accelerating positrons in plasma: the quasi-linear regime, nonlinear regime, hollow channel regime, and wake-inversion regime. Each of these regimes have attractive attributes for accelerating positrons in plasma, but they each have unique challenges that must be overcome in order to demonstrate their feasibility.

\subsection{The Quasi-Linear Regime}
Despite the assertion in the introduction that plasmas respond asymmetrically to beams of opposite charge, this asymmetry can be avoided by operating in the quasi-linear regime. In the quasi-linear regime, the beam density is smaller than the background plasma density. The low-density beam generates a longitudinally sinusoidal wakefield as it propagates through the plasma, and this sinusoidal wake has accelerating phases for both electron and positron witness beams. The acceleration of a positron beam in a quasi-linear wake was demonstrated at FACET in 2017~\cite{DochePosi}.

The challenge associated with the quasi-linear regime is that the witness beam (and drive beam in the case of PWFA) must maintain a low density as it propagates through the plasma. However, the plasma wakefield produces a strong focusing force that will squeeze the beam until the beam density is greater than the plasma density. The evolution of the beam size is called a betatron oscillation, and it can be avoided by matching the beam into the plasma. The matched beam size is determined by the plasma density, beam emittance, and beam current. The beam density is equal to the beam current divided by the matched spot size squared. This completely constrains the beam emittance, yielding a value that is intolerably large for a linear collider~\cite{DochePosi,WAn}.

Matching can be made possible at very low beam emittance by tuning the focusing/defocusing using high-order laser modes~\cite{Cormier}. This allows a low-emittance beam to propagate in the plasma wakefield without self-focusing. However, this implies a very large betafunction and emittance growth due to scattering.

Another way around this issue is to use multi-bunch, multi-pulse acceleration, as recently demonstrated by Cowley et. al.~\cite{Cowley}, or as is proposed to be done at the AWAKE facility~\cite{AWAKE}. In the multi-bunch regime, each beam carries only a fraction of the charge and therefore the beam emittance can be reduced by the same fraction. This approach requires careful consideration of collider luminostiy and efficiency, as both of these numbers decrease with decreasing bunch charge.

\subsection{The Nonlinear Regime}
The nonlinear regime is wholly asymmetric. Whereas a high-intensity laser or electron beam will create a plasma ``bubble'' with a uniform background of plasma ions, a dense positron beam ``sucks in'' the plasma electrons, leading to a highly-nonuniform plasma distribution within the volume of the beam. The nonuniform plasma distribution is accompanied by strong, nonlinear focusing forces, leading to halo formation and emittance growth~\cite{MuggliPosi}.

However, the nonlinear regime can also provide strong and uniform accelerating gradients, as demonstrated in a recent experiment at FACET~\cite{CordePosi}. In this case, the wakefield was generated by a positron drive beam, and the shape of the wakefield is strongly determined by the amount of positron charge in the tail of the bunch~\cite{DochePosi}. The evolution of the nonlinear positron beam-driven wakefield is not well understood. This topic requires a significant investment in simulations and analytic theory in order to gain a more complete understanding of a potentially attractive approach to a plasma afterburner.

Similarly, the topic of positron acceleration in a highly nonlinear wakefield driven by a laser or electron beam is also not well understood. So far, simulation studies have identified a vanishingly-small region where positrons are both focused and accelerated~\cite{LotovPosi}. These studies did not consider the effect of beam loading of the positron witness beam. Therefore, more work is needed on this topic as well. We note that there have been no experiments to date on electron beam-driven or laser-driven wakefield acceleration of positrons.

\subsubsection{Wake-Inversion}
Recently, a novel technique for producing accelerating and uniformly focusing plasma wakefields has been proposed. In this concept, a donut-shaped driver forces plasma electrons onto the axis in a controlled fashion. The driver can be a hollow laser pulse~\cite{VieiraHollow}, a hollow electron beam~\cite{JainHollow}, or the superposition of electron and positron beams~\cite{VieiraALEGRO}.

At first glance, it appears challenging to propagate ring shaped structures through a plasma while preserving azimuthal symmetry. However, some simulations have indicate that the plasma may provide a symmetrizing effect that suppresses initial inhomogeneities~\cite{VieiraALEGRO}. Further simulations are needed to establish the robustness of this proposed technique.

\subsection{The Hollow Channel Regime}
Similar to the quasi-linear regime, the hollow channel regime is a concept that symmetrizes the response of the plasma to beams of opposite charge. Unlike the quasi-linear regime, there are no constraints on the beam charge in the hollow channel because there is no on-axis plasma. This has the advantage of providing large amplitude wakefields without sacrificing beam emittance. In addition, there is no contribution to emittance growth due to scattering off of plasma particles. The hollow channel also has wave-guide like features~\cite{ChiouHollow}, which is attractive for an LWFA driven PLC.

The main drawbacks of the hollow channel approach are that they are difficult to create and the transverse wakefields are very strong, which leads to beam breakup~\cite{SchroedMulti}. Recently, the transverse wakefields have been measured in the hollow channel plasma and they are indeed quite strong~\cite{LindstromTrans}.

Future work on hollow channel plasma wakefield accelerators should focus on two critical aspects. First, they should demonstrate ``true'' hollow channels, meaning on-axis vacuum, unlike previously demonstrated hollow channels which have on-axis neutral vapor~\cite{GessnerHollow}. Possible candidates for hollow channels with an on-axis vacuum include centrifugal plasmas or cryo-cooled gases that flow around an obstacle. The second critical aspect is the suppression of the BBU instability. Possible candidates include using strong permanent magnet quadrupoles to provide external focusing, or tailoring the shape of the plasma channel to suppress transverse wakefields with respect to accelerating fields.

\section{Novel Sources of Positron Beams}
Positron beams are generated by smashing high-energy beams of electrons or photons into a high-Z target. This produces positron beams with large energy-spread and large-emittance. The beams must be cooled in a damping ring before they can be used in positron acceleration experiments. This process is inefficient and requires large, expensive infrastructure. The accelerator community as a whole would benefit from novel, compact, efficient sources of positron beams.

\subsection{Positron Beams from Laser Driven Sources}

Recent experimental results on laser-based generation of high-quality ultra-relativistic positrons are suggesting an alternative pathway towards positron beam production. For instance, fs-scale and narrow divergence positron beams in a plasma-based configuration have been recently reported~\cite{Sarri2013,Sarri2015}. In a nutshell, the positrons are generated as a result of a quantum cascade initiated by a laser-driven electron beam propagating through a high-Z solid target. For sufficiently high electron energy and thin converter targets, the generated positrons present properties that resemble those of the parent electron beam, hence the fs-scale duration, mrad-scale divergence, and micron-scale source size. The maximum positron energy attainable in this scheme is naturally dictated by the peak energy of the parent electron beam. Positrons with energy up to 0.5 GeV were produced in recent experiments. These beams have durations comparable to the positron-accelerating region of a wakefield and are naturally synchronised with a high-power laser. However, other characteristics still need to be carefully optimised, such as their non-negligible normalised emittance and the relatively low charge. 
In order to optimise these sources, it is desirable to increase the charge in the positron beam and minimise its emittance. Both goals can be achieved if the charge and maximum energy of the primary electron beam is increased, together with dedicated studies of beam manipulation and transport. In this respect, the technological and scientific developments in laser technology and electron acceleration will naturally benefit also the generation of such positron beams, a necessary ingredient towards the high-energy applications of laser-wakefield electrons and positrons.

\subsection{Positron Beams from Ultra-Cold Traps}

Electrostatic traps are the critical component of modern, low-energy, antimatter experiments. These devices can store cold positrons, antiprotons, and recently anti-hydrogen for an indefinite period of time~\cite{Alpha}. Since the trapped particles have temperatures near absolute zero, when considered as a beam ensemble, their emittance is extremely small. However, the beams are also centimeters-long, low-charge, and non-relativistic. A team from UC San Diego has studied the injection and longitudinal compression of positron beams from the trap~\cite{Weber}. If the electrostatic compression can be coupled with RF bunching, it may be possible to generate short, low emittance positron bunches from a compact source. While it is not clear that this approach can ultimately by used for a linear collider, it may still be useful for studying positron acceleration in plasma. An electromagnetic design study is need to establish the viability of this concept.

\section{Facilities Providing Positrons for PWFA/LWFA Research}
Research on positron acceleration in plasma is hamstrung by a dearth of facilities that can provide short, intense positron bunches for plasma acceleration experiments. Positron beams are generated by smashing high-energy beams of electrons or photons into a high-Z target, capturing the large energy-spread, large emittance positrons that come out, returning the positrons to a damping ring, and finally cooling and extracting the positron beam from the ring. This process is expensive and requires a large amount of infrastructure. For this reason, SLAC has been the only laboratory to provide positrons for PWFA experiments, since it as already equipped with a positron source for the SLC. 

\subsection{FACET-II}
FACET-II at SLAC is the only currently-planned facility aiming to deliver positron beams for plasma acceleration experiments. FACET-II uses some of the existing SLAC infrastructure for positron creation, but adds a new compact damping ring and a ``sailboat chicane'' for electron beam-driven positron acceleration experiments~\cite{FACETIITDR}. We note that the positron component of FACET-II is unfunded.

\subsection{Eupraxia}
Eupraxia is a proposed facility with the potential to be built at SPARC\_Lab at LNF. Eupraxia will be equipped with both a linear accelerator and multiple high-power laser systems. The availability of a high-power laser for generating positron beams means that these positrons can be subsequently by injected into a laser-driven or beam-driven wakefield.

\subsection{e-SPS}
e-SPS is a proposed facility at CERN which will use a 3.4 GeV x-band linac to deliver electrons to the SPS ring. However, because the SPS ring has a fixed cycle time, the duty factor of the x-band linac will be quite low. Therefore, e-SPS is soliciting ideas as to what can be done with the electron beam when it is not being sent to the SPS. We propose a facility capable of providing positrons for plasma acceleration studies~\cite{eSPStalk}.

\section{Contributions to this Document}

The Positron Acceleration in Plasma Working Group (PAC-WG8) received contributions from the ALEGRO community on a wide variety of topics:
\begin{itemize}
\item History of Positron Acceleration in Plasma: P. Muggli
\item Quasi-Linear Regime: W. An
\item Hollow Channel Regime: C. Lindstrom, Y. Li, A. Pukhov
\item Nonlinear Regime: K. Lotov
\item Wake-Inversion Regime: J. Viera
\item Novel Positron Sources: R. Greaves, A. Petrenko, G. Sarri, C. Surko
\item Test Facilities for Positron Acceleration in Plasma: M. Hogan, R. Walczak
\end{itemize}
We thank the community for their enthusiastic engagement on this challenging topic.
